\documentstyle{article}
\input{epsf.sty}
\font\tenbg=cmmib10 at 10pt

\def \rvecphi{{\hbox{\tenbg\char'036}}}
\def \rvecdelta {{\hbox {\tenbg\char'016}}}

\begin{document}

\title{Advective Accretion Disks and
Related Problems Including Magnetic Fields}
%: Recent Developments}

\author{G.S.Bisnovatyi-Kogan$^*$ and R.V.E. Lovelace$^+$}
\thanks{Space Research Institute, Profsoyuznaya
84/32, Moscow 117810, Russia;  GKogan@mx.iki.rssi.ru}
\thanks{
 Department of Astronomy,
Cornell University, Ithaca, NY 14853;
RVL1@cornell.edu}

\maketitle

\begin{abstract}
Accretion disk theory was
 first developed as a theory with
the local heat balance, where the
 whole energy produced by a viscous
heating was emitted to the sides of the disk.
 One of the most important
new invention of this theory
 was the phenomenological treatment of the
turbulent viscosity, known the `` alpha''
 prescription, where the $(r \phi)$
component of the stress tensor
 was approximated by
 $(\alpha p)$ with a unknown
constant $\alpha$.
 This prescription played
 the role in the accretion
disk theory as well important
as the mixing-length theory of convection
for stellar evolution.
Sources of turbulence in the accretion disk are
discussed, including
hydrodynamical turbulence, convection
and magnetic field role.
In parallel to the optically thick
geometrically thin accretion
disk models, a new branch of the optically
thin accretion disk models was
discovered, with a larger thickness
for the same total luminosity.
The choice between these solutions
should be done of the base
of a stability analysis.
   The ideas underlying the
necessity to include advection into the
accretion disk theory are presented
and first models with advection
are reviewed.
     The present status of
the solution for a low-luminous
optically thin accretion disk
model with advection is discussed and
the limits for an advection
dominated accretion flows
(ADAF) imposed by the presence
of magnetic field are analysed.
Related problems of mass ejection from accretion disks and jet formation
are discussed.
\end{abstract}

\section{Introduction} %%%%%%%%%%%%%%%%%%%%%%%%

    Accretion is served as a source
    of energy in many astrophysical
objects, including different types
of binary stars, binary X-ray sources,
most probably quasars and
active galactic nuclei (AGN).
      While first developement
of accretion theory started
long time ago (Bondi and Hoyle, 1944;
Bondi, 1952), the intensive
developement of this theory
began after discovery of first X ray sources
(Giacconi et al, 1962) and
quasars (Schmidt, 1963).
    Accretion onto stars,
including neutron stars,
terminates at an inner boundary.
   This may be the stellar surface,
or the outer boundary of a magnetosphere
for strongly magnetized stars.
   We may be sure in this case, that all
gravitational energy of the
falling matter will be transformed into heat
and radiated outward.

   The situation is quite different
for sources containing black holes, which are
discovered in some binary X-ray
sources in the galaxy, as well as in many
AGN. Here matter is falling
to the horizon, from where
no radiation arrives,
so all luminocity is formed
on the way to it.
     The efficiency of accretion
is not known from the beginning,
contrary to the accretion into a star,
and depends strongly on such
factors, like angular momentum of the
falling matter, and magnetic field
embedded into it. It was first
shown by Schwarzman (1971),
that during spherical accretion of
nonmagnetized gas the efficiency
may be as small as $10^{-8}$ for
sufficiently low mass fluxes.
He showed that presence of magnetic field in
the accretion flux matter increase
the efficiency up to about $10\%$, and
account of heating of matter due
to magnetic field annihilation in
the flux rises the efficiency up
to about $30\%$ (Bisnovatyi-Kogan,
Ruzmaikin, 1974; Mesz\'aros, 1975).
In the case of a thin disk
accretion when matter has
large angular momentum, the
efficiency is about $1/2$ of the efficiency
of accretion
into a star with a radius
equal to the radius of the last stable orbit.
    Matter cannot emit all the
    gravitational energy, part of which is
absorbed by the black hole.
In the case of geometrically thick and
optically thin accretion discs
the situation is approaching the case
of spherical symmetry, and a
presence of a magnetic field playes also
a critical role.

    Here we consider a developement
of the theory of a disk accretion,
starting from creation of a
so called ``standard model'',
and discuss recent trends,
connected with a presence of advection.
    The flow of papers
    (see e.g. references in Narayan, Barret,
McClintock, 1996; Menon, Quataert, Narayan, 1999)
considering advection dominated accretion flow
(ADAF) as a solution for many
astrophysical problems should be treated with
some caution, because of its
vague physical background.
The suggestions underlying ADAF:
ignorance of the magnetic field annihilation
in heating of accretion plasma flow,
and electron heating ONLY due to
binary collisions with protons (ions)
were critically analysed in several
papers (Bisnovatyi-Kogan
and Lovelace, 1997, 2000; Quataert, 1997,
Blackman, 1999), and the above
mentioned suggestions have been strongly
doubted.

Competition between
rapid accumulation of observational data, mainly
from Hubble Space Telescope,
X-ray satellites, and  rapid as
well construction of theoretilal models
for their explanation creates
sometimes a sutuation when the model dyes
during the time of its publication.
Some ADAF models give a good examples
of this situation. One is connected
with an explanation of the unusual
spectra of the galaxy  NGC 4258,
with a nonmonotoneous curve.
     It was
claimed (Lasota et al., 1996),
that such spectrum may be explained only
by ADAF model. Observations
(Herrnstein et al., 1998; Cannizzo et al., 1998) gave data in radio,
in optical/UV region showing existence
of spectral maximum on the place
where suspected minimum had been explained by ADAF, and posing
constraints on ADAF model. Restrictions on ADAF model had been
obtained from radio observations of elliptical galaxies (Wrobal
and Herrnstein, 2000).

   Another example is connected with
an attempt to prove the existence of
the event horison in black holes by ADAF model.
   The figure 7
from the work of Menon, Quataert
and Narayan (1997) (fig.2 from
Narayan, Garcia and McClintock, 1997)
was considered as a proof of the
existence of the event horizon of
black holes and at the same time
as a triumph of the ADAF model.
    Regretfully, the data on this figures
occures to be incomplete and a full
set of observational data
altered this picture (Chen et al., 1998).

It is, of course, dangerous to justify
any physical model by astronomical
observations without a firm physical ground.
    Here we analyse
physical processes in an optically
thin accretion flow at low accretion
rates, connected with a
presence of a small-scale magnetic field.
      We show, that accurate
account of these processes strongly
restrict the boundaries of the ADAF
solution. Namely, the efficiency of
the accretion flow cannot become less
then about 1/3 of the standard
accretion disk value.
      It makes senseless
the attempts to connect ADAF
models with existence of the
event horizon of a black hole,
or to explain why we do not see
a large luminosity from supermassive
black holes in the
nearby galactic nuclei.

It seems now, that explanation of unusual observational features,
in particular, of "underluminous galactic nuclei" is connected
with more complicated processes than a simple ADAF solution. Energy
losses connected with magneto-hydrodynamical acceleration of matter
from accretion disk, formation of jets may be very important.
These problems are considered in the last section of the review.

\section{Foundation of accretion disk theory}

\subsection{Development of the standard
model of the disk accretion}   %%%%%%%%%%%%%%%%%%%%%%%%

Matter falling into a compact object
tends to form a disk when its
angular momentum is sufficiently high.
      It happens when the matter falling
into a black hole comes from the neighboring
ordinary star companion in the binary,
or when the matter appears as a
result of a tidal disruption of
the star which trajectory of motion
approaches sufficiently close to
the black hole, so that forces of
selfgravity could be overcomed.
    The first situation is observed in
many galactic X-ray sources
containing a stellar mass black hole
(Cherepashchuk, 1996, 2000).
    A tidal disruption happens in quasars and
active galactic nuclei (AGN),
if the model of supermassive black
 hole surrounded by a dense stellar cluster
of Lynden-Bell (1969) is
true for these objects.

     The models of the accretion disk
     structure around a black hole had
been investigated by Lynden-Bell (1969),
Pringle and Rees (1972).
The modern "standard" theory of
the disk accretion was formulated in
the papers of Shakura (1972),
Novikov and Thorne (1973) and
Shakura and Sunyaev (1973).
     It is important to note, that all authors
of the accretion disk theory
from USSR were students (N.I.Shakura)
or collaborators (I.D.Novikov
and R.A.Sunyaev) of academician
Ya.B.Zeldovich, who was not
among the authors, but whose influence
on them hardly could be overestimated.

The equations of the standard disk
accretion theory were first formulated
by Shakura (1972); some corrections
and generalization to general
relativity (GR) were done by
Novikov and Thorne (1973), see also correction
to their equations in GR made by
Riffert \& Herold (1995).
The main idea of this theory
is to describe a
geometrically thin non-self-gravitating
disk of the mass $M_d$, which is much
smaller than the mass of the
black hole $M$, by hydrodynamic
equations averaged over
the disk thickness $2h$ Fig.1 shows the disk geometry.

%\begin{figure}
%\epsfysize=8cm % fix the y-dimension and scales x-dim. to y-dim.
%\hspace{3.5cm}\epsfbox{grfig0.eps} %for centering: act on hspace argument

%\label{fig0}
%\end{figure}

\subsection{Basic Equations}%%%%%%%%%%%%%%%%%%%%%

%   The equations of the thin disk
%accretion are derived from the
%hydrodynamical set of equations and their
%analytical solutions are presented
%for polytropic accretion disks
%with boundary layers near the star.
%The total angular momentum flux
%over the accretion disk and the
%thickness of the boundary layer are
%obtained self-consistently
%in these solutions.
Let us start
with Navier-Stokes
equations in Cartesian
coordinates, describing the motion
of the viscid compressible fluid
with variable viscosity $\eta$ and
bulk viscosity $\eta_B$
coefficients (Chapman and Cowling 1939)
%%%%%%%%%%%%%%%%%%%%%%%%%%%%%%%%%%%%%%%% eqn 1
\begin{eqnarray}
\rho \left( {\partial u_i \over \partial t}+
u_k{\partial u_i \over \partial x_k}  \right)\!\!&=&\!\!
-{\partial p \over \partial x_i}+\rho g_i
\nonumber\\
\!\!&+&\!\!{\partial \over \partial x_k}
{\biggl[ \eta \left ( {\partial u_i \over
\partial x_k} + {\partial u_k \over \partial x_i}-{2 \over 3}
{\partial u_l \over \partial x_l} \delta_{ik} \right)
+ \eta_B{\partial u_l \over \partial x_l} \delta_{ik}
\biggr]},
\label{F1}
\end{eqnarray}
or in vector notation,
%%%%%%%%%%%%%%%%%%%%%%%%%%%%%%%%%%%%%%%%%% eqn 2
\begin{eqnarray}
\rho {\left( {\partial {\bf u} \over \partial t} +
({\bf u \cdot \nabla })
{\bf u} \right)}\!\!&=&\!\!
-{\bf \nabla} p +\rho {\bf g}
\nonumber \\
\!\!&+&\!\! {\bf \nabla}_k
 \biggl[\eta \left(
{\bf \nabla}_k
{\bf u}+{\bf \nabla}{\bf u}_k -
{2 \over 3}({\bf \nabla \cdot u}) {\underline{\rvecdelta}}
 \right)
+\eta_B({\bf \nabla \cdot u}) {\underline{\rvecdelta}}
 \biggl].
\label{F2}
\end{eqnarray}
Here, the usual tensor definitions
are used with $\delta_{ik}$ and
$\underline{\rvecdelta}$ the unit tensor and
$g_i$ the gravitational accelleration.
   Summation over repeated indices is
assumed.
The last term in (\ref{F1}) is the divergence of the
viscosity stress tensor $\sigma_{ik}$,
which is related to the
symmetric and divergentloss deformation
tensor $ \tau_{ik} $ by the relation
%%%%%%%%%%%%%%%%%%%%%%%%%%%%%%%%%%%%%%%%%%% eqn 3
\begin{equation}
 \sigma_{ik}=2\eta \tau_{ik}
+ \eta_B{\partial u_l \over \partial x_l} \delta_{ik}~,
\qquad \tau_{ik}=
   {1 \over 2} \left({\partial u_i \over \partial x_k} + {\partial
   u_k \over \partial x_i} - {2 \over 3}{\partial u_l \over \partial
   x_l} \delta_{ik} \right)~.
\label{F3}
\end{equation}
   The bulk viscosity $\eta_B$ is connected with
finite rate of relaxation processes
(ionization, chemical or nuclear
reactions, etc). It is important only for
 rapid processes (e.g., shocks)
which are not considered here.
    Thus we take
$\eta_B=0$ everywhere below.

  For accretion flows,
it is natural to express equation ({\ref{F1})
in cylindrical
coordinates $(r,\phi,z)$.
     For making transformation to cylindrical
coordinates we need to write in
these coordinates some differendial
operators on the scalar $ a $,
vector ${\bf a}=(a_r, a_{\phi}, a_z) $
and tensor
%%%%%%%%%%%%%%%%%%%%%%%%%%% eqn 4
\begin{equation}
{\bf \underline a}=\left(\matrix{a_{rr}&a_{r \phi}&a_{rz}\cr
                a_{\phi r}& a_{\phi \phi}& a_{\phi z}\cr
                a_{zr}& a_{z \phi}& a_{zz}\cr}\right)~.
\label{F4}
\end{equation}
 \noindent{a) gradient of the scalar:}
%%%%%%%%%%%%%%%%%%%%%%%%% eqn 5
\begin{equation}
 {\bf \nabla} a = \left( {\partial a \over \partial r},
\qquad {1 \over r}
{\partial a \over \partial \phi}, \qquad
{\partial a \over \partial z} \right)~.
\label{F5}
\end{equation}
%%%%%%%%%%%%%%%%%%%%%%%%%% eqn 6
\noindent{b) divergence of vector:}
\begin{equation}
{\bf \nabla \cdot a}=
{1 \over r}{\partial(r a_r)\over \partial r}  +
{1 \over r}{\partial a_{\phi} \over \partial{\phi}} +
{\partial a_z \over \partial z}~.
\label{F6}
\end{equation}
%%%%%%%%%%%%%%%%%%%%%%%%% eqn 7
\noindent{c) curl of  vector:}
\begin{equation}
{\bf \nabla} \times {\bf a}=
\left( {1\over r}{\partial a_z
\over \partial \phi}-
{\partial a_{\phi} \over \partial z }, \qquad
{\partial a_r \over \partial z}-
{\partial a_z \over \partial r},
\qquad {\partial a_{\phi} \over \partial r} +
{a_{\phi} \over r} -
{1 \over r}{\partial a_r \over
\partial \phi} \right)~.
\label{F7}
\end{equation}
%%%%%%%%%%%%%%%%%%%%%%%% eqn 8
\noindent {d) gradient of vector:}
\begin{equation}
 {\bf \nabla a}= \left(\matrix{{\partial a_r \over
\partial r }& {\partial
 a_{\phi} \over \partial r}& {\partial a_z \over \partial r} \cr
 {1 \over r}{\partial a_r \over \partial \phi}-
{a_{\phi} \over r}&
 {1 \over r}{\partial a_{\phi} \over \partial \phi} +
{a_r \over r}&
 {1 \over r}{\partial a_z \over \partial \phi}  \cr
 {\partial a_r \over \partial z}& {\partial a_{\phi}
\over \partial z}&
 {\partial a_z \over \partial z} \cr}\right)~.
\label{F8}
\end{equation}
%%%%%%%%%%%%%%%%%%%%%%% eqn 9
\noindent {e) divergence of the tensor:}
\begin{equation}
{\bf \nabla \cdot \underline a}=
\left(\matrix
{{\partial a_{rr} \over \partial r}+
{1 \over r}{\partial a_{\phi r} \over
\partial \phi} +
{\partial a_{zr} \over \partial z} +{a_{rr}-a_{\phi \phi}
\over r}\cr {\partial a_{r \phi} \over \partial r}+
{1 \over r}{\partial
a_{\phi \phi} \over \partial \phi} +
{\partial a_{z \phi} \over
\partial z} +{a_{r \phi}+a_{\phi r} \over r} \cr
{\partial a_{rz} \over \partial r} +
{1 \over r}{\partial a_{\phi z} \over
\partial \phi} +
{\partial a_{zz} \over \partial z} + {a_{rz} \over r}
\cr}\right)~.
\label{F9}
\end{equation}
%%%%%%%%%%%%%%%%%%%%%% eqn 10
\noindent{f) Laplacian of the scalar:}
\begin{equation}
 \nabla^2  a \equiv
({\nabla \cdot \nabla}) a =
{\partial^2 a \over \partial r^2}
+{1 \over r}{\partial a \over \partial r} +
{1 \over r^2}{\partial^2 a
\over \partial \phi^2} +
{\partial^2 a \over \partial z^2} ~.
\label{F10}
\end{equation}
%%%%%%%%%%%%%%%%%%%%%%% eqn 11
\noindent{g) Laplacian of the vector:}
\begin{equation}
 \nabla^2 {\bf a} \equiv ({\bf \nabla \cdot \nabla}){\bf a} =
\left(\matrix{ {\partial^2 a_r \over \partial r^2} +
{1\over r}
{\partial a_r \over \partial r} -
{a_r \over r^2} + {1 \over r^2}
{\partial^2 a_r \over
\partial \phi^2} -
{2 \over r^2}{\partial a_{\phi} \over
\partial \phi} + {\partial^2 a_r \over \partial z^2} \cr
{\partial^2 a_{\phi} \over \partial r^2} +
{1 \over r}{\partial a_{\phi}
\over \partial r} - {a_{\phi} \over r^2} +
{1 \over r^2} {\partial^2
a_{\phi} \over \partial \phi^2} +
{2 \over r^2}{\partial a_r \over
\partial \phi} + {\partial^2 a_{\phi} \over \partial z^2} \cr
{\partial^2 a_z \over \partial r^2}+
{1 \over r}{\partial a_z \over
\partial r} +
{1 \over r^2}{\partial^2 a_z \over \partial \phi^2} +
{\partial^2 a_z \over \partial z^2} \cr} \right)~.
\label{F11}
\end{equation}

For the transformation of
equation (3) to cylindrical coordinates we have
%%%%%%%%%%%%%%%%%%%%%%%% eqn 12
\begin{equation}
 \tau_{ik}={1\over 2}\left({\bf \nabla}_k {\bf u}+
{\bf \nabla u}_k
\right) -{1 \over 3}({\bf \nabla \cdot u})
{\bf \underline{\rvecdelta}}~,
\label{F12}
\end{equation}
with
$$
2\tau_{ik}= \nonumber \\
\left(\matrix{{4 \over 3}{\partial u_r \over \partial r}-{2 \over 3}
\left( {u_r \over r}+
{1 \over r}{\partial u_{\phi} \over \partial \phi} +
{\partial u_z \over \partial z}\right)         &
{\partial u_{\phi} \over \partial r}+
{1 \over r}{\partial u_r \over
\partial \phi}-{u_{\phi} \over r}
                                               \cr
{\partial u_{\phi} \over \partial r}+
{1 \over r}{\partial u_r \over
\partial \phi}-{u_{\phi} \over r}              &
{4 \over 3} \left( {1 \over r}{\partial u_{\phi} \over \partial \phi}+
{u_r \over r} \right) -
{2 \over 3} \left( {\partial u_r \over \partial r}
+{\partial u_z \over \partial z}\right)
                                               \cr
{\partial u_z \over \partial r}+
{\partial u_r \over \partial z}                &
{1 \over r}{\partial u_z \over \partial \phi} +
{\partial u_{\phi} \over
\partial z}
                                               \cr}\right.
$$
%%%%%%%%%%%%%%%%%%%%%%%%% eqn 13
\begin{equation}
\left.\matrix{{\partial u_z \over \partial r}+
{\partial u_r \over \partial z}
\cr
{1 \over r}{\partial u_z \over \partial \phi} +
{\partial u_{\phi}\over
\partial z}
\cr
{4 \over 3}{\partial u_z \over \partial z}-
{2 \over 3}\left({\partial u_r
\over \partial r}+{1 \over r}
{\partial u_{\phi} \over \partial \phi}+
{u_r \over r}\right)    \cr}\right)
\label{F13}
\end{equation}
The second term in the brackets
on the left-hand side of equation (\ref{F2})
is a scalar product of the vector ${\bf u}$
and tensor ${\bf \nabla u}$,
which is the same in Cartesian and
curvilinear coordinates.
   The last term on the right-hand side of equation (\ref{F2})
is the divergence of the tensor $ 2 \eta \tau_{ik} $,
which is calculated using equations (\ref{F9}) and (\ref{F13}).
    As a result we get the components of  equation
(\ref{F2}) in cylindrical coordinates:
%%%%%%%%%%%%%%%% eqn 14
$$
\rho \left({\partial u_r \over \partial t}+
u_r{\partial u_r \over
\partial r} +{u_{\phi} \over r}
{\partial u_r \over \partial \phi}+
u_z{\partial u_r \over \partial z}-
{u_{\phi}^2 \over r}
 \right) =-{\partial p \over
\partial r}+\rho g_r
$$
$$
+{2 \over 3}{\partial \over \partial r}
\biggl\{\eta \biggl[{2\over r}{\partial (ru_r)
\over \partial r} -{1 \over r}{\partial u_{\phi}
\over \partial \phi}-
{\partial u_z \over \partial z} \biggr]\biggr\}-
2{u_r \over r}
{\partial \eta \over \partial r}+
{1 \over r}{\partial \over \partial \phi}
\biggl[\eta\left( {\partial u_{\phi} \over \partial r}+
{1 \over r}{\partial
u_r \over \partial \phi}-
{u_{\phi} \over r} \right) \biggr]
$$
\begin{equation}
-2{\eta \over
r^2}{\partial u_{\phi} \over \partial \phi}+
{\partial \over \partial z}
\biggl [\eta \left({\partial u_z \over \partial r}+
{\partial u_r \over \partial z}\right)\biggr]~.
\label{F14}
\end{equation}  %%%%%%%%%%%%%%%%%%%%%%%%%%%

$$   %%%%%%%%%% eqn 15
\rho \left({\partial u_{\phi} \over \partial t}+
u_r{\partial u_{\phi} \over
\partial r} +{u_{\phi} \over r}{\partial u_{\phi}
\over \partial \phi}+
u_z{\partial u_{\phi} \over \partial z}+
{u_r u_{\phi} \over r} \right)
=-{1 \over r} {\partial p \over
\partial \phi}+\rho g_\phi
$$
$$
+{1 \over r^2}{\partial \over \partial r} \biggl[\eta r^3 {\partial \over
\partial r}\left(u_{\phi} \over r \right) \biggr] +
{\partial \over
\partial r}\left({\eta \over r}{\partial u_r \over \partial \phi}\right)+
{2 \over 3r}{\partial \over \partial \phi}\biggl\{\eta \biggl[2
\left({1 \over r}{\partial u_{\phi} \over \partial \phi}+{u_r \over r}\right)
$$
\begin{equation}
-{\partial u_r \over \partial r}-{\partial u_z \over
\partial z} \biggl]\biggl\}
+{\partial \over \partial z}
+\biggl[\eta\left({1 \over r}{\partial u_z
\over \partial \phi}+{\partial u_{\phi}
\over \partial z}\right)\biggl]+
{2 \eta \over r^2}{\partial u_r \over \partial \phi} ~.
\label{F15}
\end{equation}   %%%%%%%%%%%%%%%%%%%%%%%%%%%%%%%%%%%

$$     %%%%%%%%%% eqn 16
\rho \left({\partial u_z \over \partial t}+
u_r{\partial u_z \over
\partial r} +{u_{\phi} \over r}{\partial u_z
\over \partial \phi}+
u_z{\partial u_z \over \partial z}
 \right) =-{\partial P \over
\partial z}+\rho g_z
$$
$$
+{1 \over r}{\partial \over \partial r}
\biggl[\eta r \left( {\partial u_z
\over \partial r}+{\partial u_r
\over \partial z}\right)\biggr] +
{1 \over r}{\partial \over \partial \phi}
\biggl[\eta\left(
 {1 \over r}{\partial u_z \over \partial \phi}
+{\partial u_{\phi}
\over \partial z}\right) \biggr]
$$
\begin{equation}
+{2 \over 3}{\partial \over \partial z}
\biggl[\eta \left(2 {\partial u_z
\over \partial z}-{\partial u_r \over \partial r}
-{1 \over r}{\partial
u_{\phi} \over \partial \phi}-
{u_r \over r}\right)\biggr] ~.
\label{F16}
\end{equation}

The continuity equation for the disk flow is
\begin{equation}
 {\partial \rho \over \partial t}+
{\partial (\rho u_i) \over \partial
x_i} = 0~, \quad {\rm or}\quad
{\partial \rho \over \partial t}+
{\bf \nabla} \cdot (\rho {\bf u})=0~.
\label{F17}
\end{equation}
In cylindrical coordinates this equation becomes
\begin{equation}
{\partial \rho \over \partial t}+
{\partial (r\rho u_r) \over r\partial r}+
{\partial (\rho u_{\phi}) \over
r \partial \phi}+{\partial (\rho u_z) \over
\partial z}=0~.
\label{F18}
\end{equation}

\subsection{Derivation of Thin disk Equations}

For axisymmetric disk flows the $\phi-$partial
derivatives of the different scalar quantities
vanish so that
equations (\ref{F14})-(\ref{F16}) become
%%%%%%%%%%%%%%%% eqn 19
$$
\rho \left({\partial u_r \over \partial t}+
u_r{\partial u_r \over
\partial r}+
u_z{\partial u_r \over \partial z}-
{u_{\phi}^2 \over r}
\right) =-{\partial p \over \partial r} +\rho g_r
$$
\begin{equation}
+{2 \over 3}{\partial \over \partial r}
\biggl\{\eta \biggl[{2 \over r}{\partial (ru_r)
\over \partial r} -
{\partial u_z \over \partial z} \biggr]\biggr\}-
2{u_r \over r}
{\partial \eta \over \partial r}+
{\partial \over \partial z}
\biggl [\eta \left({\partial u_z \over \partial r}+
{\partial u_r \over
\partial z}\right)\biggr]~,
\label{F19}
\end{equation}
$$
\rho \left({\partial u_{\phi} \over \partial t}+
u_r{\partial u_{\phi} \over
\partial r} +
u_z{\partial u_{\phi} \over \partial z}+
{u_r u_{\phi} \over r}
 \right)
$$
\begin{equation}
={1 \over r^2}{\partial \over \partial r}
\biggl[\eta r^3 {\partial \over
\partial r}\left(u_{\phi} \over r \right) \biggr] +
 {\partial \over \partial z}\left(\eta
{\partial u_{\phi} \over \partial z}\right)~,
\label{F20}
 \end{equation}
%%%%%%%%%%%%%%%%%%%%%%%%%%%%%%%% eqn 21
$$
\rho \left({\partial u_z \over \partial t}+
u_r{\partial u_z
\over
\partial r} +
u_z{\partial u_z \over \partial z}
  \right) =-{\partial p \over
\partial z}+\rho g_z
$$
\begin{equation}
+{1 \over r}{\partial \over \partial r}
\biggl[\eta r \left( {\partial u_z
\over \partial r}+{\partial u_r \over \partial z}
\right)\biggr] +
{2 \over 3}{\partial \over \partial z}
\biggl[\eta \left(2{\partial u_z
\over \partial z}-{\partial u_r \over \partial r}-
{u_r \over r}\right)\biggr]~.
\label{F21}
\end{equation}
The continuity equation (\ref{F18}) takes the form
\begin{equation}
{\partial \rho \over \partial t}+
{1\over r}{\partial (r\rho u_r) \over
\partial r}
+{\partial (\rho u_z) \over
\partial z}=0 ~.
\label{F22}
\end{equation}
   Many problems concerning thin accretion
disks can be solved using the above
equations in the equatorial plane or
averaged over the thickness of the disk.
   The vertical structure of the disk
is determined
by the equilibrium of the pressure gradient
and the vertical gravitational force $g_z$,
which may include self-gravitation for
sufficiently massive disks.

   Problems connected with
convective and meridonal motion in
disks need more detailed
description with the full set of equations
(\ref{F18})-(\ref{F18}).
    A complete description
requires including the magnetic force
in the Navier-Stokes equations and
the Maxwell equations for the field
evolution.

In the equatorial plane of a thin disk,
 we have
\begin{equation}
u_z=0, \quad g_z={\partial \rho \over \partial z}=
{\partial p \over
\partial z}={\partial u_r \over \partial z} =
{ \partial u_{\phi} \over
\partial z}=0 ~,
\label{F23}
\end{equation}
where mirror symmetry
about the equatorial plane  is assumed.
    However, note that
$ {\partial u_z / \partial z} \not=
0 $ even in this case.
     Neglecting $u_z$, $g_{\phi}$
and all $z$ derivatives, we
obtain the approximate equations
%%%%%%%%%%%%%%%%%%%% eqn 24
   \begin{equation}
\rho \left({\partial u_r \over
\partial t}+u_r{\partial u_r \over
\partial r}
-{u_{\phi}^2 \over r}
\right) =-{\partial p \over
\partial r}+\rho g_r +
{4 \over 3}{\partial \over \partial r}
\biggl[{\eta \over r}{\partial (ru_r)
\over \partial r}
\biggr]-
2{u_r \over r}
{\partial \eta \over \partial r}~,
\label{F24}
\end{equation}
%%%%%%%%%%%%%%%%% eqn 25
\begin{equation}
\rho r \left({\partial r u_{\phi} \over \partial t}+
 u_r{\partial r u_{\phi} \over
\partial r} \right)=
{\partial \over \partial r} \biggl[\eta r^3 {\partial \over
\partial r}\left(u_{\phi} \over r \right) \biggr]~,
\label{F25}
 \end{equation}
%%%%%%%%%%%%%%%% eqn 26
\begin{equation}
\rho g_z  ={\partial p \over
\partial z}~,
\label{F26}
\end{equation}
%%%%%%%%%%%%%%%%% eqn 27
\begin{equation}
{\partial \rho  \over \partial t}+
{1 \over r} {\partial (r \rho u_r) \over \partial r}
=0~.
\label{F27}
\end{equation}
Combining equations (\ref{F24}) and (\ref{F27}) we get
the radial equation in the form
%%%%%%%%%%%%%%%% eqn 28
\begin{equation}
{\partial \rho r u_r \over \partial t}+
{\partial \rho r u_r^2 \over
\partial r}
-\rho r({u_{\phi}^2 \over r}
+ g_r ) =-r {\partial p \over
\partial r}+
{4 \over 3}r{\partial \over \partial r}
\biggl[{\eta \over r}{\partial (ru_r)
\over \partial r}
\biggr]-
2u_r
{\partial \eta \over \partial r}~.
\label{F28}
\end{equation}
Following Shakura (1972), we write
the viscosity coefficient $\eta$
using the $\alpha-$ pre\-scrip\-tion,
\begin{equation}
 \eta=\frac{2}{3}\alpha~ \rho_0 u_{s0}~ z_0~,
\label{F29}
\end{equation}
where $u_{s0}$ is a sound velocity in
the equatorial plane ($z=0$), and
$z_0$ is half-thickness of the disk.
   Integrating the equations (\ref{F25}) and
(\ref{F27}) over the thickness
of the disk, we get an approximate
set of equations describing the
thin disk behaviour, neglecting
viscosity in the radial equation (\ref{F24})
which is taken in the symmetry plane of the disk

\begin{equation}
{\partial u_r \over \partial t} +
u_r {\partial ( u_r) \over
\partial r} -\rho_0 ( \Omega^2 r + g_r) +
{\partial P_0 \over \partial r}=0
\label{2.12}
\end{equation}

\begin{equation}
 r \Sigma {\partial j \over \partial t} +
{\dot M \over 2 \pi}
{\partial j \over \partial r} =
\frac{2}{3}\alpha {\partial \over \partial r}
\left( \Sigma u_{s0} z_0 r^3
{\partial \Omega \over \partial r}\right)
  \label{2.13}
\end{equation}

\begin{equation}
 {\partial \Sigma \over \partial t} +
{1 \over 2 \pi r}
{\partial \dot M \over \partial r}=0
\label{2.14}
\end{equation}
The average values, surface density $\Sigma$, mass flux
over the disk $\dot M $,
and the values of angular velocity $\Omega$ and
specific angular momentum $j$
are determined as follows

$$ \Sigma=\int_{-z_0}^{z_0}\,\rho dz,
\qquad \dot M=2 \pi r \int_{-z_0}^
{z_0}\,\rho u_r dz=2\pi r \Sigma u_r,
$$
\begin{equation}
\Omega = {u_{\phi} \over r},
\quad j=r u_{\phi}
\label{2.15}
\end{equation}
The equations (\ref{2.12})-(\ref{2.14}) determine the variables
($\Sigma,\, \dot M ,\, j$), where
($u_r,\, \Omega $) are expresssed from (\ref{2.15})
with

\begin{equation}
u_r,\, \dot M<0
\label{2.16}
\end{equation}
The solution of the equation of
vertical balance (\ref{F26}) permits to obtain the
variables

\begin{equation}
u_{z0}=\left(\partial P_0
\over \partial \rho_0 \right)_
S^{1/2}, \quad z_0, \quad  P_0
\label{2.17}
\end{equation}
as a functions of ($ r,\,\Sigma $),
when the equation of state is given in
the form $P(\rho)$ (e.g. a polytropic).
Here $\rho_0,\, P_0 $ are the density
and the pressure in the equatorial plane.
\hfill\break\indent
When pressure P and velocity
$u_r$ derivatives are neglected in
(\ref{2.12}),
the boundary conditions consist of the value
of the mass flux and Keplerian
angular velocity at infinity; and of a condition
on the angular velocity derivative
near the accreting body (black hole,
neutron star or other), determining
the total angular momentum flux through the
disk and zero density and pressure on the outer
boundary of the disk.

$$ \dot M(\infty)=\dot M_{\infty},\quad
\Omega(r \rightarrow \infty)=\left(GM \over
r^3\right)^{1/2}, \,
\rho(\pm z_0)=0 $$
\begin{equation}
  \Omega(r_{in})=\Omega_{in}, \quad {\rm or} \quad
\left(\partial \Omega \over \partial r\right)(r_{in})
= \left(\partial \Omega
\over \partial r\right)_{in}
\label{2.18}
\end{equation}
where $M$ is the mass of the star-disk body.
If all terms remain in the equation (\ref{2.12})
then an additional boundary
condition is necessary. It  follows
from the demand, that the solution
must go smoothly through the sonic
(singular) point in the radial motion
of the accretion disk to the
black hole, or be adjusted continuously to
the conditions at the outer
boundary on the stellar equator.
The stellar angular velocity may
be much less than the Keplerian velocity
of the accretion disk.
In that case a boundary layer is established
between the disk and the star,
which needs to be treated separately.

In the case of general equation
of state $P(\rho,T)$ a new variable the
temperature $T$ appeares,
which is connected with a heat flux $F\,$(ergs/cm$^2$/s),
radiative or convective.
Two additional equations of the
energy balance over the disk and
heat transfer to its outer boundary
must be added together with boundary conditions
for obtaining the full solution. These equations will be
considered in later sections.

For solution of nonstationary
accretion problems initial distributions
are needed

\begin{equation}
\Sigma (r), \, j(r), \, \dot M(r)
\label{2.19}
\end{equation}
in addition to the boundary conditions.
In most cases the external force is
represented by the gravity of the star,
which may be treated  as a point mass with a potential

\begin{equation}
 \phi_g=-{GM \over (r^2+z^2)^{1/2}}
\label{2.20}
\end{equation}
The components of the gravitational
accelleration $g_i$ are written as

\begin{equation}
g_i=-{\partial \phi_g \over \partial x_i}=\left(-{GMr \over
(r^2+z^2)^{3/2}}, \,
0, \,
{GMz \over
(r^2+z^2)^{3/2}}, \right)
\label{2.21}
\end{equation}
For a thin disk we can use the expansion

\begin{equation}
g_i=\biggl\{
 -{GM \over r^2}\left(1-{3 \over 2}{z^2 \over r^2} \right), \,
0, \,
 -{GMz \over r^3}\left(1-{3 \over 2}{z^2 \over r^2} \right) \biggr\}
\label{2.22}
\end{equation}

\subsection{Structure of Polytropic Accretion Disks}

For a polytropic equation of state

\begin{equation}
P=K \rho^{1+{1 \over n}},
\label{3.1}
\end{equation}
taking the main term in the gravitational force $g_z$
from (\ref{2.22})  we get
the solution of the equation (\ref{F26}) (H\~oshi, 1977)

\begin{equation}
\rho=\rho_0\left(1-{z^2 \over z_0^2}\right)^n,
\label{3.2}
\end{equation}
where the density in the equatorial plane $\rho_0$ is connected with $r$
and $z_0$ by the relation

\begin{equation}
\rho_0=\biggl[{GM \over 2K(n+1)} \biggr]^n {z_0^{2n} \over r^{3n}}.
\label{3.3}
\end{equation}
Some physical quantities from (\ref{2.15}) -(\ref{2.17})
can be expressed in terms of $r$
and $\rho_0$.

\begin{equation}
 z_0=
\biggl[{2K(n+1) \over GM} \biggr]^{1 \over 2}
\rho_0^{{1 \over 2n}}
r^{3 \over 2}=\beta_{z0}\rho_0^{1 \over 2n}r^{3 \over 2},
\label{3.4}
\end{equation}

\begin{equation}
\Sigma=
\sqrt{\pi} {\Gamma(n+1) \over \Gamma \left(n+{3 \over 2}\right)}
\rho_0 z_0 =
\beta_{\Sigma} \rho_0^{2n+1 \over 2n} r^{3 \over 2},
\label{3.5}
\end{equation}

\begin{equation}
P_0=K\rho_0^{1+\frac{1}{n}},\quad  u_{s0}=\left({n+1 \over n} K
\right)^{1/2}
\rho_0^{1 \over 2n}
=\beta_{s0}
\rho_0^{1 \over 2n},
\label{3.7}
\end{equation}
where

\begin{equation}
\beta_{\Sigma}=
\sqrt{\pi} {\Gamma(n+1) \over \Gamma \left(n+{3 \over 2}\right)}
\biggl[{2K(n+1) \over GM} \biggr]^{1 \over 2}.
\label{3.8}
\end{equation}
For the isothermal case,
corresponding to $n=\infty$,

\begin{equation}
P=K\rho
\label{3.10}
\end{equation}
we have instead of (\ref{3.2})

\begin{equation}
\rho=\rho_0 \exp \left( -{GMz^2 \over 2Kr^3} \right)=\rho_0 \exp
\left(-{z^2 \over z_i^2} \right), \quad z_i=\left(2Kr^3
\over GM \right)^{1/2}.
\label{3.11}
\end{equation}
Formally an isothermal disk has infinite
thickness $z_0$, but the density falls
exponentially with the characteristic height $z_i$ from (\ref{3.11}).
Instead of (\ref{3.5})-(\ref{3.7}) we have

\begin{equation}
\Sigma=\left(2 \pi K \over GM \right)^{1 \over 2} \rho_0 r^{3/2},
\label{3.12}
\end{equation}

\begin{equation}
P_0=K\rho_0, \quad u_{s0}=\sqrt{K}.
\label{3.14}
\end{equation}
Restrict ourself by stationary accretion disks with ${\partial \over
\partial t}=0$. Then we have from (\ref{2.14})

\begin{equation}
\dot M={\rm const}
\label{3.15}
\end{equation}
and (\ref{2.13}) is integrated, giving (Shakura, 1972)

\begin{equation}
{\dot M \over 2 \pi}(j-j_0)=\frac{2}{3}\alpha \Sigma u_{s0} z_0 r^3 {d \Omega
\over dr}.
\label{3.16}
\end{equation}
The integration constant $j_0$ after multiplication by $\dot M$
gives the total
(advective plus viscous) flux of angular momentum
within the accretion disk. A positive value of $j_0$ corresponds to a
negative total flux (due to a negative $ \dot M $), which means that
the central body accretes its angular momentum. When $j_0$ is negative
the central body decreases its total angular momentum,
while increasing its mass.

For a thin disk neglecting $u_r$ and ${\cal P}$ and
taking only the main term for $g_r$ from (\ref{2.22})
we get from (\ref{2.12})

\begin{equation}
\Omega^{(0)}=\left({GM \over r^3} \right)^{1/2}=\Omega_K,
\quad j^{(0)}=j_K=(GMr)^{1/2}.
\label{3.17}
\end{equation}
The solution (\ref{3.17}) is used up to the inner
boundary of the accretion disk at
$r=r_{in}$. The integration constant $j_0$ can be scaled by
the Keplerian angular
momentum $j_K (r_{in})$, so that

\begin{equation}
j_0=\xi j_K(r_{in})=\xi \sqrt{GMr_{in}}
\label{3.18}
\end{equation}
Substituting (\ref{3.18}) into (\ref{3.16})
we get with account of (\ref{3.4})-(\ref{3.8}) the
relation for the density in the equatorial plane

\begin{equation}
\rho_0^{(0)}=b_{\rho 0}
\biggl[-{\dot M \over r^3} \left(1-\xi \sqrt
{r_{in} \over r} \right) \biggr]^
{2n \over 2n+3},
\label{3.19}
\end{equation}

\begin{equation}
b_{\rho 0}=
\biggl[{1 \over 4 \pi^{3/2} \alpha}{\sqrt{n} \over
(n+1)^{3/2}}{\Gamma \left(n+{3 \over 2}\right) \over \Gamma (n+1)}
{GM \over K^{3/2}} \biggr]^{2n \over 2n+3}.
\label{3.20}
\end{equation}
Using (\ref{3.8}), in (\ref{3.4})-(\ref{3.7}) we get

\begin{equation}
z_0^{(0)}=b_{z0}
\biggl[-\dot M \left(1-\xi \sqrt{r_{in} \over r} \right) \biggr]^
{1 \over 2n+3} r^{
{3 \over 2}{2n+1 \over 2n+3}},
\label{3.21}
\end{equation}

\begin{equation}
\Sigma^{(0)}=b_{\Sigma }
\biggl[-\dot M \left(1-\xi \sqrt{r_{in} \over r} \right) \biggr]^
{2n+1 \over 2n+3} r^{-
{3 \over 2}{2n-1 \over 2n+3}},
\label{3.22}
\end{equation}

\begin{equation}
 P_0^{(0)}=b_{P0}
\biggl[-{\dot M \over r^3}
\left(1-\xi \sqrt{r_{in} \over r} \right)\biggr]^{\frac{2(n+1)}{2n+3}},
\label{3.23}
\end{equation}

\begin{equation}
u_{s0}^{(0)}=b_{s0}
\biggl[-{\dot M \over r^3} \left(1-\xi \sqrt
{r_{in} \over r} \right) \biggr]^
{1 \over 2n+3},
\label{3.24}
\end{equation}
where

\begin{equation}
b_{z0}=[2(n+1)]^{1/2}
\biggl[{1 \over 4 \pi^{3/2} \alpha}{\sqrt{n} \over
(n+1)^{3/2}}{\Gamma \left(n+{3 \over 2}\right) \over \Gamma (n+1)}
\biggr]^{1 \over 2n+3}
K^{n \over 2n+3} (GM)^{-{2n+1 \over 2(2n+3)}}
\label{3.25}
\end{equation}

$$b_{\Sigma}=$$
\begin{equation}
{[2\pi (n+1)]}^{1 \over 2} {\Gamma(n+1) \over \Gamma \left(n+{3 \over
2}\right)}
\biggl[{1 \over 4 \pi^{3/2} \alpha}{\sqrt{n} \over
(n+1)^{3/2}}{\Gamma \left(n+{3 \over 2}\right) \over \Gamma (n+1)}
\biggr]^{2n+1 \over 2n+3}
K^{-{2n \over 2n+3}} (GM)^{{2n-1 \over 2(2n+3)}}
\label{3.26}
\end{equation}

\begin{equation}
b_{P0}=K\,b_{\rho 0}^{1+\frac{1}{n}},\quad
b_{s0}=b_{\rho 0}^{1 \over 2n} \left(K \frac{n+1}{n}\right)^{1/2}.
\label{3.27}
\end{equation}
In the formula for viscosity coefficient (\ref{F29}) $z_0$ must be replaced
by $z_i$ from (\ref{3.11}) in the isothermal case,

\begin{equation}
\eta=\frac{2}{3}\alpha \rho u_{s0} z_i
\label{3.29}
\end{equation}
Then instead of (\ref{3.18})-(\ref{3.27}) we get a solution

\begin{equation}
P_0^{(0)}=K\, \rho_0^{(0)}={GM \over 4 \pi^{3/2} \alpha K^{3/2}}
\left(-{\dot M \over r^3}\right)
\left(1-\xi \sqrt{r_{in} \over r} \right),
\label{3.30}
\end{equation}

\begin{equation}
\Sigma^{(0)}=
{\sqrt{GM} \over \alpha 2 \pi K \sqrt {2} }
\left(-{\dot M \over r^{3/2}}\right)
\left(1-\xi \sqrt{r_{in} \over r} \right).
\label{3.31}
\end{equation}
The formulas for the values of
 $z_i$ in (\ref{3.11}) and $u_{so}$ in (\ref{3.14})
remain the same in all
approximations for isothermal case.

For accretion into a black hole
in presence of a free innner
boundary with $\xi = 1$ the zero
order approximation is not valid in
the vicinity of $r_{in}$,
where other terms in (\ref{2.12}) must be taken
into account. For accretion
onto the slowly rotating star with
angular velocity smaller than
the Keplerian velocity on the equator, the
drop of the angular velocity
from Keplerian in the disk to stellar
equatorial happens in a thin
boundary layer, which must be considered
separately with proper account
of the pressure term in (\ref{2.12}).

%%%%%%%%%%%%%%%%%%%%%%%%%%%%%%%%%%%%%%%%%%%%%%%%%%%%%%%%%%%%%%%%%%%%%
\subsection{Standard Accretion Disk Model}

\subsubsection{Equilibrium equations}

The small thickness of the disk in
comparison with its radius $h \ll r$
indicate to small importance
of the pressure gradient
$\nabla P$ in comparison with
gravity and inertia forces.
That leads to a simple
radial equilibrium equation
denoting the balance between the last two
forces occuring when the angular
velocity of the disk $\Omega$ is equal to
the Keplerian one $\Omega_K$,

\begin{equation}
\label{ref1.1}
\Omega=\Omega_K=\left(\frac{GM}{r^3}\right)^{1/2}.
\end{equation}
Note, just before a last stable
orbit around a black hole, and of course
inside it, this suggestion fails,
but in the ``standard''
accretion disk model the
relation (\ref{ref1.1}) is suggested to be fulfilled
all over the disk, with an inner
boundary at the last stable orbit.

The equilibrium equation in the
vertical $z$-direction is determined by a
balance between the gravitational
force and pressure gradient

\begin{equation}
\label{ref1.2}
\frac{dP}{dz}=-\rho\frac{GMz}{r^3}
\end{equation}
For a thin disk this differential
equation is substituted by an
algebraic one, determining the
half-thickness of the disk in the form

\begin{equation}
\label{ref1.3}
h \approx \frac{1}{\Omega_K}
\left(2\frac{P}{\rho}\right)^{1/2}.
\end{equation}
The balance of angular momentum,
related to the $\phi$ component of the
Euler equation has an integral
in a stationary case written as

\begin{equation}
\label{ref1.4}
\dot M(j-j_{in})=-2\pi r^2\,2ht_{r\phi},\quad t_{r\phi}=
\eta r\frac{d\Omega}{dr}.
\end{equation}
Here $j=v_{\phi}r=\Omega r^2$ is
the specific angular momentum,
$t_{r\phi}$ is a component of the
viscous stress tensor, $\dot M>0$ is a mass
flux per unit time into a black hole,
$j_0$ is an integration constant having,
after multiplication by $\dot M$, a physical
sence of difference between viscous
and advective flux of the
angular momentum, when $j_{in}$
itself is equal to the specific angular
momentum of matter
falling into a black hole.
In the standard theory the value of $j_{in}$
is determined separately, from physical considerations.
For the accretion
into a black hole it is suggested,
that on the last stable orbit the
gradient of the angular velocity is
zero, corresponding to zero
viscous momentum flux. In that case

\begin{equation}
\label{ref1.5}
j_{in}=\Omega_K r_{in}^2,
\end{equation}
corresponding to the Keplerian
angular momentum of the matter on the last
stable orbit. During accretion
into a slowly rotating star which
angular velocity is smaller than a
Keplerian velocity on the inner edge
of the disk, there is a maximum
of the angular velocity close to its surface,
where viscous flux is zero,
and there is a boundary layer between
this point and stellar surface.
In that case (\ref{ref1.5}) remains to
be valid.
The situation is different
for accretion discs around rapidly
rotating stars with a critical
Keplerian speed on the equator (see Section 3).
Note, that in the pioneering
paper of Shakura (1972) the integration
constant $j_{in}$ was found
as in (\ref{ref1.5}), but was taken
zero in his subsequent formulae.
Importance of
using $j_{in}$ in the form (\ref{ref1.5})
was noticed by Novikov and
Thorne (1973), and became a
feature of the standard model.

\subsubsection{Turbulent Viscosity and Instabilities}

The choice of the viscosity
coefficient is the most difficult and
speculative problem of the
accterion disk theory. In the laminar case of
microscopic (atomic or plasma)
viscosity, which is very low, the stationary
accretion disk must be very
massive and very thick,
and before its formation
the matter is collected by
disk leading to a small flux inside.
It contradicts to observations
of X-ray binaries,
where a considerable matter flux
along the accretion disk may be explained
only when viscosity coefficient
is much larger then the microscopic one.
In the paper of Shakura (1972)
it was suggested, that matter in the disk
is turbulent, what determines
a turbulent viscous stress tensor,
parametrized by a pressure

\begin{equation}
\label{ref1.7}
t_{r\phi}=-\alpha\rho v_s^2 = -\alpha P,
\end{equation}
where $v_s$ is a sound speed in the matter.
This simple presentation comes
out from a relation for a turbulent viscosity
coefficient $\eta_t\approx \rho v_t l$
with an average turbulent velocity
$v_t$ and mean free path of the
turbulent element $l$. It follows from
the definition of $
t_{r\phi}$ in (\ref{ref1.4}),
when we take $l \approx h$
from (\ref{ref1.3})

\begin{equation}
\label{ref1.8}
t_{r\phi}=\rho v_t h r \frac{d\Omega}{dr}
\approx \rho v_t v_s =-\alpha
\rho v_s^2,
\end{equation}
where a coefficient $\alpha<1$ is
connecting the turbulent and sound speeds
$v_t=\alpha v_s$.
Presentations of $t_{r\phi}$ in (\ref{ref1.7}) and
(\ref{ref1.8}) are equivalent,
and only when the angular velocity
differs considerably from the
Keplerian one the first relation to the
right in (\ref{ref1.8}) is  preferable.
That does not appear
(by definition) in the standard
theory, but may happen when advective
terms are included.

Developement of a turbulence
in the acctretion disk cannot be justified
simply, because a Keplerian disk
is stable in linear approximation to
the developement of axially symmetric perturbations,
conserving the angular momentum, and against the
Reyleigh-Taylor effect.
It was suggested by Ya.B.Zeldovich,
that in presence of very large Reynolds number
${\rm Re}=\frac{\rho v l}{\eta}$
the amplitude of perturbations
at which nonlinear effects become important
is very low, so in this situation
turbulence may develope due to
nonlinear instability even when
the disk is stable in linear approximation.
Another source of viscous stresses
may arise from a magnetic field,
but it was suggested by Shakura (1972),
that magnetic stresses cannot
exceed the turbulent ones.

Magnetic plasma instability as a
source of the turbulence in the
accretion discs has been
studied extensively in last years (see
review of Balbus and Hawley, 1998).
  They used an instability of the
uniform magnetic field parallel
to the axis in differentially rotating
disk, discovered by Velikhov (1959).
It could be really important in
absence of any other source
of the turbulence, but it is hard to belive
that there is no radial or azimuthal
component of the magnetic field
in matter flowing into the
accretion disk from the companion star.
In that case the field amplification
due to twisting by a differential
rotation take place without
necessity of any kind of instability.
Explanation of viscosity in accretion disks, based on
developement of plasma instabilities, have been considered by
Coppi (2000).

It was shown by Bisnovatyi-Kogan
and Blinnikov (1976, 1977), that inner
regions of a highly luminous
accretion discs where pressure is dominated
by radiation, are unstable to
vertical convection. Developement of this
convection produce a turbulence,
needed for a high viscosity. Other
regions of a standard accretion
disk should be stable to developement of a
vertical convection, so other
ways of a turbulence exitation are needed
there.

For Keplerian angular velocity the angular momentum per unit mass
$j=\omega\,r^2 \sim r^{1/2}$ is groving outside. In this respect it is
similar to the viscid flow between two rotating cylinders (Taylor
experiment), when the inner cylinder is at rest. There are however
important differences between these two problems. There are two boundaries
on cylinder surfaces with adhesion conditions, and  free boundaries in
the case of accretion disk. The radial equilibrium in the fluid between
cylinders is supported by the balance between centrifugal, and pressure
gradient forces, while the last is substituted by the gravity in the
accretion disks. Finally, the fluid between cylinders
in the laboratory experiments is usually strongly subsonic, determining
a non-compressible medium, while the Keplerian motion in the thin accretion
disk is highly supersonic. Nevertheless, it is hoped that the
fundamental theory checked in the laboratory will answer the astrophysical
questions.

Phenomenological analysis of the Taylor experiment, and the onset of
turbulence in the "stable" case of the inner cylinder at rest had been
done by Zeldovich (1981). There are two characteristic
specific energies in the problem. One is the energy, necessary for
performing the ring exchange with conservation of the angular momentum of
each ring
$E_s=2\frac{\omega}{r}\frac{d(\omega\, r^2)}{dr} (\Delta r)^2$;
here the exchange is supposed to happen for rings with small distance
$\Delta r$ between them. This enegry is positive in the "stable", and
negative in the case unstable to Reyleigh-Taylor effect. Another energy
$E_t$ produced after merging of these two rings with conservation of the
total angular momentum
$E_t=\frac{1}{4}r\left(\frac{d\omega}{dr}\right)^2 (\Delta r)^2$ is always
positive. The ratio of these two energies is defined by Zeldovich (1981)
as Taylor number

\begin{equation}
\label{v1}
Ty=\frac{E_s}{E_t}=4\frac{d[(\omega r^2)^2]/dr}{r^5(d\omega/dr)^2}.
\end{equation}
The boundary $Ty=0$ separate regions stable and unstable to Reyleigh-Taylor
effect. In the case of close cylinder radiuses $R_{out}-R_{in} \ll R=
\frac{1}{2}(R_{out}+R_{in})$
we get with account of (\ref{v1})

\begin{equation}
\label{v2}
\omega_{out} \approx \omega_{in} \approx \omega,\,\,
R_{out}\approx R_{in} \approx R,\,\,
R_{out}-R_{in}=t,\,\, \frac{d}{dr}=\frac{1}{t},\,\,
Ty\approx 8\frac{t}{R}.
 \end{equation}
This case at $t \rightarrow 0$ corresponds
to the flat Couette flow, which is formally stable at all $Re$ numbers
(Schlichting, 1964) when $\frac{dv}{dr}$,
is constant ($v$ is the fluid velocity), in contradiction with
the experiment. Account of the curvature of the curve $v(r)$, even
small, is very important here, and gives the instability, similar
to the case with the free outer boundary, also corresponding to $Ty=0$,
when the instability starts at $Re=\frac{\rho v t}{\eta} \approx 2000$.
Note, that account of a
compressibility leads to the instability even in the case of the flat
Couette flow Glatzel (1989). The maximum of instability region corresponds to
Mach number $Ma \approx 4.9$, when unstable mode appeares at $Re \approx 84$.

Taylor experiments with rotating cylinders has shown that at $Ty>0$ the flow
is becoming turbulent, but the Reynolds number $Re$, corresponding
to the boundary of stability is increasing with increasing $Ty$.
For the case of the Keplerian disk $\omega \sim r^{-3/2}$ we have
$Ty=\frac{16}{9}$, formally corresponding to $\frac{t}{R}=\frac{2}{9}=0.222$,
for which the experimental number of the critical
$Re\approx 2\times 10^5$ (Zeldovich, 1981).
Experimental dependence of the critical value $Re=f(Ty)$ is characterized by
the Couette-flow critical value $Re(0)=2000$, and can be approximated by
the dependence $Re \approx 10^4 Ty^2$ for $Ty>0.6$, $\frac{t}{R}>0.075$.
The last dependence was interpreted theoretically by Zeldovich (1981) by
introduction of the {\it split} regime of the turbulence  between cylinders.

In the laminar viscid flow between cylinders the exact
solution (Schlichting, 1964)
gives for $\omega_{in}=0$ the rotational velocity distribution as

\begin{equation}
\label{v3}
v(r)=\frac{R_{out}^2
\omega_{out}}{R_{out}^2-R_{in}^2}\left(r-\frac{R_{in}^2}{r}\right)\approx
\omega_{out} \frac{R}{t} (r-R_{in})=\omega_{out}\frac{q}{t}=\omega_{out}x
 \end{equation}
$$q=r-R_{in},\,\,\, x=\frac{q}{t}$$
The local values of $Re'$ and $Ty'$ numbers, defined by fluid parameters near
the inner cylinder, have been introduced by Zeldovich (1981) as

\begin{equation}
\label{v4}
 Re'=\frac{\rho vx}{\eta}=x^2\,Re \, \,\,\,\, Ty'=8\frac{x}{R}=x\,Ty.
 \end{equation}
It was suggested by Zeldovich (1981) that turbulence
between rotating cylinders
is installed, when one of the inequalities is fulfilled

\begin{equation}
\label{v5}
{\bf a.}\,\,\,\, Re=f(Ty)\qquad {\rm or}\qquad {\bf b.}\,\,\,\, Re'=(Ty').
 \end{equation}
If the relation (\ref{v5}b) is fulfilled first the global condition
of instability is not reached, and the turbulence is started in the local
region near the inner cylinder, and was called as a split regime. Assuming
that (\ref{v5}b) is fulfilled first we get

\begin{equation}
\label{v6}
 Re= \frac{Re'}{x^2}=\frac{f(Ty')}{x^2}= Ty^2 \frac{f(y)}{y^2},\quad
y=x\,Ty.
 \end{equation}
Suppose now, that the function $\phi(y)=\frac{f(y)}{y^2}$ has a minimum at

\begin{equation}
\label{v7}
y=y_m, \quad \phi_m= \frac{f(y_m)}{y_m^2}.
 \end{equation}
Than, taking into account (\ref{v3}) we get the relation $Ty>y_m$,
so
\begin{equation}
\label{v8}
\phi(Ty)>\phi(y_m), \quad Re=Ty^2 \phi(y_m)<f(Ty),
 \end{equation}
the global criterion of the turbulence is not fulfilled, and the
split regime is realized. The turbulence is started in the region
$x<x_m=y_m/Ty$. So, the criterion of the beginning of the split
regime of the convection has a quadratic dependence on $Ty$ in accordance
with the experimental results.

It is noticed by Zeldovich (1981) that a boundary of
the split regime of turbulence
does not necessary mean the beginning of the turbulence in the accretion
disk, where the condition (\ref{v5}a) is probably needed. Two examples of
different functions are considered by Zeldovich (1981). For the function

\begin{equation}
\label{v9}
 f(y)=A \left(1-\frac{y}{y_0}\right)^{-1}
 \end{equation}
the absolute instability (\ref{v5}a) exist only at $Ty<y_0$, and the
minimum is reached at

\begin{equation}
\label{v10}
y=\frac{2}{3} y_0,\quad \phi_m=\frac{f(y_m)}{y_m^2}=
\frac{27}{4} \frac{A}{y_0^2}.
 \end{equation}
From the condition of the plane Couette frow it follows $A=2000$, and from
Taylor experiments, assuming there the split regime turbulence and assuming
(\ref{v8}), we get $\phi_m=10^4$, $y_0=1.16$. Note,that for the Keplerian
accretion disk we have $Ty=\frac{16}{9}>1.16$.

Another function from Zeldovich (1981) is

\begin{equation}
\label{v11}
f(y)=A\exp(\frac{y}{y_0}),
 \end{equation}
where $A=2000$, similar to (\ref{v9}), and $y_0=0.3$ from the Taylor
experiment. Here $Re=f(16/9)=2000 \exp(1.78/0.3)\approx 8 \times 10^5$ for
the beginning of the absolute instability to the turbulence developement.

The evaluations above show, that the problem of the hydrodynamic
instability of the Keplerian accretion disk is far from a full
understanding, and there are arguments, both experimental and theoretical,
supporting the hydrodynamic origin of the accretion disk turbulence.
The cirtain answer could be obtained from the full stability analysis
of the Keplerian disk, similar to the one made for the viscous accretion
tori in cylindrical approximation, made
by Kleiber and Glatzel (1999) for a constant
specific angular momentum distribution.

With alpha- prescription
of viscosity the equation of angular
momentum conservation is
written in the plane of the disk as

\begin{equation}
\label{ref1.9}
\dot M(j-j_{in})=
4\pi r^2 \alpha P_0 h.
\end{equation}
When angular velocity is far
from Keplerian the relation
(\ref{ref1.4}) is valid with
a coefficient of a turbulent viscosity

\begin{equation}
\label{ref1.10}
\eta=\frac{2}{3}\alpha\rho_0 v_{s0} h,
\end{equation}
where values with the index ``0''
denotes the plane of the disk.

\subsubsection{Heat Balance}

In the standard theory a heat
balance is local, what means that all
heat produced by viscosity in the
ring between $r$ and $r+dr$ is
radiated through the sides
of disk at the same $r$.
   The heat production
rate $Q_+$ related to the
surface unit of the disk is written as

\begin{equation}
\label{ref1.11}
Q_+=h\,t_{r\phi}r\frac{d\Omega}{dr}=
\frac{3}{8\pi}\dot M \frac{GM}{r^3}
\left(1-\frac{j_{in}}{j}\right).
\end{equation}
Heat losses by a disk depend on
its optical depth. The first standard disk
model of Shakura (1972) considered
a geometrically thin disk as an optically
thick in a vertical direction.
That implies enegry losses $Q_-$ from the disk
due to a radiative conductivity,
after a substitution of
the differential eqiation
of a heat transfer by an algebraic relation

\begin{equation}
\label{ref1.12}
Q_- \approx \frac{4}{3} \frac{acT^4}{\kappa \Sigma}.
\end{equation}
Here $a$ is the usual radiation
energy-density constant,
$c$ is a speed of light,
$T$ is a temperature in the disk plane,
$\kappa$ is a matter opacity,
and a surface density

\begin{equation}
\label{ref1.13}
\Sigma=2\rho h,
\end{equation}
here and below
$\rho,\, T,\,P$ without the
index "0" are related to the disk plane.
The heat balance equation
is represented by a relation

\begin{equation}
\label{ref1.14}
Q_+=Q_-,
\end{equation}
A continuity equation in the
standard model of the stationary accretion flow
is used for finding of a
radial velocity $v_r$

\begin{equation}
\label{ref1.14a}
v_r=\frac{\dot M}{4\pi rh\rho}=
\frac{\dot M}{2\pi r\Sigma}.
\end{equation}
Equations (\ref{ref1.1}),(\ref{ref1.3}),
(\ref{ref1.9}),(\ref{ref1.13}),
(\ref{ref1.14}), completed
by an equation of state $P(\rho,T)$ and relation
for the opacity
$\kappa=\kappa(\rho, T)$
represent a full set of equatiions
for a standard disk model.
   For power low equations of state of an ideal gas
$P=P_g=\rho {\cal R} T$ (${\cal R}$
is a gas constant), or radiation pressure
$P=P_r=\frac{aT^4}{3}$,
and opacity in the form of electron scattering
$\kappa_e$, or Karammer's formula $\kappa_k$,
the solution of a standard
disk accretion theory is obtained
analytically (Shakura, 1972;
Novikov, Thorne, 1973; Shakura, Sunyaev, 1973).
Checking the suggestion of a
large optical thickness confirms a
self-consistency of the model.
    One of the shortcoming of the analytical
solutions of the standard model lay in the fact, that
solutions for different regions
of the disk with different
equation of states and opacities are not matched
to each other.

\subsubsection{Optically thin solution}

Few years after appearence of
the standard model it was found that
in addition to the opically
thick disk solution there is another branch
of the solution for the disk
structure with the same input parameters
$M,\,\dot M,\,\alpha$ which
is also self-consistent and has a small
optical thickness
(Shapiro, Lightman, Eardley, 1976). Suggestion of
the small optical thickness
implies another equation of energy losses,
determined by a volume emission

\begin{equation}
\label{ref1.15}
Q_- \approx q\, \rho\,h,
\end{equation}
where due to the Kirchoff law the
emissivity of the unit of a volume $q$ is
connected with a Plankian averaged opacity
$\kappa_p$ by an approximate relation
$q \approx acT_0^4 \kappa_p$.
Note, that Krammers formulae for opacity
are obtained after Rosseland averaging of the
frequency dependent absorption coefficient.
In the optically thin limit the
pressure is determined by a gas $P=P_g$.
Analytical solutions are obtained
here as well, from the same
equations with volume losses and gas pressure.
In the optically thin solution
the thickness of the disk is larger then
in the optically thick one, and density is lower.

While heating
by viscosity is determined mainly
by  ions, and cooling is determined
by electrons, the rate of the energy
exchange between them is important for
a thermal structure of the disk.
The energy balance equations are written
separately for ions and electrons.
For small accretion rates and lower
matter density the rate of energy
exchange due to binary collisions is
so slow, that in the thermal
balance the ions are much hotter then the
electrons.
That also implies a high
disk thickness and brings the
standard accretion theory
to the border of its applicability.
Nevertheless, in the highly
turbulent plasma the energy exchange
between ions and electrons may
be strongly enhanced due to presence
of fluctuating electrical fields,
where electrons and ions gain the
same energy. In such conditions
difference of temperatures between ions and
electrons may be negligible.
Regretfully, the theory of relaxation
in the turbulent plasma is
not completed, but there are indications
to a large exhancement of
the relaxation in presence of plasma turbulence,
in comparison with the
binary collisions (Quataert, 1997).

\subsection{
Accretion disk structure from
equations describing continuously
optically thin and
optically thick disk regions}

In order to find equations
of the disk structure
 valid in both limiting cases
 of optically thick and optically thin disk,
and smootly describing transition
between them, use Eddington
approximation for obtaining formulae
for a heat flux and for a radiation
pressure (Artemoma et al., 1996).
Suppose that disk is geometrically
thin and has a constant density along
$z$- axis. Defining $S_r$ as the
energy density of the radiation, $F_{rad}$
as radiation flux in $z$-direction,
$P_{rad}$ as radiation pressure we
write momentum equations for averaged
$\kappa_p$ in the form (Bisnovatyi-Kogan, 1989)

\begin{equation}
\label{ref11.1}
 {dF_{rad} \over dz}=
 -\rho c \kappa_p aT^4
 \left({S_r \over aT^4} - 1\right),
\end{equation}
\begin{equation}
\label{ref11.2}
 c{dP_{rad} \over dz}=-\kappa_e \rho F_{rad}.
\end{equation}
Consider the case when scattering
opacity $\kappa_e$ is much larger then
 absorption opacity $\kappa_p$,
 and suggest that heat production rate
 is proportional to the mass
 density $\rho$.
 Then, neglecting the flux in
 the radial direction we get

\begin{equation}
\label{ref11.3}
 F_{rad}=2{F_{0} \over \Sigma_{0}}\rho z,
\end{equation}
 where $F_{0}$ is the flux from the
 unit surface of the disk at $z=h$.
 Substituting (\ref{ref11.3})
 into (\ref{ref11.1}) we get

\begin{equation}
\label{ref11.5}
  S_r=3P_{rad}=
  aT^4\left(1-{2F_{0} \over
  c\kappa_p aT^4\Sigma_0}\right).
\end{equation}
 Using (\ref{ref11.3}) in (\ref{ref11.2}) we get

\begin{equation}
\label{ref11.6}
 c{dP_{rad} \over dz}=
 -2\kappa_e{F_{0} \over \Sigma_{0}}\rho^2 z.
\end{equation}
 Introducing the scattering optical depth

 $$\tau=\int_z^{\infty} \kappa_e \rho dz=
 \kappa_e \rho(h-z)=
 \tau_{0}-\kappa_e \rho z,
$$
\begin{equation}
\label{ref11.7}
  \tau_{0}=\kappa_e \rho h=
  {1 \over 2}\kappa_e \Sigma_0,
\end{equation}
 we rewrite (\ref{ref11.6}) in the form

\begin{equation}
\label{ref11.8}
  c{dP_{rad} \over d\tau}=
  2{F_{0} \over\Sigma_{0}}{\tau_0-
 \tau \over \kappa_e}.
\end{equation}
 Solve (\ref{ref11.8}) with
 the following boundary condition

\begin{equation}
\label{ref11.9}
 F_{rad}\vert_{\tau=0}=
 F_{0}={cS_r \vert_{\tau=0} \over 2}=
 {3cP_{rad}\vert_{\tau=0} \over 2},
\end{equation}
 resulting in

\begin{equation}
\label{ref11.10}
 cP_{rad}=F_0\left({2 \over 3}+
 \tau-{\tau^2 \over 2\tau_0} \right).
\end{equation}
 In the symmetry plane of the
 disk at $\tau=\tau_0$ we have

\begin{equation}
\label{ref11.11}
 cP_{rad,0}=F_0\left({2 \over 3}+
 {\tau_{0}\over 2}\right).
\end{equation}
 Using (\ref{ref11.11}) in
 (\ref{ref11.5}) we get in
 the symmetry plane, where $T=T_0$

\begin{equation}
\label{ref11.12}
 F_0=caT_0^4 \left(2+{3 \tau_0 \over 2} +
 {1 \over \tau_{\alpha 0}}
   \right)^{-1},
\end{equation}
where absorption optical depth

 $$\tau_{\alpha}=\int_z^{\infty} \kappa_p \rho dz=
 \kappa_p \rho(h-z)=
 \tau_{\alpha 0}-\kappa_p \rho z,
$$
\begin{equation}
\label{ref11.13}
  \tau_{\alpha 0}=\kappa_p \rho h=
  {1 \over 2}\kappa_p \Sigma_0.
\end{equation}
Introducing effective optical depth

\begin{equation}
\label{ref11.14}
  \tau_{*}=\left(\tau_{0}\tau_{\alpha 0}\right)^{1/2},
\end{equation}
we get finally the expressions
for the vertical energy flux from the disk
$F_0$ and the radiation pressure in the
symmetry plane as

\begin{equation}
\label{ref11.15}
 F_{0}={2acT_0^4 \over
 3\tau_{0}}\left(1+{4 \over 3\tau_{0}}+
 {2 \over 3\tau_{*}^2}\right)^{-1},
\end{equation}
\begin{equation}
\label{ref11.16}
 P_{rad,0}={aT_0^4 \over 3}
 {1+{4 \over 3 \tau_{0}}\over 1+
 {4 \over 3\tau_{0}}+
 {2\over 3\tau_{*}^2}}.
\end{equation}
At $\tau_0 \gg \tau_* \gg 1$
we have (\ref{ref1.12}) from (\ref{ref11.15}).
In the optically thin limit
$\tau_* \ll \tau_0 \ll 1$ we get

\begin{equation}
\label{ref11.17}
 F_{0}=acT_0^4 \tau_{\alpha 0}, \quad
 P_{rad,0}={2 \over 3}acT_0^4 \tau_{\alpha 0}.
\end{equation}
Using $F_0$ instead of $Q_-$
and equation of state
$P=\rho {\cal R} T+P_{rad,0}$,
the equations of accretion
disk structure together with equation

\begin{equation}
\label{ref11.19}
Q_+=F_0
\end{equation}
with $Q_+$ from (\ref{ref1.11}),
have been solved numerically by Artemova et al. (1996).
It occures that two solutions,
optically thick and optically thin, exist
separately when luminosity is
not very large. Two solutions intersect at
$\dot m=\dot m_b$, where

\begin{equation}
\label{ref11.20}
\dot m=\frac{\dot M c^2}{L_{\rm Edd}}, \quad
L_{\rm Edd}=\frac{4\pi c GM}{\kappa_e},
\end{equation}
and there is
no global solution for accretion disk at
$\dot m > \dot m_b$ (see Figure \ref{art1}).
It was concluded by Artemova et al (1996),
that in order to obtain a
global physically meaningful solution
at $\dot m > \dot m_b$,
account for an advective
term in (\ref{ref11.19}) is needed.

\begin{figure}
\epsfysize=8cm % fix the y-dimension and scales x-dim. to y-dim.
\hspace{3.5cm}\epsfbox{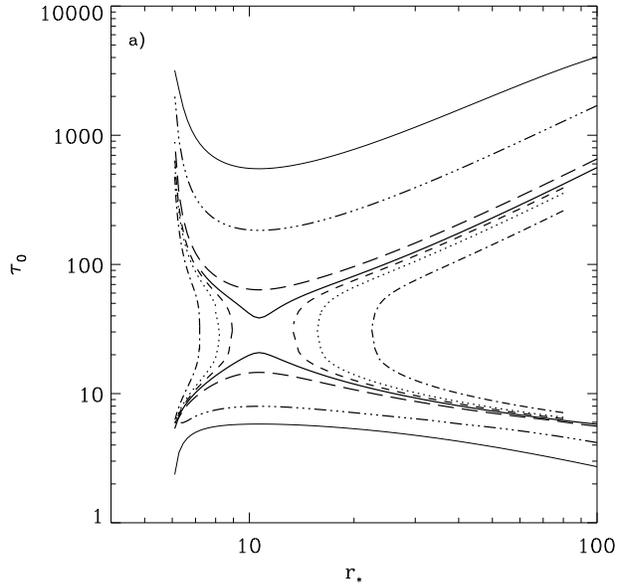} %for centering: act on hspace argument
\caption[h]{
 The dependences of the optical
 depth $\tau_0$ on
radius, $r_*=r/r_{g}$, for the case $M_{BH}=10^8\;M_\odot$,
$\alpha=1.0$ and different values of $\dot m$. The thin solid,
dot-triple dash, long dashed, heavy solid, short dashed, dotted
and dot-dashed curves correspond to $\dot m=1.0, 3.0, 8.0, 9.35,
10.0, 11.0, 15.0$, respectively. The upper curves correspond to
the optically thick family, lower curves correspond to the
optically thin family.}
\label{art1}
\end{figure}

\subsection{Accretion discs with advection}

Standard model gives somewhat nonphysical behaviour near the inner
edge of the accretion disk around a black hole.
%where temperature
%goes to zero, and density to infinity.
For high mass fluxes when
central regions are radiation-dominated
($P \approx P_r, \,\, \kappa
\approx \kappa_e$), the radial dependence
follows relations (Shakura, Sunyaev, 1973, Bisnovatyi-Kogan, 1999)

\begin{equation}
\label{ref31.1}
\rho \sim r^{3/2}{\cal J}^{-2} \rightarrow \infty,
\quad T \sim r^{-3/8},
\end{equation}
$$\quad h \sim {\cal J} \rightarrow 0,
\quad \Sigma \sim r^{3/2}{\cal J}^{-1}
\rightarrow \infty,
\quad v_r \sim r^{-5/2}{\cal J} \rightarrow 0,
$$
where limits relate to the inner
edge of the disk with $r=r_{in}$,

\begin{equation}
\label{ref31.2}
{\cal J}=1-\frac{j_{in}}{j}=
1-\sqrt{\frac{r_{in}}{r}}.
\end{equation}
At smaller $\dot M$, when near the inner edge
$P \approx P_g, \,\, \kappa \approx \kappa_e$,
there are different type
of singularities

\begin{equation}
\label{ref31.3}
\rho \sim r^{-33/20}{\cal J}^{2/5} \rightarrow 0,
\quad T \sim r^{-9/10}{\cal J}^{2/5}\rightarrow 0,
\end{equation}
$$\quad h \sim r^{21/20}{\cal J}^{1/5} \rightarrow 0,
\quad \Sigma \sim r^{-3/5}{\cal J}^{3/5} \rightarrow 0,
\quad v_r \sim r^{-2/5}{\cal J}^{-3/5} \rightarrow \infty.
$$
This results from the local
form of the equation of the
thermal balance (\ref{ref1.14}). It is
clear from physical ground,
that when a local heat production due
to viscosity goes to zero,
the heat brought by radial motion of matter
along the accretion disk
becomes more important.
   In the presence of
this advective heating (or cooling term,
depending on the radial entropy $S$ gradient)
written as

\begin{equation}
\label{ref3.1}
Q_{\rm adv}=-\frac{\dot M}{2\pi r}T \frac{dS}{dr},
\end{equation}
the equation of a heat balance is modified to

\begin{equation}
\label{ref3.2}
Q_+ - Q_{\rm adv}=Q_-.
\end{equation}
   In order to describe self-consistently
   the structure of the accertion disk
we should also modify the radial
disk equilibrium equation, including pressure
and inertia terms

\begin{equation}
\label{ref3.3}
r(\Omega^2-\Omega_K^2)=
\frac{1}{\rho}\frac{dP}{dr}-v_r\frac{dv_r}{dr}.
\end{equation}
Appearence of inertia term leads
to transonic radial flow with a
singular point. Conditions of
a continious passing of the solution through
a critical point choose a
unique value of the integration constant $j_{\rm in}$.

Simplified solutions with inclusion of the advective terms into
the vertically averaged equations
describing accretion disks was obtained by
Paczy\'nski \& Bisnovatyi-Kogan (1981).
This approach with some modifications have been used
by many researchers to study transonic accretion flows
around black holes (Muchotrzeb \& Paczy\'nski 1982;
Matsumoto et al. 1984; Abramowicz et al. 1988; Beloborodov 1998).
The importance of transonic nature of accretion flows
on the disk structure has been emphasized
by H\"oshi \& Shibazaki (1977), Liang \& Thompson (1980) and
Abramowicz \& Zurek (1981), and later studied in more details
by Abramowicz \& Kato (1989), see also Kato et al. (1998).

The problems at finding the numerical solution of advective disk structure
are connected with the possible non-uniqueness
of global solutions at $\alpha> 0.01$
and the non-standard behavior of a type of a singular point.
It was reported by Matsumoto et al. (1984), Abramowicz et al. (1988)
that in the case of viscosity prescription (\ref{ref1.1})
the singular point changes its type
from a saddle to node when increasing $\alpha$.
The presence of the nodal-type singular point leads to creating of
possibility of multiple solutions as the authors have claimed.

In the paper of Artemova et al. (2001) it was shown that the mentioned
problems have been created by several inconsistencies
in the preceding studies. Some problems are connected with
an inaccurate averaging of the equations over the disk thickness,
another ones appear due to incomplete investigation of
the singular points (Abramowicz et al. 1988).
It was found  that in the case of viscosity prescription (\ref{ref1.7})
a set of equations describing the vertically averaged advective
accretion disks has {\it two} singular
points, independing on $\alpha$ and accretion rate.
Calculations of the advective disk structure are performed
in Paczy\'nski-Wiita potential (Paczy\'nski \& Wiita 1980)

\begin{equation}
\label{ad1}
\Phi(r)=-{GM\over r-2r_g},
\end{equation}
where $M$ is the black hole mass and
$2r_g=2GM/c^2$
is the gravitational radius.
The disk self-gravity is neglected.
Note, that the multiplicity of singular points in
solutions for accretion flows
in Paczy\'nski-Wiita potential (\ref{ad1}) was revealed by Fukue (1987),
Chakrabarti \& Molteni (1993) and Chakrabarti (1996) in somewhat different
context. Another type of accretion flows with multiple singular points
was found by Dullemond \& Turolla (1998) and Turolla \&
Dullemond (2000), who considered accretion disks experienced a phase
transition.
It was shown by Artemova et al. (2001),
that at $\alpha< 0.01$ the inner and
outer (with respect to the black hole location)
singular points are of a saddle
type, and only one integral curve (`separatrix')
which crosses the inner point
simultaneously crosses the outer one.  This separatrix corresponds to the
unique global solution which is determined by two parameters, $\alpha$ and
$\dot{m}=\dot{M}c^2/L_{Edd}$, for the given black hole mass.
At larger $\alpha> 0.1$
the inner singular point changes its type to a node, while the outer point
remains of a saddle-type.  There is still one integral curve which goes
continuously through both the singular points providing the unique global
solution.

In the case of viscosity prescription in the first relation of
(\ref{ref1.8})

\begin{equation}
\label{ad2}
t_{r\phi} = \eta r \frac{d\Omega}{dr} = \rho \nu r \frac{d\Omega}{dr}
\end{equation}
it was found that
there is only one singular point which is always a saddle,
and only one physical solution which passes through this point exists.
Solutions which correspond to both forms of viscosity
(\ref{ref1.7}) and (\ref{ad2})
are very close quantitatively at low $\alpha$ limit, $\alpha< 0.1$.

A numerical method was developed by Artemova et al. (2001)
to solve a set of
equations describing the vertically averaged advective accretion disks.
The method is based on the standard relaxation technique and
explicitly uses conditions at the inner singular point and its vicinity.
These conditions had been obtained by expanding the solution
into power series around the singular point.
Such a modification of the method allowd to construct solutions which
smoothly pass the singular points and satisfy the regularity
conditions at these points
with high computer precision in wide range of parameters
$\alpha$ and $\dot{m}$.

The equations describing the radial disk structure
are written for the midplane density $\rho$, pressure $P$, radial velocity
$v$ and angular velocity $\Omega$. They consist of
the mass conservation equation taken the form,

\begin{equation}
\label{ad3}
\dot{M}=4\pi r h \rho v,
\end{equation}
$\dot{M}>0$, and $h$ is the disk half-thickness,
which is expressed in terms of the isothermal sound
speed $c_{\rm s}=\sqrt{P/\rho}$ of  gas,

\begin{equation}
\label{ad4}
h={c_{\rm s}\over\Omega_K}.
\end{equation}
The equations of motion in the radial (\ref{ref3.3})
and azimuthal direction

\begin{equation}
\label{ad5}
{\dot{M}\over 4\pi}{dj\over dr}+{d\over dr}(r^2 h t_{r\phi})=0,
\end{equation}
where $\Omega_K$ is the Keplerian angular velocity,
$\Omega_K^2=GM/r(r-2r_g)^2$,
$j=\Omega r^2$ is the specific angular momentum and $t_{r\phi}$ is the
($r$, $\phi$)-component of the viscous stress tensor.
Other components of the stress tensor are assumed to be negligiblly small.
The vertically averaged energy conservation equation is written
in (\ref{ref3.1}),(\ref{ref3.2})
\footnote{The vertical averaging in equation (\ref{ref3.2}) have been made
differently by different authors (compare e.g.
Shakura \& Sunyaev 1973, and Abramowicz et al. 1988). Our choice of
the coefficients in (\ref{ad6}),(\ref{ad6a}), may be not the optimal one.
Aposteriory analysis had shown that using the factor 4 instead of 2
in the denominator of (\ref{ad6})
would be more consistent choice, but this change has a
little influence on our numerical results.},
with the advective term in the form

\begin{equation}
\label{ad6}
Q_{\rm adv}=-{\dot{M}\over 2\pi r}\left[{dE\over dr}+P{d\over dr}
\left({1\over\rho}\right)\right],
\end{equation}
and

\begin{equation}
\label{ad6a}
Q_{+}= h t_{r\phi}r{d\Omega\over dr}, \quad
Q_{-}={2 a T^4 c \over 3 \kappa \rho h},
\end{equation}
are the viscous dissipation rate and
the cooling rate per unit surface, respectively,
$T$ is the midplane temperature.
The equation of state for accretion matter consisted of
a gas-radiation mixture is

\begin{equation}
\label{ad7}
c_{\rm s}^2={\cal R}T+{1\over 3}{aT^4\over\rho},
\end{equation}
where ${\cal R}$ is the gas constant.
The specific energy of the mixture is

\begin{equation}
\label{ad8}
E={3\over 2}{\cal R}T+{aT^4\over\rho}.
\end{equation}
Both prescriptions of viscosity (\ref{ref1.7}) and (\ref{ad2})
have been used in calculations with
$\nu$ in the form

\begin{equation}
\label{ad9}
\nu={2\over 3}\alpha c_{\rm s} h.
\end{equation}
Note that in the limit $\Omega\longrightarrow\Omega_K$ both the prescriptions
(\ref{ref1.7}) and (\ref{ad2}) coincide.
Integration of equation (\ref{ad5}) gives

\begin{equation}
\label{ad10}
r^2 h t_{r\phi}=-{\dot{M}\over 4\pi}(j-j_{\rm in}),
\end{equation}
where the integration constant $j_{\rm in}$ has meaning
of the specific angular momentum of
accreting matter near the black hole horizon.
The value of $j_{\rm in}$ is chosen to obtain the global transonic
solution with the subsonic part at large radii and the supersonic part
in vicinity of the black hole horizon.
In the case of viscosity prescription (\ref{ref1.7})
the expression (\ref{ad10})
results in an algebraical equation, and the radial structure of
the accretion disks is described by two
first order differential equations (\ref{ref3.3}) and (\ref{ref3.2}).
In general formulation, these two equations require to fix
two parameters as boundary conditions to determine the solution.
In the case of viscosity prescription (\ref{ad2})
expression (\ref{ad10}) results in
additional first order differential equation, so
we have to fix three parameters as boundary conditions.

We show here two examples of numerical results obtained by
Artemova et al. (2001).
Each calculated model is characterized by the unique value of $j_{\rm in}$.
Figures 3 and 4
show the dependences of $j_{\rm in}$
on the accretion rate $\dot{m}=\dot{M}c^2/L_{\rm Edd}$
for three values of $\alpha=0.01$, $0.1$
and $0.5$ in the case of viscosity prescription
(\ref{ref1.7}) and (\ref{ad2}) respectively.  At low
$\dot{m}< 1$ the value of $j_{\rm in}$ is independent of $\dot{m}$ and
weakly varies with $\alpha$.  In the low $\alpha$ case, $\alpha=0.01$ and
$0.1$, the values of $j_{\rm in}$ are close to the minimum value of the
Keplerian angular momentum, $(j_K)_{\rm min}=3.6742$.  At high $\dot{m}>
0.1$ the values of $j_{\rm in}$ deviate from $(j_K)_{\rm min}$ to larger
or smaller values depending on $\alpha$.  In the case of $\alpha=0.01$ and
$0.1$ one can see only minor differences between
models with different forms of
viscosity.  But, for large
$\alpha=0.5$ the difference in values of $j_{\rm in}$ increases.

Figures 5 and 6
show locations of the inner singular points $(r_{\rm s})_{\rm in}$
as a function of $\dot{m}$
for different values of $\alpha$ and both viscosity prescriptions.
Similar to the case of $j_{\rm in}$ discussed above the models
at low $\dot{m}$ show a weak dependence of
$(r_{\rm s})_{\rm in}$ on $\dot{m}$.
In the low $\alpha$ models (squares and circles in Figures
3-6) the values of $(r_{\rm s})_{\rm in}$ are close to the
location of the black hole last stable orbit at $r=6r_g$.
At high $\dot{m}> 0.1$
the values of $(r_{\rm s})_{\rm in}$ are decreasing functions of $\dot{m}$
in the case of low $\alpha=0.01$ and $0.1$,
and non-monotonically behave in the
case of $\alpha=0.5$ (triangles in Figures 3-6.

In the case of viscosity prescription (\ref{ref1.7}) the solutions have the
outer singular point in addition to the inner one.
The change of value of
$\beta=\frac{\cal{R}T}{c_s^2}$
form 1 to 0 corresponds to the change of a state from the
gas pressure to radiative pressure dominated one.  The thin disks with
$\beta\simeq 1$ are locally stable, whereas parts of the disk in which
$\beta\simeq 0$ are thermally and viscously unstable
(Pringle, Rees, \& Pacholczyk 1973).
At large $\dot{m}> 100$ the instability
can be suppressed by the advection effect (Abramowicz et al. 1988).
It was shown by Artemova et al. (2001), that significant deviations
of $j_{\rm in}$ and $(r_s)_{\rm in}$ in Figures 3-6
from the constant happens when $\beta$ beguns to deviate from the unity.
The analysis of the critical points had shown that
they are either saddle or nodal-type.
In Figures~3-6
the saddle-type points are indicated by the solid
squares, circles and triangles. The nodal-type points are represented by the
corresponding empty dots in the same figures.  In the case of viscosity
prescription (\ref{ref1.7})
the inner singular points can be saddles or
nodes depending on values of $\alpha$
and $\dot{m}$.  Note that the change of type from a saddle to nodal one does
not introduce any features in the solutions.  The outer singular points are
always of a saddle-type.
In the case of viscosity prescription (\ref{ad2}) the
solutions have only one (inner) singular point which is always of a
saddle-type.

The models with $\dot{m}< 16$ and low $\alpha< 0.1$
have values of $r_{\rm s}$ and $j_{\rm in}$ which are
very close to the location of the last stable orbit,
$r_{\rm in}=6r_g$, and
value of $j_{\rm in}=(j_K)_{\rm min}$ assumed in the
standard model (Shakura \& Sunyaev 1973).
The radial structures of the models are also very close
to the ones for the standard model
in the same range of $\dot{m}$ and $\alpha$.
Such a good coincidence means that
the advective terms in equations (\ref{ref3.2}) and (\ref{ref3.3})
are negligiblly small in the considered models.
However, the high $\alpha$ models show quite
significant deviation from the standard model for all
$\dot{m}$ (see Figures~3-6).

At high accretion rates, $\dot{m}> 16$, the effect of advection
becomes significant, and at $\dot{m}=100$ the
advective part of the energy flux
is $\sim 2$ times larger than the luminosity.

\bigskip

%\begin{figure}
%\epsfysize=8cm % fix the y-dimension and scales x-dim. to y-dim.
%\hspace{3.5cm}\epsfbox{fad1.eps} %for centering: act on hspace argument

%\label{fad1}
%\end{figure}

%\begin{figure}
%\epsfysize=8cm % fix the y-dimension and scales x-dim. to y-dim.
%\hspace{3.5cm}\epsfbox{fad2.eps} %for centering: act on hspace argument

%\label{fad2}
%\end{figure}

%\begin{figure}
%\epsfysize=8cm % fix the y-dimension and scales x-dim. to y-dim.
%\hspace{3.5cm}\epsfbox{fad3.eps} %for centering: act on hspace argument

%\label{fad3}
%\end{figure}

%\begin{figure}
%\epsfysize=8cm % fix the y-dimension and scales x-dim. to y-dim.
%\hspace{3.5cm}\epsfbox{fad4.eps} %for centering: act on hspace argument

%\label{fad4}
%\end{figure}

\subsection{Convection in accretion discs}

\subsubsection{Vertical structure and convective instability in the
radiation-dominated disk region}

Let us consider in more details the vertical structure of the accretion disk.
Following Shakura(1972),  Shakura and Sunyaev(1973)
let us suggest that viscous heat production per unit
mass in the accretion disk is constant, what is used also in section 1.6.
For variable vertical density the equation of vertical heat flux
is different from (\ref{ref11.3}),
and has a differential form
with $F_0 \equiv Q_+$ from (\ref{ref1.11}):

\begin{equation}
\label{con1}
\frac{dF_{rad}}{dz}=2F_0\frac{\rho}{\Sigma_0}=
\frac{3}{4\pi} {\dot M}\frac{GM}{r^3}(1-\frac{j_{in}}{j})
\frac{\rho}{\Sigma_0},
\end{equation}
what gives after integration

\begin{equation}
\label{con2}
F_{rad}=2F_0\frac{\Sigma}{\Sigma_0}.
\end{equation}
Here ${\Sigma}$ is the mass in the layer $(0,z)$, so that
${\Sigma}(h)=\Sigma_0/2$, $d\Sigma=\rho\,dz$.
The equation of a heat conductivity in the optically thick disk in
vertical direction is written as

\begin{equation}
\label{con3}
F_{rad}=-\frac{ac}{3\kappa \rho}\frac{dT^4}{dz}=
-\frac{ac}{3\kappa}\frac{dT^4}{d\Sigma}.
\end{equation}
We consider here only regions where electron scattering dominates,
so $\kappa=\kappa_e$.
Solving (\ref{con3}) together with (\ref{con2}) we obtain

\begin{equation}
\label{con4}
T^4=T_0^4\left(1-\frac{3\kappa F_0 \Sigma^2}{ac \Sigma_0 T_0^4}\right).
\end{equation}
At the surface of the disk

\begin{equation}
\label{con4a}
\Sigma=\Sigma_0/2, \qquad T=T_s,\qquad
F_0=\frac{acT_s^4}{4},
\end{equation}
like in black-body radiation, so we obtain
from (\ref{con4})

\begin{equation}
\label{con5}
T_0^4=T_s^4\left(1+\frac{3\kappa \Sigma_0}{16}\right).
\end{equation}
Assuming high optical depth $\Sigma_0 \kappa \gg 1$, and using (\ref{con5})
in (\ref{con4}) we obtain the temperature distribution over the thickness of
the disk in the form

\begin{equation}
\label{con6}
T^4=T_0^4\biggl[1-\left(\frac{2\Sigma}{\Sigma_0}\right)^2\biggr].
\end{equation}
Note, that equation (\ref{ref1.12}) follows from (\ref{con5}) at high
optical depth, after using black-body connection between $T_s$ and $F_0$.

Consider now a vertical density dependence in the accretion disk.
In the radiation-dominated region with $P\approx P_r=\frac{aT^4}{3}$
we obtain from the vertical equilibrium equation (\ref{ref1.2})

\begin{equation}
\label{con7}
\frac{a}{3\rho}\frac{dT^4}{dz}=-\frac{GMz}{r^3}.
\end{equation}
Using (\ref{con2}),(\ref{con3}),(\ref{con7}), we obtain

\begin{equation}
\label{con8}
\Sigma=\frac{GMc \Sigma_0}{2r^3\kappa F_0}z.
\end{equation}
After differentiating, we obtain that in the radiation-dominated disk
the density does not depend on the vertical coordinate $z$

\begin{equation}
\label{con9}
\rho=\frac{GMc \Sigma_0}{2r^3\kappa F_0}.
\end{equation}
In the constant density layer with falling temperature the entropy

\begin{equation}
\label{con10}
S=\frac{4}{3}\frac{aT^3}{\rho}
\end{equation}
is falling outside in the direction of the gravitation force. It
corresponds to convectively unstable situation, what was noticed first
by Bisnovatyi-Kogan and Blinnikov (1976, 1977).

For outer region of the accretion disk with $P_g \gg P_r$, $\kappa=\kappa_e$,
there is no analytic solution for a vertical disk structure. We may
however to check the convective instability in another way.
In the gas-pressure dominated region
the entropy

\begin{equation}
\label{con11}
\frac{S}{\cal R}=\log(\frac{T^{3/2}}{\rho})+{\rm const}, \quad
\frac{dS}{\cal R}=\frac{3}{2}\frac{dT}{T}-\frac{d\rho}{\rho},
\end{equation}
and with $P=P_g={\cal R}\rho T$ it
follows from the vertical equilibrium equation (\ref{ref1.2})

\begin{equation}
\label{con12}
\frac{dT}{T}+\frac{d\rho}{\rho}=-\frac{GMz}{r^3P}d\Sigma.
\end{equation}
We obtain from (\ref{con6})

\begin{equation}
\label{con13}
\frac{dT}{T}=-\frac{1}{2}\frac{T_0^4-T^4}{T^4}\frac{d\Sigma}{\Sigma}.
\end{equation}
Using (\ref{con12}),(\ref{con13}) in (\ref{con11}) we obtain

\begin{equation}
\label{con14}
\frac{dS}{\cal R}=-\frac{5}{2}\frac{dT}{T}+\frac{GMz}{r^3P}d\Sigma
= \left(\frac{GMz}{r^3P}\Sigma
-\frac{5}{4}\frac{T_0^4-T^4}{T^4}\right)\frac{d\Sigma}{\Sigma}
\end{equation}
It is easy to prove that entropy is growing with height in the region near
the equatorial plane $z \ll h$.
Here we have $\Sigma\approx \rho_0 z \ll \Sigma_0$, and it follows from
(\ref{con6}),(\ref{con13}),(\ref{con14})

\begin{equation}
\label{con15}
\frac{dS}{\cal R} \approx \left(\frac{GM\Sigma_0^2}{r^3P_0 \rho_0}-5\right)
\frac{\Sigma \, d\Sigma}{\Sigma_0^2}.
\end{equation}
Calculating the last expression in brackets, using the accretion disk
solution (see e.g. Shakura \& Sunyaev (1973)), we obtain

$$
\frac{dS}{\cal R}= (8-5)\frac{\Sigma \, d\Sigma}{\Sigma_0^2} >0,
$$
what means stability relative to convection.
Similarily we prove a stability relative to convection in the region close to
the surface of accretion disk, where
$\Sigma=\frac{\Sigma_0}{2}$, $T=T_s$, $z=h$, $P=P_s$. From the
theory of stellar atmospheres (see e.g. Bisnovatyi-Kogan (2001)) we have
a condition at the photosphere, corresponding to the optical depth
$\tau=2/3$, in the form
$P_s=\frac{2}{3}\frac{g}{\kappa}$, where

\begin{equation}
\label{con15a}
g=\frac{GMh}{r^3}
\end{equation}
is the gravitational acceleration at the disk surface.
Using these relations and (\ref{con5}) in (\ref{con14}) we get

$$
\frac{dS}{\cal R} \approx \left(\frac{GM\Sigma_0 h}{2r^3P_s}
-\frac{5}{4}\frac{T_0^4}{T_s^4}\right)\frac{d\Sigma}{\Sigma}
$$
$$
=\left(\frac{GM h}{r^3P_s\kappa}
-\frac{15}{32}\right)\kappa d\Sigma=
\left(\frac{3}{2}-\frac{15}{32}\right) \kappa d\Sigma >0
$$
So, the accretion disk in a standard model
is convectively unstable in the radiation dominated
region, which is the nearest to the black hole, and is stable
in the adjusting region. The most outer gas-pressure dominated regions
with $\kappa_p \gg \kappa_e$ are also stable relative to
convection until the gas is fully ionized. In the colder
regions with incomplete
ionization a behaviour of the accretion disk
becomes more complicated, with
a non-unique solution, and convective instability
(Cannizzo, Ghosh \& Wheeler, 1982).

\subsubsection{Structure of convective accretion disk region}

In a standard model with a radiative heat conductivity in vertical direction
the radiation-dominated region exist only when (Shakura and Sunyaev, 1973)

\begin{equation}
\label{con16}
\frac{L}{L_{\rm Edd}} \ge \frac{1}{50}
\left(\frac{\alpha M}{M_{\odot}}\right)^{-1/8}.
\end{equation}
Existence of convection changes the structure of the radiation-dominated
region, and condition (\ref{con16}). Let us estimate a part of the
energy flux carried out by convection, using a mixing length model
of the convection (Bisnovatyi-Kogan and Blinnikov, 1976).

The convective heat flow $Q_{conv}$ (ergs/cm$^2$/s), the convection velocity
$v$, and the excess $\Delta\nabla T$ of the temperature gradient above  the
adiabatic gradient (we shall take the mixing length $l$ to be the
half-thickness $z_0$ of the disk) are given by the equations

\begin{equation}
\label{con17}
Q_{conv} = c_p \rho v (z_0 /2) \Delta\nabla T ,
\end{equation}
where
$$
c_p={\cal R}\biggl[\frac{5}{2}+20\frac{P_r}{P_g}
+\left(\frac{4P_r}{P_g}\right)^2\biggr],
$$
\begin{equation}
\label{con18}
v = \left[ \frac{Q_{conv}(1 + 4P_{r}/P_g)}{2\rho c_p T}
\frac{GM}{r^3}\right]^{1/3} z_0^{2/3},
\end{equation}

\begin{equation}
\label{con19}
\Delta\nabla T = \left(\frac{4Q_{conv}}{\rho c_p}
\right)^{2/3} \left[ \frac{T}{(1 + 4P_r/P_g)GM}\right]^{1/3}
 r\,z_0^{-5/3}.
\end{equation}
Hence we find that the excess
$\Delta\nabla T$ in the temperature gradient
does not exceed $20\%$ of  $\nabla T$, while
the convection velocity is $v = 3 \cdot 10^8$ cm/sec for $r = 10 r_g$,
what is close to the velocity $v_s$ of the sound in this region.
The heat flow is carried mainly by convection $(Q \approx Q_{conv})$,
because of a high matter density, and the radiative flux is
$\sim 20\%$ of the total flux. Convection is smoothing out the entropy
in vertical direction, so with a precision $\sim 20\%$ we may
consider the entropy as a constant in the vertical direction.

Bisnovatyi-Kogan and Blinnikov (1977) had obtained an
analytic solution for the
isentropic convective accretion disk in the radiation-dominated region.
When $P_r \gg P_g$ the surface pressure at the photosphere with
$\tau = 2/3$ is equal to

\begin{equation}
\label{con20}
P_s=\frac{4g}{3\kappa}=\frac{4GMh}{3\kappa r^3}.
\end{equation}
The effective temperature at the disk surface is connected with the
surface radiative pressure in (\ref{con20}), and with the surface
radiative flux in (\ref{ref1.11}) as

\begin{equation}
\label{con21}
\frac{aT_s^4}{3}=P_s=\frac{4GMh}{3\kappa r^3};\qquad
\frac{acT_s^4}{4}=Q_+=\frac{3}{8\pi}{\dot M}\frac{GM}{r^3}
\left(1-\frac{j_{in}}{j}\right).
\end{equation}
The disk thickness is obtained from comparison of two expressions in
(\ref{con21}) as

\begin{equation}
\label{con22}
h=\frac{3\kappa}{8\pi c}{\dot M}
\left(1-\frac{j_{in}}{j}\right).
\end{equation}
In the radiative-dominated region the equation of state at constant
entropy, using (\ref{con10}), is writetn as

\begin{equation}
\label{con23}
P=\frac{aT^4}{3}=\left(\frac{3 S^4}{256 a}\right)^{1/3} \rho^{4/3} \equiv
  K_r \rho^{4/3}.
\end{equation}
Integration of the equation of vertical equilibrium (\ref{ref1.2}) gives

\begin{equation}
\label{con24}
\rho^{1/3}=\rho_0^{1/3}-\frac{GMz^2}{8K_r r^3}.
\end{equation}
Using the surface boundary condition
$$
\rho_s \equiv \rho(h)=\left(\frac{P_s}{K_r}\right)^{3/4}
=\left(\frac{4GMh}{3\kappa r^3 K_r}\right)^{3/4}
$$
we determine the equatorial density $\rho_0$ from (\ref{con24}).
Assuming $P_0 \gg P_s$ and $\rho_0 \gg \rho_s$, we obtain

\begin{equation}
\label{con25}
\rho_0^{1/3}=\frac{GMh^2}{8K_r r^3}, \qquad
P=K_r \left(\frac{GMh^2}{8K_r r^3}\right)^4
\left(1-\frac{z^2}{h^2}\right)^4
\end{equation}
$$
=P_0\left(1-\frac{z^2}{h^2}\right)^4, \qquad
\rho=\rho_0\left(1-\frac{z^2}{h^2}\right)^3.
$$
The equatorial pressure is obtained from the angular momentum equation
(\ref{ref1.4}) and (\ref{con22}) as

\begin{equation}
\label{con26}
P_0=\frac{2c\Omega_K}{3\kappa \alpha},\qquad
T_0^4=\frac{2c\Omega_K}{a\kappa \alpha},
\end{equation}
and from the boundary condition (\ref{con4a}) we
obtain

\begin{equation}
\label{con27}
P_s=\frac{4F_0}{3c}, \qquad
\frac{P_s}{P_0}=2\alpha\frac{\Omega_K h}{c} \ll 1,
\end{equation}
what justifies the previous assumption.
We used here the relation following from (\ref{ref1.11}) and (\ref{con22})

\begin{equation}
\label{con28}
F_0=\frac{c\Omega_K}{\kappa} h.
\end{equation}
Note, that we used here the exact angular momentum equation (\ref{ref1.4})
in the plane of the disk, instead of the approximate equation obtained by
averaging over the thickness, used by Bisnovatyi-Kogan and Blinnikov (1977).
In this approach the equations (\ref{con26})-(\ref{con28}) are valid
always in the radiation-dominated region, both radiative and convective.
The convection does not change the pressure and temperature distribution
over the thickness of the disk, and the thickness inself,
changing only its density distribution.
The temperature distribution in the radiative disk (\ref{con6})
at constant density, and corresponding pressure distribution are identical to their
distribution in the convective disk (\ref{con25}).

The density and surface density in the radiative disk are found from
(\ref{con4a}), ({\ref{con5}), (\ref{con26}) and (\ref{con28}) as

\begin{equation}
\label{con29}
\rho_0=\frac{4c}{3\kappa\alpha}\frac{1}{\Omega_K h^2}, \qquad
\Sigma_0=2\rho_0 h
=\frac{8c}{3\alpha}\frac{1}{\kappa \Omega_K h}.
\end{equation}
In the isentropic convective disk the density distribution is given in
(\ref{con25}), the equatorial density $\rho_0$,
surface density $\Sigma_0$,
constant $K$ and entropy $S$ are found from (\ref{con25}),(\ref{con26})
as

\begin{equation}
\label{con30}
\rho_0=\frac{16c}{3\kappa\alpha}\frac{1}{\Omega_K h^2}, \qquad
\Sigma=\frac{32}{35}\rho_0 h
=\frac{512c}{105\kappa\alpha}\frac{1}{\Omega_K h}, \qquad
\end{equation}
$$
K=\left(\frac{3\alpha\kappa}{2c\Omega_K}\right)^{1/3}
\frac{(\Omega_K h)^{8/3}}{16}, \qquad
S=\left(\frac{a\alpha\kappa}{2c\Omega_K}\right)^{1/4}
\frac{\Omega_K^2 h^2}{2}.
$$
It follows from (\ref{con29}),(\ref{con30}), that in the equatorial plane
the density of the convective disk is 4 times larger, than in the
radiative one, and in the surface density the difference is
only 64/35 times.
Increase of the convective disk density at the same temperature
narrows the radiative-dominated region, which may exist
now at luminosities $\sim 2$ times larger than in the radiative one
(\ref{con16}).

\subsubsection{Numerical simulations of convection in accretion disks}

2-D numerical simulations had confirmed in general the conclusion
about the isentropic vertical structure of the disk in convectivly
unstable regions.
Fujita and Okuda (1998) had obtained that the excess
$\Delta\nabla T$ in the temperature gradient does not exceed $\sim 1\%$,
what is even closer to the adiabatic structure, than followed from
estimations of Bisnovatyi-Kogan and Blinnikov (1976). It probably
happend, because in the computations
convective cells generally stretch from the disk mid-plane to the
disk surface, increasing the efficiency of the convective mixing in
vertical direction, relative to the model of isotropic convection used in
estimations.
 2-D calculations made by Agol et al. (2000) also had shown a satisfactory
correspondence with the analytical solution of the isentropic
convective disk structure of Bisnovatyi-Kogan and Blinnikov (1977).

2-D numerical simulations of accretion of highly viscous nonradiating
gas had been performed by Igumenshchev (2000), and Igumenshchev and
Abramowicz (1999,2000).
Ideal gas with an adiabatic index $\gamma$ was considered, so
that pressure was connected with the density and specific internal energy
$\varepsilon$ in the form

\begin{equation}
\label{con31}
P=(\gamma-1)\rho\varepsilon.
\end{equation}
Calculations have been done in a range of viscous parameters
$\alpha=0.01 - 1$, and $\gamma=\, 4/3, \, 3/2, \, 5/3$.
Convective instability or large scale circulation occured at smaller
$\alpha$ and $\gamma$.
At $\gamma=4/3$ all models with $\alpha \le 0.3$ are unstable;
at $\gamma=3/2$ and 5/3
the instability takes place at $\alpha \le 0.1$, and
$\alpha \le 0.03$ respectively.
The calculations have been performed
also with account
of a turbulent heat conductivity, characterized by non-dimensional
Prandtl number

\begin{equation}
\label{con32}
Pr=\frac{\nu}{\chi},
\end{equation}
where $\chi=\lambda/\rho$ is the thermometric conductivity, and kinematic
turbulent viscosity coefficient $\nu$ is given in (\ref{ad9}).
Stabilization of the flow had been obtained at $Pr=1$ for all $\gamma$
with $\alpha \ge 0.1$, and for $\alpha=0.03$ at $\gamma=5/3$.

The high-viscosity models at $\alpha=1$ form powerful bipolar outflows.
The pure inflow and large-scale circulation patterns occure in the
moderate viscosity models $\alpha=0.1-0.3$. All models with bipolar
outflows and pure inflows are steady. The models with large scale
circulation could be either steady or unsteady, depending on values
of $\alpha$ and $\gamma$. All convective models are unsteady.
The self-similar solution
for accretion flows with bipolar outflows
(advection-dominated inflows-outflows ADIOs) proposed by
Blandford and Begelman (1999) have not been confirmed in the
numerical simulations of Igumenshchev and Abramowicz (2000).

3-D numerical simulations of viscous non-radiating accretion flows made
by Igumenshchev et al. (2000) have confirmed qualitatively
and quantitatively those obtained in 2-D simulations. It was shown
that convective eddies are nearly axisymmetric and transport angular
momentum inward, tending to smooth out the angular momentum distribution.
This is in contrast with a small scale turbulence, acting as a viscosity,
transporting angular momentum outside, and tending to smooth out
the angular velosity distribution.

\subsubsection{Hot corona and Cyg X-1 variability}

In the convective disk the flow $Q_{ac}$ of acoustic energy
in the vertical direction 1s given by Bierman and L\"ust (1960) as
\begin{equation}
Q_{ac} \simeq \rho v^3 (v/v_s)^5 \simeq 10^{21-22}
{\rm erg} \cdot {\rm sec}^{-1} \cdot {\rm cm}^2,
\end{equation}
and is a quantity of the same order as the total energy
flux for characteristic parameters of Cyg X-1
(Bisnovatyi-Kogan and Blinnikov, 1976).

Acoustic waves generated in
the convection zone escape into optically thin layers,
induce variable soft X rays in the photosphere, and also are
responsible for variable heating of the corona. Comptonization of the
photospheric radiation by hot electrons under conditions where both
temperature and density are variable should lead to flux variations in the
hard range, h${\nu} \ge $ 5 keV. This X-ray fluctuation mechanism, involving
the emergence of waves into transparent layers, evidently is of the same
wave nature as the variability of the ultraviolet excess in stars
experiencing intensive convection, such as T Tauri and UV Ceti.

 The waves escaping into the transparent layers and producing
variable radiation in the photosphere and corona occupy a rather
narrow frequency band. Physically, the reason for this circumstance is
the following.
A media with a high radiation pressure and a non-uniform distribution
of plasma along the z coordinate (across the disk) serve as efficient
filters, picking out a characteristic frequency range from the broader
spectrum generated by convection and turbulence. Waves of low
frequency and a wavelength exceeding the scale height of the
atmosphere will not escape outside but will induce oscillations of
the coronal atmosphere as a whole. On the other hand, under conditions
where radiation pressure predominates, high-frequency waves
experience a severe damping because of radiative friction, and their
role in heating the corona will be insignificant.

 The propagation of acoustic waves through
a medium with strong radiation pressure, followed by their escape into
the atmosphere had been investigated in details by Bisnovatyi-Kogan and
Blinnikov (1978a,b; 1979), and by Agol and Krolik (1998) for MHD waves.
Waves emerging into the transparent layers will have a phase
and group velocity equal to the velocity of sound in gas, v$_g = (\gamma P_g
/\rho)^{1/2} = (\gamma{\cal R}T)^{1/2}$, where $\gamma = $5/3 or $\gamma =$ 1
 according as scattering or absorption predominates, ${\cal R}$ is the gas
constant, and the temperature  T  of the equilibrium atmosphere is
  approximately equal to the temperature T$_{\rm s}$ of the photosphere.
   The characteristic frequency $\omega_{\rm c}$ of waves
emerging into the atmosphere is given by the expression

\begin{equation}
\label{cor1}
\omega_{\rm c}= \left(\frac{\gamma}{{\cal R}T}\right)^{1/2}
g \left(1 - \frac{F}{F_c}\right) {\rm sec}^{-1} .
\end{equation}
For the accretion disk model, the gravitational acceleration g at radius
r is given by (\ref{con15a}), and
 characteristic thickness $h$ is given by (\ref{con22}).
The quantity
$F/F_{\rm c} = \kappa F/gc$  represents the ratio of the
 radiation pressure force to the
gravity; $\kappa$ is the opacity, including both absorption and
scattering.
In a spherically symmetric star of luminosity $L$ and radius $R$ we
have
$g = GM/R^2$, $F = L/4 \pi R^2$, and $F/F_c = L/L_c$,
where $L_c = 4 \pi cGM/\kappa$
is the Eddington limiting luminosity.

Characteristic frequencies $\omega_c$ from (\ref{cor1})
of fluctuations
of Cyg X-l in the convective accretion-disk model
of Bisnovatyi-Kogan and Blinnikov (1977),
with a black hole mass M = 10 M$_{\odot}$,
 luminosity between 0.1 L$_c$ and 0.3 L$_c$,
corresponding to the photospheric temperature $T_s$,
lay in the range 5-40 msec. If a corona
with T$_{cor}$ $\approx \,\, 10^2$ T$_s$ is present,
 waves whose frequency is $\omega = \omega_{\rm c}$ or even
little lower will be able to escape. This frequency range
is in a good accord with the observed time-scales of
variability (Bolt, 1977).

Due to strong radiative damping of the sound waves, only small part
of acoustic energy flux of order

\begin{equation}
\label{cor2}
Q_{cor} \sim (P_g/P_r)Q_{ac}
\end{equation}
 is expended
in heating the outer layers with an optical depth $\tau < 1$.
Another mechanism is acting to heat these layers
(Bisnovatyi-Kogan and Blinnikov, 1977).
A particle in a region near the surface of the disk that is
 transparent to radiation will be subject to the influence of radiation from
the whole disk, and not only to the local radiation pressure
 gradient. The force of the radiation pressures will
accelerate particles in the transparent region above the
disk. Calculations of the equations of motion for particles
subject to radiative, centrifugal, and gravitational forces
show that in the region with $\tau < 1$ the particles (protons
and electrons) will acquire vertical velocities corresponding to the temperatures
$(1-8)\cdot 10^8$ K of protons for
$L \approx 0.1 L_c$.
Turbulent relaxation of the particle motions in
the corona will tend to equalize the mean energies of the
ions and electrons and to form a quasi-Maxwellian
particle velocity distribution. Thus the combined action of the
two heating mechanisms we have described will produce
around the accretion disk a hot corona with $T_e \approx 10^9$ K
for $L \approx 0.1L_c$. The existence of an analogous corona has
been postulated phenomenologically by Price and Liang (1977).

Let us estimate the density $\rho_{cor}$ of the corona and
the amount of material it contains. Since the gas
 pressure varies continuously with transition from the
photosphere to the corona, we readily find that at the base of
the corona

\begin{equation}
\label{cor3}
\rho_{cor} \simeq \rho_s T_s/T_{cor} \simeq 10^{-2} \rho_s
\approx 10^{-5}\, {\rm g/cm}^3,
\end{equation}
where $\rho_s$, $T_s$ are the density and temperature in the
photosphere of the disk for the parameters of the source
Cyg X-1 in the region of maximum energy release. The
surface density of the corona is determined by the
condition $\tau _{es} \approx 1$ and is $\approx\,\, 2 {\rm g/cm}^2$,
which is more than an
order of magnitude lower than the density of the opaque
disk.

The existence of a hot corona can easily explain the
peculiarities of the Cyg X-1 spectrum during its variations
(Bisnovatyi-Kogan and Blinnikov, 1976).
The soft $X$-rays
at $h\nu \leq 7$ keV, which comprise $\approx 70\%$ of the total flux,
are formed in the photosphere of the opaque disk. Some of
the radiation $(\approx 10\%)$ passing through the hot corona,
experience
comptonization and transforms into a hard radiation up to
$h\nu \approx 3kT_e \approx 200$ keV, which amounts up
to $\approx 30\%$ of the total flux.
In Cyg X-1 the total luminosity varies around $L = 0.1 L_c$.

The changes in the spectrum as the luminosity of
Cyg X-1 varies (Holt et al., 1975) exhibit the following characteristic
behavior. As the total energy flux rises, the radiation in
the soft part of the spectrum ($h\nu \leq 7$ keV) increases, but
in the hard range ($h\nu \geq 10$ keV) it remains almost
constant, or perhaps may even decrease slightly. We shall
assume that the variations in the luminosity are associated
with variations in the power of accretion. As the mass
flow ${\dot M}$ rises in the region with $P_r \gg P_g$ the fraction of
the acoustic flow (\ref{cor2}) used for heating of the corona
decreases:

\begin{equation}
\label{cor4}
Q_{cor}=(P_g/P_r)Q_{ac} \sim \dot M^{-1}.
\end{equation}
For $L \approx 0.1L_c$, when acoustic heating predominates,
the rise in $\dot M$ may cause some decrease in the heating
of the corona and in the amount of hard radiation. At the
same time the flux in the soft range is determined by the
radiation of the disk photosphere and increases
$\sim \dot M$. In strong bursts of luminosity,
when $L$ reaches about $0.3L_c$, the heating begins to
be governed by radiation-pressure forces, so the
temperature of the corona and thereby also the power of
the hard radiation should increase along with the rise of $\dot M$
and total energy flux.
Similar explanation of spectral transitions in Cyg X-1
was considered by Poutanen et al. (1997).

Even if a corona with $T \simeq 10^9$ K is present in the disk
accretion model, radiation at $E_{\gamma} \ge 200$ keV
cannot be produced, and more energetic electrons are needed.
Radiation with energies $E_{\gamma}$ up to 5 MeV was
oserved by CRGO from Cyg X-1 (McConnell et al., 2000).
It was noted by Bisnovatyi-Kogan and Blinnikov (1976), that explanation of
hard X-ray and gamma radiation from Cyg X-1 would need a presence of
a magnetic field in the accretion disk.
 Two ways to form very energetic electrons had been considered.
Both required the presence of a magnetic field in the disk.
A magnetic field could exist in the disk either through
twisting of the lines of force by differential rotation
(Shakura and Sunyaev, 1973),
or through infall onto the black hole of magnetized
material having a small angular momentum
(Bisnovatyi-Kogan and Ruzmaikin, 1976). In the latter
case a poloidal magnetic field would be generated.

In the binary system containing the source Cyg X-1,
some of the material flowing from the giant star is
dispersed in space near the system. The attraction of the
black hole will not only produce an accretion disk.
A small proportion of material having a low angular momentum
will fall into the black hole and be decelerated in the
disk. If this deceleration takes place at radii of
$(10-30)r_g$ and if a thin collisionless shock wave is formed
(as is very likely in the presence of an azimuthal magnetic field)
wherein the kinetic energy is transformed into thermal
energy and $T_e \approx T_i$, then hot electrons with
$T= 10^{10} - 10^{12}$ K
will appear. The inverse Compton mechanism of interaction
of the disk radiation with these electrons can
lead to the generation of hard radiation with $h\nu \approx
200-2000$ keV or even higher energies.

Another mechanism for producing fast particles is
analogous to the pulsar process. If magnetized matter
with low angular momentum falls into the black hole (in
addition to the disk accretion), a strong poloidal
magnetic field will arise (Bisnovatyi-Kogan and Ruzmaikin, 1976).
 By analogy to pulsars (Goldreich and Julian, 1969),
 rotation will generate an electric field of strength $E \approx
-(v/c)B$ in which electrons are accelerated to energies $\varepsilon \approx
R(v/c)Be \approx 3\cdot 10^4 [B/(10^7 gauss)]$ Mev
 where $v/c \approx 0.1$ and
$R \approx 10^7$ cm is the characteristic scale.
 In a field $B \approx 10^7$
 gauss, such electrons will generate synchrotron radiation
 with energies up to
$\approx 10^5$ keV. Just as in pulsars, it would be possible here for
$e^{+}e^{-}$  pairs to be formed and to participate in the synchrotron
radiation.

\section{Boundary layers between disks and rotating Stars}

\subsection{Boundary Layers}

\subsubsection{Structure of the boundary layer.
Fitting with the disk solution}

Inside the boundary layer (BL) variables change considerably over the small
thickness of the layer $H_b \ll r_{in}$. Matter has no room to accelerate
in radial direction, so the radial velocity term in
(\ref{2.12}) is negligible, but
the pressure term is comparable with the gravitational and centrifugal forces
(Regev, 1983; Papaloizou and Stanley, 1986; Regev and Hougerat, 1988).
The thickness $H_b$ of the BL is smaller then its vertical
size $z_0$, which also remains small. The adopted
inequalities for the boundary
layer parameters

\begin{equation}
\label{4.1}
H_b \ll z_0 \ll r_{\ast}
\end{equation}
will be confirmed by the results. The radius of the star $r_{\ast}$
differs from the radius, at which ${\partial \Omega \over \partial r}=0 $,
by the very small value $H_b$. In the asymptotic consideration of the BL
we use $r_{\ast}$ as an inner boundary for the disk solution

\begin{equation}
\label{4.2}
r_{in}=
r_{\ast}.
\end{equation}
The variable $x$

\begin{equation}
\label{4.3}
 r=r_{\ast}+\delta x,\quad  \delta={H_b \over r_{\ast}} \ll 1
\end{equation}
is used inside BL instead of $r$. The inner solution within BL is looked for
in the region $0<x< \infty$, while the outer solution
(\ref{3.18})-(\ref{3.23}) is
valid in $r_{\ast}<r<\infty $. According to the method of matched
asymptotic expansion (MAE) (see Nayfeh, 1973)
the inner and outer solutions are fitted so that
values for the outer solution at $r=r_{\ast}$ are equal to the
corresponding values of the
inner solution at $x=\infty$. This condition is
valid asymptotically at $\delta \rightarrow 0 $.

Consider a stationary accretion disk BL where equations
(\ref{2.12}), (\ref{3.15}),(\ref{3.16})
are valid with the integration constant $\xi_b$ for
$\xi$ in (\ref{3.18}). The thickness of the disk remains small in the
boundary layer, so relations (\ref{3.1}) -(\ref{3.8})
for the polytropic case and
(\ref{3.10})-(\ref{3.14}) for the
isothermal case are valid. Accounting only for the main
terms in the asymptotic expansion inside BL, we get the
equations from (\ref{2.12}), (\ref{3.16}) (Regev,1983)

\begin{equation}
\label{4.4}
{d P_0 \over dx}=-\Omega_{K \ast}^2 H_b \rho_0
(1-\omega^2),
\end{equation}

\begin{equation}
\label{4.5}
{d\omega \over dx}=-{\dot M H_b \over 2 \pi \Sigma \nu_b r_{\ast}^2}
(\xi_b-\omega),
\end{equation}
where
\begin{equation}
\label{4.6}
\omega = {\Omega \over \Omega_{K \ast}},\quad \Omega_{K \ast}^2=
{GM \over r_{\ast}^3}.
\end{equation}
The equation (\ref{4.4}) was obtained from the exact equation of radial
equilibrium (\ref{2.12}), instead of the approximate equation averaged
over the disk thickness, used originally by Regev (1983). In the MEA
approximation these two equations are identical (Shorokhov, 2000), in
the case of an advective disk the difference is essential, and the exact
equation (\ref{2.12}) should be used.
The viscosity coeffisient $\eta$ is expressed in BL
through the coefficient of kinematical viscosity $\nu_b$

\begin{equation}
\label{4.7}
\eta = \rho_0 \nu_b.
\end{equation}
While
the radial extend of the BL $H_b$ is much less then its vertical size $z_0$
the formula (\ref{F29})
cannot be used for a viscosity coefficient approximation.
Inside BL we thus use the $\alpha$ approximation in the form

\begin{equation}
\label{4.8}
\nu_b=\alpha_b u_{s0} H_b.
\end{equation}

 \subsubsection{Polytropic case}

This case was investigated by Bisnovatyi-Kogan (1994).
It is convenient to use the variables ($\Sigma, \omega$) in the equations
(\ref{4.4}) ,(\ref{4.5}) . We get from (\ref{3.5}) -(\ref{3.7})

\begin{equation}
\label{4.9}
\rho_0=d_{\rho 0} \Sigma^{\frac{2n}{2n+1}}, \quad
P_0=d_{P0} \Sigma^{2(n+1)\over 2n+1},
\end{equation}

\begin{equation}
\label{4.10}
u_{s0}=d_{s0} \Sigma^{1 \over 2n+1},
\end{equation}

\begin{equation}
\label{4.12}
d_{\rho 0 }=
\biggl[{1 \over \sqrt{2 \pi}}
{\Gamma\left(n+{3 \over 2}\right) \over \Gamma(n+1)}\biggr]^{2n \over 2n+1}
\biggl[ {GM \over r_{\ast}^3 K(n+1)}\biggr]^{n \over 2n+1},\,\,\,
d_{P0}=Kd_{\rho 0}^{\frac{n+1}{n}},
\end{equation}

\begin{equation}
\label{4.11}
d_{s0}=\biggl[\frac{n+1}{n} K\biggr]^{\frac{1}{2}}d_{\rho 0}^{\frac{1}{2n}}=
\end{equation}
$$
\left({n+1 \over n}K\right)^{1/2}
\biggl[{1 \over \sqrt{2 \pi}}
{\Gamma\left(n+{3 \over 2}\right) \over \Gamma(n+1)}\biggr]^{1 \over 2n+1}
\biggl[ {GM \over r_{\ast}^3 K(n+1)}\biggr]^{1 \over 2(2n+1)}.
$$
Write for convenience the dimension of $d_{s0}$

\begin{equation}
\label{4.13}
[d_{s0}]=[v][\rho r]^{-{1 \over 2n+1}}.
\end{equation}
From the matching conditions in MAE we need to get Keplerian angular
velocity $\Omega = \Omega_{K \ast}, \quad \omega=1 $ from the inner solution
at $x \rightarrow \infty$ in order to fit the outer solution
(\ref{3.17}) at the
inner boundary $r=r_{\ast}$. This implies $\xi_b=1$ for the constant in
(\ref{4.5}) .
We get after transition to the variable $\Sigma$
in (\ref{4.4}), (\ref{4.5})

\begin{equation}
\label{4.14}
{d \Sigma \over dx}=-{d_{\rho 0} \over d_{P0}}{2n+1 \over 2(n+1)}
\Omega_{K \ast} ^2 H_b(1-\omega^2)\Sigma^{2n-1 \over 2n+1},
 \end{equation}

\begin{equation}
\label{4.15}
{d \omega \over dx}=-{\dot M \over 2 \pi \alpha_b r_{\ast}^2 d_{s0}}
(1-\omega) \Sigma^{-2{n+1 \over 2n+1}}.
\end{equation}
Dividing (\ref{4.14}) on (\ref{4.15}) we get

\begin{equation}
\label{4.16}
 D_b{d\Sigma \over d \omega}=-(1+\omega) \Sigma^{4n+1 \over 2n+1}
\end{equation}
with

\begin{equation}
\label{4.17}
D_b=-{2(n+1) \over 2n+1}{d_{P0} \over d_{\rho 0} d_{s0}}{\dot M \over 2 \pi
\alpha_b H_b}{1 \over r_{\ast}^2 \Omega_{K \ast}^2}.
\end{equation}
The solution of (\ref{4.16}) which fit the boundary condition on the stellar
surface $\Sigma=\Sigma_{\ast}$ at $\omega=\omega_{\ast}$.
Taking into account, that the surface density rapidly grows inter the star
we may put with sufficient accuracy $\Sigma_{\ast}=\infty $ and obtain
the solution, using (\ref{4.11}),
(\ref{4.12}) and (\ref{4.17}) in the form

\begin{equation}
\label{4.19}
\Sigma=
d_{\Sigma}{\sqrt{K \Omega_{K \ast}^{1
\over n}} \over (\Omega_{K \ast} r_{\ast})^
{2n+1 \over n}}\left(-{\dot M \over \alpha_b H_b}\right)^{2n+1 \over 2n}
\biggl[(\omega-\omega_{\ast})\left(1+{\omega+\omega_{\ast} \over 2}
\right)\biggr]^{-{2n+1 \over 2n}},
\end{equation}
where

\begin{equation}
\label{4.20}
d_{\Sigma}=\sqrt{n+1} n^{-{2n+1 \over 4n}}(2 \pi)^{-{4n+3 \over 4n}}
\biggl[{\Gamma \left(n +{3 \over 2}
\right) \over \Gamma(n+1)} \biggr]^{1 \over 2n}.
\end{equation}
For fitting the inner BL solution and
the outer solution for the accretion disk we must
make the surface densities (\ref{4.19}) at $\omega=1$ and (\ref{3.21}) at
$r=r_{in}$ equal.
This fitting uniquely determines the outer integration constant
$\xi$. After some algebraic calculations with account of
(\ref{3.25}),(\ref{4.20}), we get

\begin{equation}
\label{4.21}
1-\xi=
d_n \alpha
\alpha_b
^{-{2n+3 \over 2n}}
\left(r_{\ast} \over H_b \right)
^{{2n+3 \over 2n}}
\biggl[-{\dot M K^n \over r_{\ast}^2 (\Omega_{K \ast} r_{\ast})^{2n+1}}
\biggr]
^{3 \over 2n}
\bigl[(1-\omega_{\ast})(3+\omega_{\ast}) \bigr]^{-{2n+3 \over 2n}},
\end{equation}
where

\begin{equation}
\label{4.22}
 d_n=
2^{{4n+3 \over 2n}}
 \sqrt{\pi}
(2\pi)^{-{2n+9 \over 4n}}
n^{-{4n+3 \over 4n}} (n+1)^{3 \over 2}
\biggl[{\Gamma \left(n +{3 \over 2}
\right) \over \Gamma(n+1)} \biggr]^{3 \over 2n}.
\end{equation}
The value of $H_b$ is still not determined. It must be
found from equation (\ref{4.15}), where $\Sigma$ is
substituted from the solution (\ref{4.19}).
Before doing this we estimate by order of magnitude the values of
${u_{s0} \over u_K}$ from (\ref{3.3}) and ${u_r \over u_K}$
from (\ref{3.16})

\begin{equation}
\label{4.23}
u_K=\Omega_K r, \quad
{u_{s0} \over u_K} \sim {z_0 \over r}, \quad
{u_r \over u_K} \sim \alpha
\left({z_0 \over r}\right)^2 \left(1-\xi \sqrt{r_{in} \over
r} \right)^{-1}.
\end{equation}
The equation (\ref{4.15})
contains a nonphysical logarithmic divergency because of
using the MAE method. For an approximate estimation of the value of $H_b$
we use a characteristic thickness over which the
$\omega$ variation happens, and for this,
using the definition (\ref{4.3}) , the following relation may be  written

\begin{equation}
\label{4.25}
-{2 \pi \alpha_b r_{\ast} d_{s0}
\over\dot M}
\Sigma^{2{n+1 \over 2n+1}}\vert_{\omega=1}=1.
\end{equation}
With the help of (\ref{4.11}),(\ref{4.19}),(\ref{4.20})
we get from (\ref{4.25})

\begin{equation}
\label{4.26}
{H_b \over r_{\ast}} \approx d_H
\alpha_b^{-{1 \over n+1}}
\biggl[-{\dot M K^n \over r_{\ast}^2 (\Omega_{K \ast} r_{\ast})^{2n+1}}
\biggr]
^{1 \over n+1}
\bigl[(1-\omega_{\ast})(3+\omega_{\ast}) \bigr]^{-1},
\end{equation}

\begin{equation}
\label{4.27}
d_H=2
(2\pi)^{-{3 \over 2(n+1)}}
n^{-{2n+1 \over 2(n+1)}} (n+1)^{n \over n+1}
\biggl[{\Gamma \left(n +{3 \over 2}
\right) \over \Gamma(n+1)} \biggr]^{1 \over n+1}.
\end{equation}
Using (\ref{4.26}),(\ref{4.27}) in (\ref{4.21})
we finally get the expression for $(1-\xi)$

\begin{equation}
\label{4.28}
1-\xi \approx d_{\xi} \alpha
\alpha_b^{-{2n+3 \over 2(n+1)}}
\biggl[-{\dot M K^n \over r_{\ast}^2 (\Omega_{K \ast} r_{\ast})^{2n+1}}
\biggr]
^{1 \over 2(n+1)},
\end{equation}

\begin{equation}
\label{4.29}
d_{\xi}=d_n d_H^{-{2n+3 \over 2n}}=
2 \sqrt{\pi}
(2\pi)^{-{2n+5 \over 4(n+1)}}
n^{1 \over 4(n+1)} (n+1)^{n \over 2(n+1)}
\biggl[{\Gamma \left(n +{3 \over 2}
\right) \over \Gamma(n+1)} \biggr]^{1 \over 2(n+1)}.
\end{equation}
Estimating the values by order of magnitude, with account of (\ref{4.23}) we
get, combining (\ref{4.26}) and (\ref{4.28})

\begin{equation}
\label{4.30}
(1-\xi) \sim {\alpha \over \alpha_b}{z_0 \over r},
\end{equation}

\begin{equation}
\label{4.31}
{H_b \over r_{\ast}} \sim {1 \over 1-\omega_{\ast}} \left( z_0 \over
r \right)^2.
\end{equation}
Thus we get a complete analytical solution for the accretion disk structure
with the boundary layer near the star for a
viscosity inside the layer from (\ref{4.8})
and a polytropic equation of state everywhere.
Analytical solutions for
an accretion disk with boundary layer for some simplifying assumptions
about the viscocity coefficient and temperature distribution were
obtained directly without expansions by Colpi at.al.(1991) and
Glatzel (1992). The MAE method, used by Regev (1983) for the solution of the
accretion disk BL problem permits to obtain a self-consistent analytical
solution for a disk with boundary layer for a viscosity coefficient in the
usual $\alpha$ representation.

 \subsubsection{Isothermal case}

For an isothermal disk with the same viscosity (\ref{4.8})  we get,
using (\ref{3.10}) -(\ref{3.12}),
(\ref{4.4}) and (\ref{4.5})

\begin{equation}
\label{4.33}
\frac{d_{\rho 0}}{d_{P0}}=\frac{1}{K},\quad
{d\Sigma \over dx }=-{\Omega_{K \ast}^2 \over K} H_b \Sigma (1-\omega^2),
\end{equation}

\begin{equation}
\label{4.34}
{d\omega \over dx}=-{\dot M (1-\omega) \over 2 \pi \alpha_b \sqrt{K}
\Sigma r_{\ast}^2}.
\end{equation}
Dividing the last two equations, and using the
approximate boundary condition $\Sigma_{\ast}=
\infty$, we get the solution (Shakura and Sunyaev, 1988)

\begin{equation}
\label{4.36}
\Sigma=\left( -{\dot M \sqrt {K} \over 2 \pi \alpha_b H_p}\right)
{1 \over r_{\ast}^2 \Omega_{K\ast}^2}
\biggl[(\omega-\omega_{\ast})\left(1+{\omega+\omega_{\ast} \over 2}
\right)\biggr]^{-1}.
\end{equation}
From (\ref{4.25}),(\ref{4.36}),(\ref{4.34})
we obtain the characteristic scale, which
we identify with $H_b$

\begin{equation}
\label{4.37}
{H_b \over r_{\ast}} \approx
{2K \over r_{\ast}^2\Omega_{K\ast}^2}
\bigl[(1-\omega_{\ast})(3+\omega_{\ast}) \bigr]^{-1}.
\end{equation}
Matching (\ref{3.30}) and (\ref{4.36}) with account of
(\ref{4.37}) like for plytrope,
we get the expression for the integration constant

\begin{equation}
\label{4.38}
1-\xi={\sqrt{2K} \over r_{\ast}\Omega_{K\ast}}
{\alpha \over \alpha_b}.
\end{equation}
Comparing (\ref{4.37}),(\ref{4.38}) with (\ref{4.30}),
(\ref{4.31}) we see that the order of magnitude
estimates for the polytropic case become exact the for isothermal case apart
a numerical coefficients close to unity.

In the isothermal case there is a simple solution of equation
(\ref{4.34}) when
(\ref{4.36}) is substituted for $\Sigma$,

\begin{equation}
\label{4.40}
(1-\omega)^{-2}
(\omega-\omega_{\ast})^{3+\omega_{\ast} \over 1+\omega_{\ast}}
(2+\omega +\omega_{\ast} )^{-{1-\omega_{\ast} \over 1+\omega_{\ast}}}=
\exp {\biggl[
2{H_b \over K}
\Omega_{K\ast}^2 x
(1-\omega_{\ast})(3+\omega_{\ast})\biggr]}.
\end{equation}
We can see from (\ref{4.40}) that the scale (\ref{4.37})
is naturally appears under
the exponent. It is clear that this solution makes physical sence
only over this characteristic  scale (thickness
of BL) where $ 0<x<r_{\ast}$.
%%%%%%%%%%%%%%%%%%%%%%
It is proved (Nayfet, 1973), that the region of applicability
of the inner and outer solutions, obtained by MAE method, do
overlap and the formula, constructed from the inner (i) and outer (e)
solutions, gives a good interpolation also for the intermediate
region. The formula, describing the function $f$ in the whole region,
has a structure

\begin{equation}
\label{4.41}
f=f_i+f_e-(f_i)_e.
\end{equation}
Here $(f_i)_e$ is the value of the inner solution on the outer edge
[equal to the value of the outer solution on the inner edge $(f_e)_i$].
The solution for $\Sigma(r)$ is constructed using $\Sigma^{(0)}(r)$ from
the relation (\ref{3.5}) for the outer solution, and the solution of the
equation (\ref{4.14}) with account of (\ref{4.3}),(\ref{4.19}),(\ref{4.20})
for $\Sigma_i (r)$. The choice of $\xi$ from (\ref{4.28}) gives the
equality $(\Sigma_i)_e$=$(\Sigma_e)_i$. The solution for $\Omega(r)$ is
constructed using $\Omega_K(r)$ from (\ref{3.17}) for the external, and
the solution of (\ref{4.15}), with account of
(\ref{4.3}), (\ref{4.6}), (\ref{4.19}), for the internal
solution $\Omega_i(r)$. It is clear from the above consideration that
$(\Omega_i)_e$=$(\Omega_e)_i$=$\Omega_{K*}$.

The existence of the overlapping regions, where both solutions (external and
internal) are valid was first proved by Prandtl
(1905, cited by Nayfet, 1973) for the problem of the boundary layer near
the body in the flow of the rapidly moving viscid fluid, for which he
had invented the MAE method. The interpolation formulae of the type
(\ref{4.41}) had been introduced by Vasil'eva (1959, cited by
Nayfet, 1973), and is considered as a uniformly valid approximation
in the whole region, due to existence of the overlapping regions.
The direct proof of the validity of (\ref{4.41}), for the cases where
general analytic solutions exist, can be found in the books of
Nayfet (1981) and Zwillinger (1992). These books, as well as
Nayfet (1973), contain examples of the application of MAE method,
including quasi-linear equations of the type (\ref{3.15}), (\ref{3.16}),
and also references to many other applications of the MAE method and its
detailed mathematical investigation.

Using the relation $\dot M=2\pi r_* \Sigma u_r$ for the estimation
of the radial velocity $u_r$ in the boundary layer, we get, from
(\ref{4.9})-(\ref{4.12}), (\ref{4.19}), (\ref{4.26}) for polytropic
[or from (\ref{4.36}), (\ref{4.37}) for isothermal] cases, the estimation

\begin{equation}
\label{4.42}
u_r/u_{s0}\, \sim \alpha_b.
\end{equation}
This means that the solution obtained,
where radial velocity was neglected, is valid only for sufficiently
small viscosity, with $\alpha_b \ll 1$.

 \subsubsection{Boundary layers with account of thermal processes.
Heat production in the boundary layer}

Rapid braking of rotational velocity in the boundary layer
during accretion into a slowly rotating star leads to
strong heating and large energy release. It is expected, that
matter in the boundary layer is hotter than in the disk, and
is emitting radiation with a larger effective temperature.
The equations of a heat balance in the boundary layer are
different from the similar equations in the accretion disk
(\ref{ref3.1}), (\ref{ref3.2}), (\ref{ad6a}). The component of the
stress tensor should be written on the form
(\ref{ad2}) only, because in the boundary layer $\Omega$ deviates
strongly from the keplerian value. The turbulent viscosty coefficient
should be scaled by a smaller value of
boundary layer thickness $H_b$, instead
of the disk thickness $h$, like in (\ref{4.7}), (\ref{4.8}).
Due to thinness of the boundary layer the radial heat flux
inside it should be taken into account, and is usually more
important than the losses through the sides of the disk
$Q_-$ in (\ref{ad6a}). From (\ref{ad6a}), (\ref{ad2}), and
using the momentum conservation equation (\ref{ad10}),
we obtain the expression for the viscous heat production rate
in the half-thickness of the boundary layer
per unit disk surface in the form

\begin{equation}
\label{blt1}
Q_+=h t_{r\phi}r \frac{d\Omega}{dr}
=h\eta \left(r\frac{d\Omega}{dr}\right)^2
\approx \frac{GM{\dot M}}{4\pi r_*^2}(\omega-1)\frac{d\omega}{dr}.
\end{equation}
Here, like in the previous section, we take in the
boundary layer $\xi_b=1$, $r=r_*$, $\omega=\Omega/\Omega_{K*}$.
It is convenient in the boundary layer to define the value

\begin{equation}
\label{blt2}
Q_t=4\pi r Q_+ \approx 4\pi r_* Q_+
= \frac{GM{\dot M}}{r_*}(\omega-1)\frac{d\omega}{dr},
\end{equation}
determining the heat production by the viscosity in the whole ring of the boundary
layer per unit of its radial thickness in the accretion disk.
Let us consider the  radiative heat
conductivity along the radius as the main term in the heat
balance of the boundary layer. Defining the radial heat flux through the
optically thick boundary layer $\Phi$ as

\begin{equation}
\label{blt3}
\Phi=-\frac{4ac T^3}{3\kappa \rho}\frac{dT}{dr}\,4\pi r h,
\end{equation}
we write the heat balance equation as

\begin{equation}
\label{blt4}
\frac{d\Phi}{dr}=Q_t
=\frac{GM{\dot M}}{r_*}(\omega-1)\frac{d\omega}{dr}.
\end{equation}
Integrating this equation, we obtain

\begin{equation}
\label{blt5}
\Phi=\frac{GM{\dot M}}{r_*}\left(\omega-\frac{\omega^2}{2}
-\omega_A + \frac{\omega_A^2}{2}\right)
=\frac{GM{\dot M}}{r_*}(\omega-\omega_A)
\times \left(1-\frac{\omega}{2} - \frac{\omega_A}{2}\right).
\end{equation}
We consider here the layer at $\omega=\omega_A$, where the temperature
has a maximum, so the radial flux is zero in this point, and had
opposite direction at both sides of this layer.
A position of this layer should be found from time-dependent evolutionary
calculation of the boundary layer structure. If, for example, accretion
is started into the cold neutron star, that the heating wave is started from
the boundary layer, outside in the accretion disk,
and inside the neutron star, heating it.
In this case the level with $\omega=\omega_A$ is moving with time
to the boundary of the star, and for larger times when the whole star
becomes almost isothermal, this level coincides with the star boundary.
If accretion is going on into the hot neutron star, the
heat production in the boundary layer may be comparable with the
heat flux from the star, which is equal to $\Phi_*$ on the star
surface with $\omega=\omega_*$.
Than we obtain instead of (\ref{blt5}), the solution

\begin{equation}
\label{blt6}
\Phi=\Phi_* +
\frac{GM{\dot M}}{r_*}(\omega-\omega_*)
\times \left(1-\frac{\omega}{2} - \frac{\omega_*}{2}\right).
\end{equation}
It follows from (\ref{blt6}), taking into account that the angular
velocity is keplerian at the outer boundary of the boundary layer
$\omega_{out}=1$, that the total energy flux from the boundary
layer due to heating
by viscosity $Q_{bl}$ is equal to (Popham and Narayan, 1995)

\begin{equation}
\label{blt7}
Q_{bl}= \frac{GM{\dot M}}{2r_*}(1-\omega_*)^2.
\end{equation}
The relation (\ref{blt7}) may be obtained also from the
conservation laws. When a mass $dm$ crosses the boundary layer the
change of its energy is

\begin{equation}
\label{blt7a}
dE_m= \frac{1}{2} dm \Omega^2_{K*} r_*^2 (1-\omega_*^2).
\end{equation}
From the angular momentum conservation law the angular momentum
lost by the accreting matter

\begin{equation}
\label{blt7b}
dJ_m= dm\Omega_{K*}r_*^2 (1-\omega_*)
\end{equation}
is absorbed by the star, so that

\begin{equation}
\label{blt7c}
dJ_*= I_* \Omega_{K*} d\omega_*=dJ_m.
\end{equation}
Obtaining from two last equations
$I_* d\omega_*=dm r_*^2 (1-\omega_*)$, we obtain the kinetic energy
absorbed by the star in the form

\begin{equation}
\label{blt7d}
dE_*= I_* \Omega_{K*}^2 \omega_* d\omega_*
=dm \Omega_{K*}^2 r_*^2 \omega_* (1-\omega_*).
\end{equation}
The heat produced in the boundary layer is defined as

\begin{equation}
\label{blt7e}
dE_{bl}=dE_m-dE_*
= \frac{1}{2}dm \Omega_{K*}^2 r_*^2 (1-\omega_*)^2,
\end{equation}
what gives (\ref{blt7}) with account of

\begin{equation}
\label{blt7f}
Q_{bl}=\frac{dE_{bl}}{dt},\quad
{\dot M}= \frac{dm}{dt},\quad
\Omega_{K*}^2=\frac{GM}{r_*^3}.
\end{equation}

\subsubsection{Explicit models of the accretion disk with a
            boundary layer}

Set of equations, describing stationary accretion disk together with a
boundary layer was solved numerically by several authors.
The equations of radial and azimuthal motion, and continuity
equation coincide with the
corresponding equations of the advective disk structure
(\ref{ref3.3}), (\ref{ref1.4}) with $\Omega$ - derivative
description of the viscosity, and
a continuity equation (\ref{ad3}). Contrary to
the advective disk structure at accretion into a black hole
with a transonic radial motion, the accretion disk around a star
with a boundary layer has always a subsonic radial velocity
(\ref{4.42}), with a Mach number $\sim \alpha_b$, where
$\alpha_b$ is characterizing the viscosity in the boundary layer
(\ref{4.7}),(\ref{4.8}). The viscosity coefficient describing
continuously the disk and boundary layer should include a
transition from the thickness scale $h$ in the disk to the smaller
value of the boundary layer thickness $H_b$ in the layer itself.
The expression having these properties was used by Popham and
Narayan (1995)

\begin{equation}
\label{blt8}
\nu\, = \,
\frac{2}{3}\left[\frac{1}{(\alpha \,h)^2}+
\frac{(2\,dP_0/dr)^2}{(3\alpha_b\, P_0)^2}\right]^{-1/2}.
\end{equation}
The coefficient $\left(\frac{2}{3}\right)$ is included 2
times into (\ref{blt8}) to make a continuous transition to the
(\ref{ad9}) in the disk where $h \ll H_b= \frac{P_0}{dP_0/dr}$,
and to (\ref{4.8}) in the boundary layer where $h \gg H_b$.
The equation of a thermal balance should include the radiative
heat transfer $\Phi$ in radial direction from (\ref{blt3}),
which exceeds all other cooling
sources inside the boundary layer. Taking $Q_+$ and $Q_-$ from
(\ref{ad6a}), advective heating $Q_{adv}$ from (\ref{ref3.1}),
and $\Phi$ from (\ref{blt3}), we obtain the equation of the heat
balance valid in the whole disk, including boundary layer

\begin{equation}
\label{blt9}
Q_+\,=\,Q_-\,+\,Q_{adv}\,+\,\frac{1}{4\pi r}\frac{d\Phi}{dr},
\end{equation}
which may be written as

\begin{equation}
\label{blt10}
{\dot M}\,T \frac{dS}{dr}\,=
\frac{d\Phi}{dr}\,+\,{\dot M}(j-j_{\rm in})\frac{d\Omega}{dr}
+\,\frac{2aT^4c}{3\kappa \rho h} 4\pi r.
\end{equation}
The thickness of the disk $h$ is everywhere determined by vertical
equilibrium equation (\ref{ref1.3}).
Popham and Narayan (1995) used a bit more complicated equations of
a heat transfer. Instead of $Q_-$ and $\Phi$ from above
they considered expressions following from simplified solution of
the radiative transfer equation permitting to distinguish between
absorption and scattering opacities, similar to the description in
Section 1.6, which is valid also in the optically thin case.
The equation of state $P=P_g+P_r$ takes into account deviations of
the radiation pressure $P_r$ from the equilibrium value, similar
to (\ref{ref11.16}).

The boundary conditions used by Popham and Narayan (1995)
required angular velocity to be equal to the stellar one
$\Omega_*$ at the inner boundary at stellar radius $r_*$, and
to the keplerian angular velocity at the outer boundary, specified
at $R_{out}=100R_*$. The boundary conditions for the vertical heat
flux had been equivalent to the $Q_-$ energy losses from
(\ref{ad6a}). The radial heat flux from the star $\Phi_* > 0$ was
specified on its surface, what is equivalent to the situation
described by (\ref{blt6}).

Calculations have been performed for cataclismic binaries
where a disk accretion goes onto a
 white dwarf star. Other similar calculations (Popham et
al., 1996; Collins et al., 1998) considered white dwarfs or
pre-main-sequence stars. Boundary layers in the accretion disks
around a neutron star had not been considered in explicit
calculations, and had been treated either
analytically, or by MAE method (Regev, 1983).
Calculations of Popham and Narayan (1995) revealed the existence
of a dynamical boundary layer,
where there is a rapid drop of the angular velocity and of a
velocity of the radial motion, inside the thermal boundary layer.
There is a strong radial acceleration of matter
when the flow approaches the dynamical boundary layer, and the density is
rapidly increasing inside this layer. The heat flux from the boundary
layer is propagating radially into the accretion disk and is
radiated vertically on the scale of the order of the disk
thickness.

Time-dependent calculations for accretion disks with boundary
layers around pre-main-sequence stars had been performed in 1-D
approximation by Godon (1996a,b), using nonflux thermal boundary
condition $dT/dr=0$ at $r=r_*$, what is corresponding to $\Phi_*=0$
in (\ref{blt6}).
The results of 1-D calculations are qualitatively  in accordance with the
stationary model of Popham and Narayan (1995).
The thermal boundary condition used above does not take into account
a possible important heat flux from or inside the star. It seems
more consistent to use the thermal boundary condition in the
following simplified form

\begin{equation}
\label{blt11}
\frac{dT_*}{dt}=-\frac{\Phi_*}{C_*},
\end{equation}
where $C_*$ is the corresponding heat capacity of a star, and
$\Phi_*$ is a heat flux on the star boundary from (\ref{blt3}),
which may be positive or negative, depending on the temperature
$T_*$. In the case of a neutron star or a white dwarf the heat
capacity $C_*$ may be estimated from the existing models of their
structure, and the heat capacity of the isothermal star could be
considered  as a rough approximation.

 Time-dependent 2-D model had been calculated by Kley (1991).
These
calculations show oscillations in time of luminosity
which amplitude depends on the choice of input parameters.
These oscillations  could
be connected with instabilities, but numerical effects are also
cannot be excluded.

\subsubsection{Models of accretion disk with a boundary layer
constructed by MAE method}

Matched asymptotic expansion (MAE) method was used for description of
boundary layers in hydrodynamics (Prandtl, 1905), and was first
applied for accretion disks by Regev (1983). The
structure of the accretion disk and boundary layer with a
polytropic equation of state is solved analytically by MAE method
(Bisnovatyi-Kogan, 1994), and is described in Section 3.1. Account
of thermal processes with the equation of state $P(\rho, T)$ makes
the problem more complicated, and analytic solution was not
obtained. The equilibrium equations describing the outer solution for the disk
may be taken from the standard model without advection with
$\Omega=\Omega_K$, valid until the inner boundary of the outer
solution at $r=r_*$. Contrary to the standard model, where the
integration constant $j_{\rm in}$ was taken as $j_{\rm in}=\xi
\Omega_{K*}\,r_*^2$ with $\xi=1$, in the outer MAE solution for the
disk the constant $\xi$ is slightly less than unity, and should be
found from matching of the inner and outer solutions, like in the
polytropic case (see Section 3.1.2).

The thermal equations in the disk may be taken in the form
(\ref{blt10}), where advective term may be neglected, and
(\ref{blt3}). The term with $\Phi$ in this equation should take
into account the heat flux coming from the boundary layer into the
accretion disk, which is given by the boundary value
$\Phi(r_*)=\Phi_{bl}$. The constant $\Phi_{bl}$ should be found
from the matching to the inner solution. The second boundary
condition for solving two differential equations (\ref{blt10}),
(\ref{blt3}) determines approach of the solution at large $r$ to
the standard one with a very small $\Phi(r)$. It is possible to
write the thermal balance equation in the disk in a more simple
way. It is evident from the physical model, and also is confirmed
by calculations of Popham and Narayan (1995), that the heat flux
from the boundary layer $\Phi_{bl}$ coming into the disk is
radiated from the disk on the scale of the order of the disk
thickness $h_*$ near the boundary layer.
Instead of solving the differential equation, we
may model the radiation flux from the boundary layer by an
additional heating term, which is distributed over the distance
$h_*$ from the inner boundary by the following profiling

\begin{equation}
\label{blt12}
\frac{d\Phi}{dr} \approx \frac{\Phi_{bl}}{h_*}
\exp{\left(-\frac{r - r_*}{h_*}\right)}.
\end{equation}
Integration of (\ref{blt12}) over
the radius between $r_*$ and $\infty$
gives the additional energy emitted by the disk due to
the heat flux from the boundary layer equal to $\Phi_{bl}$, and
this flux is emitted in the region $\delta r \sim h_*$, attaching
the boundary layer. In this description there is one additional
matching condition defining  $\Phi_{bl}$, in comparison with the
polytropic case.

To obtain the inner MAE solution we should solve the equilibrium
equations (\ref{4.4}) and (\ref{4.5}) together with the equations
of the heat production and heat transfer. The constant $\xi_b=1$
here, like in the polytropic case from the matching condition for
$\omega=1$ at $x=\infty$, equal to the inner keplerian value of
the outer solution. Matching the densities of the inner and outer
solutions determines the inntegration constant $\xi$ of the outer
solution. Application of MAE to the solution of the boundary layer
with thermal processes meets with a definite problem. Using
(\ref{blt5}) in (\ref{blt3}), we write the equation of a heat
transfer in the boundary layer in a form

\begin{equation}
\label{blt13}
\Phi=
-4\pi r h\,\frac{4ac T^3}{3\kappa \rho}\frac{dT}{dr}
=\frac{GM{\dot M}}{2r_*}[(1-\omega_A)^2
-(1-\omega)^2].
\end{equation}
It follows from (\ref{blt13}) that the heat flux in the boundary
layer consists of the constant value $\Phi_0$ in the first term to the
left, and variable part $\sim (1-\omega)^2$. Existence of the
constant flux leads to non-physical behaviour of the temperature
at large $x$ in MAE method, leading to unrestricted fall of the
temperature.
Therefore, in the papers of Regev (1983),
Regev and Bertout (1995) the solution was obtained at $\Phi_0=0$,
what determines that all heat produced in the boundary layer heats
the star and does not go into the accretion disk, what is opposite
to the condition used by Popham and Narayan (1995). As was
mentioned above, the maximum of temperature may correspond to the
intermediate level in the boundary layer, and heat flux from it
may heat simultaneously the star and the boundary layer.
In the direct calculations it corresponds to changing of
boundary condition, but was not realized. In MAE method solution
cannot be obtained directly at non-zero $\Phi_0$. Additional
modification of MAE method are needed for its account.

Note, that large deviations of the $\xi$ value from unity, $\xi$=0.61, obtained by
Regev (1983) for $\omega_{\ast}=0.3$, could indicate the step into the region,
where the asymptotic expansion
begins to be violated.
When the stellar angular velocity
approaches a critical value, the boundary
layer becomes thick (see (\ref{4.26})
and the MAE method fails.

\subsection{Accretion Disks Around Rapidly Rotating Stars}

Investigations of low-mass-X-ray binaries (LMXB)
containing a neutron star have lead
to the conclusion that the accreting object can rotate rapidly
with a surface angular velocity close to Keplerian value.
The question of accretion onto a rapidly rotating star
also arises in cataclysmic variables, in which the primary
 is a white dwarf.
When the stellar angular velocity reaches its critical value the
maximum of the angular velocity and the boundary layer disappear,
and a regime of accretion is changing. Disk accretion into a
rapidly rotating star, with the Keplerian angular velocity at the
equator, had been investigated by Popham and Narayan (1991),
Paczynski (1991), Colpi et al. (1991). Self-consistennt regime was
found by Bisnovatyi-Kogan (1993a), in which star remains to rotate
rapidly during the accretion process.

\subsubsection{Self-consistent model }

When a star reaches the critical rotation rate for
which the the surface
angular velocity becomes Keplerian, the flux of angular momentum
into the star leads to differential rotation inside
the star. It is clear that disk properties far away from the star do
not change when the star reaches
the critical rotation rate, so that the disk remains
Keplerian. The properties of the intermediate zone between the star
and the disk are determined by the stellar rotational velocity law,
but we do not need
to know exactly the properties of this transition zone
in order to build a self-consistent model.
We only have to know that during accretion onto a rapidly rotating star,
the transition layer is smooth, all values are monotonous and
there is no enhanced heat production in this layer, in contrast to the
case of boundary layers between the accretion disk and a slowly rotating
star.

The evolution of the star under mass accretion may be
characterized by the variations of two values: the mass $M$ and the total
angular momentum $J$ and by one function, $j(r)$, determining the angular
momentum distribution within the star; $j(r)$
is determined by the viscosity law inside the star.
For a given mass flux
$\dot M$, determined by conditions far
away from the star, the star gradually
increase its mass and angular momentum.
When the
stellar rotation rate is much smaller than the Keplerian limit
the specific angular momentum
of matter accreting onto the star is approximately equal to
$\dot Mv_{\rm Ke} R_{\rm se}$,
($v_{\rm K}=(GM/r^3)^{1/2}$ is the Keplerian velocity, $R_{\rm se}$-
the stellar equatorial radius, index "e" referes to
values at the stellar equator).
When the stellar rotational velocity at the surface is
smaller than the Keplerian value, the angular velocity
of the disk has a maximum near $R_{\rm se}$; at this  point,
the viscous flux is zero
and the momentum flux is determined only by the convective term
$\dot Mj_{\rm e}$, $j=v_{\phi}r $.
The maximum of the rotational velocity disappears, when the star
rotation is close to the critical velocity and the
(negative) viscous momentum flux becomes important.

The demand of self-consistency during accretion onto a rapidly rotating
star may be formulated as the condition that {\bf the star absorb
the accreted matter with a specific angular momentum such that the
star remains in the state of critical rotation}. Depending on
the viscosity law, a critically rotating star may have different
distributions $j(r)$ and in general, the structure of the star
gradually changes with time. If viscosity inside the star is
high, the star rotates almost rigidly; this case
has been considered in the above mentioned papers.
If the star-disc system can be described by a polytropic
equation of state, there is a self-similar solution for such a system
which does not depend on mass.

\subsubsection{ Self-similar solution for a polytropic
         star-disc system in the case of a rigidly rotating star }

The equations describing the structure of a rotating star with
a polytropic equation of state (\ref{3.1})
the equilibrium equations
in spherical coordinates are (Chandrasekhar 1989)

\begin{equation}
\label{s2}
 {1\over \rho} {\partial P\over \partial r}+ {\partial \Phi
\over \partial r} - \Omega^2 r(1-\mu^2) =0
\end{equation}
\begin{equation}
\label{s3}
 {1\over \rho} {\partial P\over \partial \mu}+{\partial \Phi
\over \partial \mu}+ \Omega^2 r^2\mu =0
\end{equation}
and the equation for the gravitational potentional $\Phi$ is

\begin{equation}
\label{s4}
{1 \over r^2} {\partial \over \partial r } \left ( r^2 {\partial
\Phi \over \partial r}\right) + {1 \over r^2}{\partial \over \partial
\mu}\left((1-\mu^2){\partial \Phi \over \partial \mu}\right)=
4\pi G\rho
\end{equation}
Here $\mu=\cos\theta$ and $ \Omega$ is the
angular velocity of the stellar matter. In
general $\Omega$ is not constant throughout the star, and for a
polytropic star in equilibrium depends only on the cylindrical radius
$r\sin\theta$.
Excluding $\Phi$ from Eqs.(2)-(4), we get

$${1\over r^2}{\partial \over \partial r} \left( {r^2 \over \rho}
{\partial P \over \partial r} \right ) + {1 \over r^2} {\partial
\over \partial \mu} \left( {1-\mu^2 \over \rho}{\partial P \over
\partial \mu} \right) +$$
\begin{equation}
\label{s5}
 \quad 4\pi G\rho-{1 \over x}{d \over dx}
(x^2\Omega^2)=0
\end{equation}
where $x=r\sin \theta = r\sqrt{1-\mu^2}$, $\Omega=\Omega(x)$.
Introduce the nondimensional polytropic variables

\begin{equation}
\label{s6}
 \xi=r/r_{\ast}, \quad \Theta^n= \rho/\rho_{\rm c}, \quad
\omega=\Omega/\Omega_{\ast},
\end{equation}
where $\rho_{\rm c}$ is the central density of the star and $r_{\ast}$
and $\Omega_{\ast}$
are given by

\begin{equation}
\label{s7}
r_{\ast}=\left( {(n+1)K\rho_{\rm c}^{{1\over n}-1}\over 4\pi G}
\right)^
{1/2}, \quad \Omega_{\ast}= \sqrt{2\pi G\rho_{\rm c}}
\end{equation}
In the nondimensional variables (6)-(7) Eq.(5) for $\Omega$=
const. can be written in the form

\begin{equation}
\label{s8}
{1 \over \xi^2}{\partial \over \partial \xi} \left( \xi^2{\partial
\Theta \over \partial \xi} \right) + {1 \over \xi^2}
{\partial \over \partial \mu}
\left( (1-\mu^2){\partial \Theta \over \partial \mu} \right) +
\Theta^n - \omega^2=0
\end{equation}
The solution of Eq.(8) is uniquely determined by the boundary
conditions

$$\Theta=1,\quad {\partial \Theta \over \partial \xi} =0 \quad
{\rm at} \quad \xi=0, $$
\begin{equation}
\label{s9}
\sqrt{1-\mu^2} {\partial \Theta \over \partial \mu}=0 \quad
{\rm at} \quad \mu=0, \quad \mu=\pm 1 ,
\end{equation}
and by the value of $\omega$; one also requires
that the solution must satisfy the
original equilibrium Eqs.(2),(3).
The latter condition results from the fact that Eqs.(2) and (3)
were differentiated in order to obtain (8), so that a solutions
of (8)
need not satisfy (2),(3). It is in fact necessary to find a
gravitational potential $\Phi$ from (4) which, when written in nondimensional
form $ \phi = \Phi/ 4\pi G\rho_{\rm c} r_{\ast}^2 $ satisfies the
scaling conditions of the self-similar solution.
For compressible matter with $n>0$
a solution exists only if
$\omega < \omega_{\rm c}$;
for $\omega = \omega_{\rm c}$
the centrifugal force on the equatorial boundary of the star exactly
balances  gravity (see the proof in
Bisnovatyi-Kogan \& Blinnikov 1981). Numerical calculations have been
done, which gave the structure of rotating polytropic
stars for different $\omega$ up to the critical value $\omega_c$ (James 1964;
Blinnikov 1975; Ipser \& Monagan 1981; see also Tassoul 1978).
\hfill\break\indent
For a given $n$ the structure of a critically rotating star does not
depend on its mass. By definition, the mass $M$ and the total
angular momentum $J$ are given by

\begin{equation}
\label{s10}
 M=2\pi \rho_{\rm c} r_{\ast}^3 \int_{-1}^1 d\mu
\int_0^{\xi_{\rm out}(\mu)}
\Theta^n(\xi,\mu)\xi^2 d\xi,
\end{equation}
\begin{equation}
\label{s11}
 J=2\pi \rho_{\rm c} r_{\ast}^5 \omega \sqrt{2 \pi G\rho_{\rm
c}}\int_{-1}^1
(1-\mu^2) d\mu
\int_0^{\xi_{\rm out}(\mu)} \Theta^n(\xi,\mu) \xi^4 d\xi.
\end{equation}
Introducing nondimensional values of
the mass ${\cal M}_n$ and of the total angular
momentum ${\cal J}_n$, defined as

\begin{equation}
\label{s12}
{\cal M}_n={1 \over 2} \int_{-1}^1 d\mu \int_0^{\xi_{\rm out}(\mu)}
\Theta^n(\xi,\mu)\xi^2 d\xi,
\end{equation}
\begin{equation}
\label{s13}
{\cal J}_n={\omega \over 2} \int_{-1}^1 (1-\mu^2) d\mu
\int_0^{\xi_{\rm out}(\mu)} \Theta^n(\xi,\mu) \xi^4 d\xi
\end{equation}
M and J may be written as

\begin{equation}
\label{s14}
M=4\pi \rho_{\rm c} r_{\ast}^3 {\cal M}_n,
\end{equation}
\begin{equation}
\label{s15}
J=4\pi \rho_{\rm c} r_{\ast}^5 \sqrt{2 \pi G\rho_{\rm c}}{\cal
J}_n.
\end{equation}
The average
specific angular momentum of the star $j_{\rm s}$  and the derivative
along the critical states
$j_{\rm a}=(dJ/dM)_c$ are written, using (7),(14),
as

\begin{equation}
\label{s16}
j_{\rm s}=J/M=
r_{\ast}^2 \sqrt{2\pi G\rho_{\rm c}}{\cal J}_n/{\cal M}_n \quad
j_{\rm a}={5-2n \over 3-n}j_{\rm s}.
\end{equation}
Using (14) one may write $\rho_{\rm c}$ and $r_{\ast}$,
which appear in (7),
as a finction of $M$

\begin{equation}
\label{s17}
\rho_{\rm c}=\left( {4\pi G \over (n+1)K} \right)^{3n \over 3-n}
\left( {M \over 4 \pi {\cal M}_n} \right)^{2n \over 3-n},\,\,\,\,
r_{\ast}=\left( {4\pi G \over (n+1)K} \right)^{n \over n-3}
\left( {M \over 4 \pi{\cal M}_n} \right)^{1-n \over 3-n}.
\end{equation}
The specific angular momentum of matter $j_{\rm e}$ at the stellar equator
$r_{\rm e}$
may be written as
\begin{equation}
\label{s18}
j_{\rm e}=\Omega r_{\rm e}^2=
r_{\ast}^2 \sqrt{2\pi G\rho_{\rm c}} \xi_{\rm e}^2 \omega , \quad
\xi_{\rm e}=r_{\rm e}/r_{\ast}
=\xi_{\rm out}(\pi/2).
\end{equation}
From a comparison of (16) and (18), it follows that the ratio

\begin{equation}
\label{s19}
{j_{\rm a} \over j_{\rm e}}=
{5-2n \over 3-n}{j_{\rm s} \over j_{\rm e}}=
{5-2n \over 3-n}{{\cal J}_n \over{\cal M}_n
\xi^2_{\rm e} \omega}
\end{equation}
does not depend on the stellar mass M, and is a function of n and $\omega$
only.
The nondimensional values of the angular velocity $\omega_{\rm c}$,
the equatorial radius $\xi_{\rm out}(\theta=\pi/2)=\xi_{\rm e}$,
the mass ${\cal M}_n$, the momentum of inertia around the rotational axis
${\cal I}_n={\cal J}_n/\omega_{\rm c} $, the average
specific angular momentum of the star $\zeta_{\rm s}$ ,
the nondimensional derivative $\zeta_a$

\begin{equation}
\label{s20}
\zeta_{\rm s}=j_{\rm s}/(r_{\ast}^2 \sqrt{2\pi G\rho_{\rm c}})=
{\cal J}_n/{\cal M}_n , \quad
\zeta_{\rm a}={5-2n \over 3-n} \zeta_{\rm s},
\end{equation}
and the angular momentum of matter on the stellar equator

\begin{equation}
\label{s21}
\zeta_{\rm e}=j_{\rm e}/(r_{\ast}^2 \sqrt{2\pi G\rho_{\rm c}})=
\xi^2_{\rm e} \omega_{\rm c}
\end{equation}
for several polytropic indices $n$ in the state of critical rotation
are given in table 1, which is based on calculations of James(1964),
Blinnikov(1975) and Ipser and Managan(1981).
The ratio of the equatorial and polar
radii $\xi_{\rm e}/\xi_{\rm p}$ and the nondimensional parameters
for nonrotating polytropes  (outer radius $\xi_{0n}$,
mass ${\cal M}_{0n}$ and momentum of inertia around the symmetry axis
${\cal I}_{0n}$, Chandrasekhar 1957; James 1964;
Blinnikov 1975) are also given.\\
\medskip
\begin{table}
\caption{Parameters of non-rotating and critically
 rotating polytropic stars }
\medskip
%{\small
{\footnotesize
\begin{tabular}
{ll@{\,\,\,\,}l@{\,\,\,\,}l@{\,\,\,\,}l@{\,\,\,\,}l@{\,\,\,\,}l@{\,\,\,\,}
l@{\,\,\,\,}l@{\,\,\,\,}l@{\,\,\,\,}l@{\,\,\,\,}}
\hline
n&$\xi_{n0}$&${\cal M}_{n0}$&${\cal I}_{n0}$&$\omega_{\rm
c}$&$\xi_{\rm e}$&
${\xi_{\rm e} \over \xi_{\rm p}}$&${\cal M}_n$&${\cal
I}_n$&$\zeta_{\rm s}$&${\zeta_{\rm a}
\over \zeta_{\rm e}}=\beta $\\
\hline
0.5   &2.7528  &3.7871  &      &0.389    &       &2.51  &        &
      &      &          \\
      &        &        &      &(0.367)  &       &      &        &
      &      &          \\
0.6   &        &        &      &0.362    &       &2.29  &        &
      &      &          \\
      &        &        &      &(0.354)  &       &      &        &
      &      &          \\
0.808 &        &        &      &0.326    &4.7652 &1.917 &5.0248  &
24.43 &1.585 &0.3304    \\
1.0   &$\pi$   &$\pi$   &8.104 &0.289    &4.8265 &1.792 &4.289   &
18.73 &1.262 &0.2805    \\
1.5   &3.65375 &2.71406 &7.413 &0.209    &5.36   &1.626 &3.2138  &
12.18 &0.792 &0.176     \\
2.0   &4.35287 &2.41105 &7.074 &0.147    &6.307  &1.555 &2.6518  &
9.62  &0.533 &0.0894    \\
2.5   &5.35528 &2.18720 &7.013 &0.09965  &7.7623 &1.522 &2.30563 &
8.49  &0.367 &0.0       \\
3.0   &6.89685 &2.01824 &7.234 &0.0639   &10.123 &1.535 &2.0743  &
8.125 &0.250 &$-\infty$ \\
\hline
\end{tabular}}
\label{st1}
\end{table}
In nondimensional variables
it is  obvious that the derivative $\zeta_{\rm a}$ of the
angular momentum of the star along the sequence of critically rotating
states is always smaller than the
equatorial value
$\zeta_{\rm e}$;
$\zeta_{\rm e}$
is equal to the Keplerian value of a critically rotating
non-spherical star.
The viscous flux of specific angular momentum
in the accretion disk around a critically rotating star
$j_{\rm v}$
is therefore negative; near the surface of the star
$j_{\rm v}$ is equal to

\begin{equation}
\label{s22}
 j_{\rm v}=j_{\rm a}-j_{\rm e}
\end{equation}
as one requires that the total angular momentum flux into the star is equal to
$\dot M j_{\rm a}$ for self-consistency. In the self-consistent model of
accretion disk around the rapidly rotating
polytropic star the value of the constant
$\xi$ in the formulae of the Section 1.4 should be taken equal to
$\zeta_{\rm a}$ from the Table \ref{st1}.

\subsubsection{ General (nonpolytropic) case of disc accretion
              into a rapidly rotating star }

In realistic cases, thermal processes are very important in the disc,
polytropic approximation is not good, and equations of the thermal balance
from Section 1.5 should be taken into account. Transition to the rapidly
rotating star is accompanied by a substantial change of the integration
constant in (\ref{ref1.4}), (ref{ref1.11}),
what leads to impotrant observational consequences.

The total luminosity of the accretion disc
is equal to

\begin{equation}
\label{s70}
 L=\int_{r_{\rm i}}^{\infty} 4\pi F r dr = ({3\over 2}-\beta)
\dot M{GM\over r_{\rm i}}.
\end{equation}
When $\beta=1$ the star accrets matter with Keplerian angular momentum,
and only half of the gravitational energy of the accreted matter
is radiated from the disc,
according to the virial theorem. The fate of the remaining half
is different. In the case of a black hole, it is absorbed,
increasing the mass and angular momentum
of the black hole without any significant
radiation; however matter falling from the last stable orbit
to the horizon still emits radiation (Bisnovatyi-Kogan \& Ruzmaikin
1974).

In the case of accretion onto a (slowly rotating) neutron star
there are several possibilities.
If the radius of the neutron star  $r_{\rm s}$
is smaller than $3r_{\rm g}$, then $r_{\rm i}=3r_{\rm g}$;
the disk luminosity is the same as in the case of a black hole,
and the remaining
gravitational energy, including the part gained during free fall onto the
neutron star surface, is emitted close to the neutron star
surface. If $r_{\rm s}>3r_{\rm g}$, then $r_{\rm i}=r_{\rm s}$,
the disc luminosity is
${GM\dot M / 2r_{\rm s}}$ and an equal amount of  energy
is emitted in the boundary layer
near the stellar surface, where the accreting matter
converts its kinetic energy into heat.

The situation changes gradually while the
central object absorbs angular
momentum. The black hole moves from Schwarzschild to extreme Kerr
solution; the
radius of the last stable orbit decreases from $3r_{\rm g}$
to $r_{\rm g}$, and the radius of the horizon
changes from $r_{\rm g}$ to $r_{\rm g}/2$.
The efficiency of the disc energy emission increases from
$5.7 \%$ to $42 \%$ (Novikov \& Thorne 1973). The changes go in an
opposite
direction when a neutron star accelerates its rotation. Let us consider
only the case $r_{\rm s}>3r_{\rm g}$. The
luminosity of the disc decreases because the
stellar radius increases. The fraction of energy emitted by
the neutron star also decreases. It changes from ${GM\dot M
/ 2r_{\rm s}}$ in the case of a nonrotating star to the difference between
rotational energy of matter in the disc and at the stellar surface.

A rapid change in the efficiency of energy release and disc
luminosity happens when the star rotation reaches its maximum
value. The energy emitted near the stellar surface tends towards
zero, but the luminosity of the disc suffers a drastic change. The
star does not absorb any longer all of the angular momentum coming
from the disc, but absorbs only the fraction required to maintain
the star in a critically rotating state. The specific angular
momentum of matter accreted by the star is equal to ${(dJ
/dM)_{\rm c}}$ and the remaining part is carried away in the disc
by viscous stresses (see sec.2). Viscosity carries not only
angular momentum, but also energy, so that the energy production
and luminosity of the disk rapidly increase, from the value
corresponding to $\beta=1 $ to the value corresponding to
${\beta=j_{\rm a}/j_{\rm e}}$ according to (\ref{s70}); for
polytropic stars the values of $\beta$ are given in the last
column of table \ref{st1}. This implies a rapid increase of the
total luminosity by a factor $2\div 3$, depending on the stellar
structure. This rapid increase is easy to understand if one
remembers that when a star accelerates its rotation, part of the
gravitational energy is converted into rotational energy without
heat production. When a star reaches the limiting rotation, the
growth rate of its rotational energy strongly diminishes and a
correspondingly larger fraction of gravitational energy is
transformed into heat. The minimum of luminosity of an accreting
neutron star is reached when the stellar angular velocity is
slightly below the critical value. A rapid increase in luminosity
must be accompanied by a corresponding hardening of the emitted
spectrum because the energy release is increased, raising the
effective temperature in the inner part of the disc. Such events
may be expected in objects like LMXB or cataclysmic variables. The
young T Tauri stars are born rapidly rotating, so their value of
$\beta$ is small from the beginning and close to that of a
polytrope  $n=1.5$ with $\beta=0.176$ given by Table \ref{st1}.
Note, that whereas the standart solution of accretion disk
(Shakura, 1972; Shakura and Sunyaev, 1973; Novikov and Thorne,
1973) is not valid in regions close to the inner edge in the case
of a black hole (see Paczy\'nski \& Bisnovatyi-Kogan 1981), the
self-consistent solution for accretion onto a rapidly rotating
star, may be used down to the very stellar radius, which is the
inner edge of the disc.

\section{Accretion flows in presence of magnetic fields}

\subsection{Magnetic Field Amplification
in Quasi-Spherical Accretion}
%%%%%%%%%%%%%%%%%%%%%%%%%%%%%%%%%% section 5
   Matter flowing into a black
hole from a companion star or from
the interstellar medium is
very likely to be magnetized.
   Due to the more rapid increase
of a magnetic energy in comparison
with any other kind of energy, the
dynamical and any other role of
the magnetic field is becoming more
and more important when matter
flows inside.
     Schwartzman (1971) was the
first to emphasize the importance
of the magnetic field
 for an energy release during
 accretion into a black hole.
   He showed
that due to more rapid growth of
magnetic energy its density $E_M$ approaches
a density of a kinetic energy $E_k$,
and proposed a hypothesis of
{\it equipartition} $E_M \approx E_k$,
supported by continious
annihilation of the magnetic field, for
the smaller radiuses of the flow where
the main enegry production happens.
   This hypothesis is
usually accepted in the modern
picture of accretion (see e.g Narayan and Yu,
1994). Schwartzman (1971)
considered an averaged quasistationary picture
with local equipartition,
another variant of magnetic accretion, where
equipartition and magnetic field
annihilation is accompanied by formation
of shock waves was considered by
Bisnovatyi-Kogan and Sunyaev (1972),
see also Chang and Ostriker (1985).
    More accurate account of the heating of
matter by magnetic field
annihilation was done by Bisnovatyi-Kogan and
Ruzmaikin (1974), where exact
nonstationary solution for a field
apmlification in the radial
accretion flow was also obtained.

   Consider the general
problem of the field amplification in
the accretion flow where gas has
an angular momentum and field annihilation
is approximated by a finite ``turbulent''
conductivity.
    This approach should
reproduce a smooth averaged flow,
considered by Schwartsman.
    In the known
stationary velocity field
${\bf v}(r,\theta)$ the magnetic
field ${\bf B}$ is
described by Maxwell
equations in the MHD
approximation (without the displacement
current), with a finite conductivity
$$
{\bf \nabla \times B} =
{4 \pi \over c} {\bf j}~,\quad
{\bf \nabla \times E } =
-\frac{1}{c}\frac{\partial {\bf B}}{\partial t},
\quad{\bf \nabla \cdot B} = 0~,
$$
\begin{equation}
\label{ref2.1}
{\bf j} =\sigma \left({\bf E}
+{1\over c} {\bf v \times B}\right)~,
\end{equation}
where $\sigma$ is the plasma
conductivity, ${\bf j}$ is the electrical
current density, and ${\bf E}$
is the electrical field strength.
Consider an axisymmetrical
pictire in a spherical coordinate system
$r,\,\, \theta,\,\, \phi$
where all $\phi$ derivative are zero,
but all vectors ${\bf v,\,\,B,\,\,E,\,\,j}$
may have nonzero all
three components.
The equation containing only
magnetic field, following
from (\ref{ref2.1})

\begin{equation}
\label{ref2.2}
\frac{\partial {\bf B}}{\partial t}=
{\bf \nabla \times [v \times B] }+
\frac{c^2}{4 \pi \sigma}{\nabla^2}{\bf  B}.
\end{equation}
These equations are completed
by the condition of zero divergence of the
magnetic field, which for the
choosen case has a view

\begin{equation}
\label{ref2.6}
\frac{1}{r^2}\frac{\partial}{\partial r}
\left(r^2 B_r\right)+
\frac{1}{r\sin\theta}
\frac{\partial}{\partial \theta}\left(\sin\theta
B_{\theta}\right)=0.
\end{equation}

\subsubsection{Time-dependent amplification of a Magnetic field   %%%%%%%%%%%%%
in Spherical Accretion}

For a spherical accretion
with ${\bf v}=(v_r,\, 0,\, 0)$ the equation
(\ref{ref2.2})
%-(\ref{ref2.5})
reduces to

\begin{equation}
\label{ref2.7}
\frac{\partial B_r}{\partial t}=
\frac{1}{r\sin\theta}
\frac{\partial}{\partial \theta}
(\sin\theta v_r B_{\theta})
+\frac{c^2}{4 \pi \sigma}
\end{equation}
$$
\times\left[\frac{1}{r^2}
\frac{\partial}{\partial r}\left(r^2
\frac{\partial B_r}{\partial r}\right)+
\frac{1}{r^2\sin\theta}
\frac{\partial}{\partial \theta}\left(\sin\theta
\frac{\partial B_r}{\partial \theta}\right)-\frac{2B_r}{r^2}-
\frac{2}{r^2 \sin\theta}\frac{\partial}{\partial \theta}\left(
\sin\theta B_{\theta}\right)\right],
$$

\begin{equation}
\label{ref2.8}
\frac{\partial B_{\theta}}{\partial t}=-\frac{1}{r}
\frac{\partial}{\partial r}(r v_r B_{\theta})
\end{equation}
$$
+\frac{c^2}{4 \pi \sigma}
\left[\frac{1}{r^2}\frac{\partial}
{\partial r}\left(r^2
\frac{\partial B_{\theta}}{\partial r}\right)+
\frac{1}{r^2\sin\theta}
\frac{\partial}{\partial \theta}\left(\sin\theta
\frac{\partial B_{\theta}}{\partial \theta}\right)-
\frac{B_{\theta}}{r^2 \sin^2\theta}+
\frac{2}{r^2}\frac{\partial B_r}{\partial \theta}\right],
$$

\begin{equation}
\label{ref2.9}
\frac{\partial B_{\phi}}{\partial t}=-\frac{1}{r}
\frac{\partial}{\partial r}(rv_r B_{\phi})
\end{equation}
$$
+\frac{c^2}{4 \pi \sigma}
\left[\frac{1}{r^2}\frac{\partial}{\partial r}\left(r^2
\frac{\partial B_{\phi}}
{\partial r}\right)+\frac{1}{r^2\sin\theta}
\frac{\partial}{\partial \theta}\left(\sin\theta
\frac{\partial B_{\phi}}{\partial \theta}\right)-
\frac{B_{\phi}}{r^2 \sin^2\theta}\right].
$$
For initial nonstationary stage
of the field amplification, when field
dissipation is negligible,
analytical solution may be obtained
(Bisnovatyi-Kogan \& Ruzmaikin, 1974).
  From (\ref{ref2.7}) with account
  of (\ref{ref2.6}), and from
(\ref{ref2.8}),(\ref{ref2.9})
we get

\begin{equation}
\label{ref2.10}
\frac{d( r^2 B_r)}{dt}=0, \quad
\frac{d(r v_r B_{\theta})}{dt}=0, \quad
\frac{d(r v_r B_{\phi})}{dt}=0,
\end{equation}
where the full Lagrangian
derivative is determined as

\begin{equation}
\label{ref2.11}
\frac{d}{dt}= \frac{\partial}
{\partial t}+v_r\frac{\partial}{\partial r}.
\end{equation}
A solution of (\ref{ref2.10})-
(\ref{ref2.11}) is determined by 4 first
integrals of the characteristic equations

\begin{equation}
\label{ref2.12}
C_1=t-\int\frac{dr}{v_r},
\quad C_2=r^2B_r, \quad C_3=rv_rB_{\theta},
\quad C_4=rv_rB_{\phi}.
\end{equation}
For a free fall case with $v_r=
-\sqrt{\frac{2GM}{r}}$ we get

\begin{equation}
\label{ref2.13}
C_1=t+\frac{2}{3}\frac{r^{3/2}}{\sqrt{2GM}}.
\end{equation}
The initial condition problem
at $t=0$ is solved separately for
poloidal and toroidal fields.
   For initially uniform field
$B_{r0}=B_0\cos\theta,
\,\, B_{\theta 0}=-B_0\sin\theta$
we obtain the integrals

\begin{equation}
\label{ref2.14}
C_1=\frac{2}{3}\frac{r^{3/2}}{\sqrt{2GM}},\quad
C_2=r^2B_0\cos\theta,
\quad C_3=\sqrt{2GMr}B_0\sin\theta.
\end{equation}
The relation between the
integrals found at initial moment after
excluding $r(C_1)$ is valid
at any time, so finally we get from
(\ref{ref2.12}) the solution
(Bisnovatyi-Kogan, Ruzmaikin, 1974)

\begin{equation}
\label{ref2.15}
B_r=\frac{B_0\cos\theta}{r^2}
\left(r^{3/2}+\frac{3}{2}t\sqrt{2GM}\right)^{4/3},
\end{equation}
$$\quad B_{\theta}=
-\frac{B_0\sin\theta}{\sqrt r}\left(r^{3/2}+
\frac{3}{2}t\sqrt{2GM}\right)^{1/3}.$$
The radial component of the
field is groving most rapidly.
It is $\sim r^{-2}$
for sufficiently large times,
$\sim t^{4/3}$ at given sufficiently
small radius, and is groving
with time everywhere.

For initially dipole magnetic field
$B_{r0}=\frac{B_0\cos\theta}{r^3},
\,\, B_{\theta 0}=
-\frac{B_0\sin\theta}{2 r^3}$ we obtain
the following solution

\begin{equation}
\label{ref2.16}
B_r=\frac{B_0\cos\theta}{r^2}\left(r^{3/2}+
\frac{3}{2}t\sqrt{2GM}\right)^{-2/3},
\end{equation}
$$\quad B_{\theta}=
-\frac{B_0\sin\theta}{2\sqrt r}\left(r^{3/2}+
\frac{3}{2}t\sqrt{2GM}\right)^{-5/3}.$$
Here the magnetic field is
decreasing everywhere with time, tending to
zero.
  That describes a pressing of a
  dipole magnetic field to a
stellar surface.
   The azymuthal stellar magnetic
field if confined
inside the star.
   When outer layers of
   the star are compressing with
a free-fall speed, then for
initial field distribution
$B_{\phi 0}=B_0 r^n f(\theta)$
the change of magnetic field with time is
described by a relation

\begin{equation}
\label{ref2.17}
\quad B_{\phi}=-\frac{B_0\,f(\theta)}{\sqrt r}
\left(r^{3/2}+
\frac{3}{2}t\sqrt{2GM}\right)^{1/3+n}.
\end{equation}

\subsubsection{Amplification of magnetic Field
in Rotating Accretion Flows}

     In a quasi-spherical accretion flow
of a perfectly conducting plasma,
$d ({\bf B}\cdot d{\bf S})/dt = 0$,
where $d{\bf S}$ is a surface area
element and $d/dt$ is the derivative
following the flow
(${\bf v}= v_r \hat{\bf r} +
v_\phi \hat{\rvecphi~}$,
in spherical coordinates).
     With $d{\bf S}
= r^2 d\Omega \hat{\bf r}$,  $v_r \approx
v_r(|{\bf r}|)$, and
$d\Omega$ a fixed solid angle increment,
one finds that $B_r \propto 1/r^{2}$
independent of the toroidal velocity $v_\phi$.
    Writing $B_r = B_0 (r_{out}/r)^2$,
the radius at which
$\rho v_K^2/2 = B_r^2/(8\pi)$
is
%%%%%%%%%%%%%%%%%%%%%%%%%%%%%%%%%%
\begin{equation}
\label{l1}
r_{equi} =\left({ g B_0^2 r_0^4
\over \sqrt{GM} \dot{M}}\right)^{2/3}~,
\end{equation}
where $v_K \equiv (GM/r)^{1/2}$ and
the accretion speed is assumed
$v_r = -g v_K$,
with $g={\rm const} \le 1$.
   For example, for a black hole
 mass $M=10^6 M_\odot$, $\dot{M}=
0.1 M_\odot$/yr, $B_0 = 10^{-5}$G,
$r_0=1$pc, and $g= 0.1$, we
find $r_{equi} \approx 6 \times 10^{14}$ cm,
which is much larger than the Schwarzschild
radius, $r_S \approx 3 \times 10^{11}$ cm.
    Thus the magnetic field resulting
from flux-freezing is dynamically important
in quasi-spherical accretion
flows (Shwartsman 1971),
and this is independent of
dynamo processes or MHD instabilities.
      Further accretion for $r<r_{equi}$ is
possible {\it only} if magnetic flux is destroyed
by reconnection and the magnetic
energy  is dissipated.
   Figure 7 shows a sketch of
the {\it instantaneous} poloidal magnetic field
configuration.

%\begin{figure}
%\epsfysize=8cm % fix the y-dimension and scales x-dim. to y-dim.
%\hspace{3.5cm}\epsfbox{adaffig1.eps} %for centering: act on hspace argument

%\label{figu7}
%\end{figure}

%%%%%%%%%%%%%%%%%%%%%%%%%%%%%%%%%%%%%%%%%%

%\begin{figure}
%\epsfysize=8cm % fix the y-dimension and scales x-dim. to y-dim.
%\hspace{3.5cm}\epsfbox{adaffi2b.eps} %for centering: act on hspace argument

%\label{figu8}
%\end{figure}

%%%%%%%%%%%%%%%%%%%%%%%%%%%%%%%%%%%%%%%%%%

    For example, for a flow
with time-averaged velocity
 ${\bf v} = v_r \hat{\bf r}+
v_\phi \hat{\rvecphi~}$,
an approximate solution for the
{\it time-averaged}
magnetic field for infinite conductivity
can readily be obtained in the vicinity of
the equatorial plane.
  In this
region it is reasonable to assume
${\bf B}=B_r\hat{\bf r} +
B_\phi\hat{\rvecphi~}$;  that is,
$B_\theta$ is neglected.
   Thus

$$
{\bf \nabla \cdot B}~=~0=~{1\over r^2}
{\partial \over \partial r}
\left( r^2 B_r \right)
+{1 \over r \sin \theta}
{\partial  \over \partial \phi}
\left( B_\phi \right)~.
$$
Near the equatorial plane, $\sin\theta
\approx 1$.
This equation is then satisfied
by taking

\begin{equation}
\label{l2}
B_r= {{\rm const}\over r^2}
\exp \left(i\int^r dr~k_r+im\phi\right)~,
\end{equation}
\begin{equation}
\label{l3}
B_\phi =f(r)
\exp \left(i\int^r dr~k_r+im\phi \right)~,
\end{equation}
with the actual fields given by the real
parts and
with

\begin{equation}
\label{l4}
{\bf k \cdot B}=k_r B_r + k_\phi B_\phi =0~,
\end{equation}
where ${\bf k}\equiv \hat{\bf r}k_r +
\hat{\rvecphi~}k_\phi$, $k_\phi \equiv m/r$ is
the azimuthal wavenumber, and
$m=\pm1, \pm2,$ etc.
    The stationary magnetic
field satisfies ${\bf \nabla \times}
({\bf v \times B}) =0$, and this
requires

$$
{\bf v \times B} =
\hat{\theta}(v_\phi B_r- v_rB_\phi) =0~,
\quad{\rm or}\quad
B_\phi =(v_\phi /v_r) B_r~.
$$
That is, ${\bf v} \propto {\bf B}$
and ${\bf k \cdot v}=0$.
   Owing to equation (3),
\begin{equation}
\label{l5}
k_r = -(v_\phi/v_r) k_\phi~.
\end{equation}
    In ADAF  accretion models,
$v_\phi/v_r =$ const,  so
that $k_r \propto 1/r$.
    Thus, $|B_r| =|v_r/v_\phi| |B_\phi|
\propto 1/r^2$, and $f(r) \propto 1/r^2$.

   The nature of the field
is most easily obtained from
the flux function
$rA_\theta$ (with ${\bf A}$
the vector potential), where
$$
B_r =-{1\over r^2}
{\partial (r A_\theta) \over \partial \phi}~,
\quad \quad B_\phi = {1\over r} {\partial
(rA_\theta) \over \partial r} ~,
$$
for $\sin \theta \approx 1$.
Note that $({\bf B\cdot \nabla})(rA_\theta)
=0$.
     Thus the frozen-in field
is described by
\begin{equation}
\label{l6}
rA_\theta = {\rm const}~ \exp\left(
im \left[~\bigg|{v_\phi \over v_r}\bigg| \ln r
+\phi \right]
\right)~,
\end{equation}
where it is assumed that
$v_\phi >0$,  $v_r<0$, and $v_\phi/v_r=$const.
    The fact that $|rA_\theta|=$const
corresponds
to the conservation of the ingoing
or outgoing flux
in a given tube.
    A given field line satisfies
\begin{equation}
\label{l7}
r ={\rm const}~\exp\left(-\phi\bigg|{v_r \over v_\phi}
\bigg| \right)~.
\end{equation}
Figure 2 shows the nature of the
equatorial field for $m=2$.

  Consider now the
problem of magnetic field
amplification in
an accretion flow which
is {\it not} infinitely
conducting.
  The flow is described by the
magnetohydrodynamics (MHD) equations,
%%%%%%%%%%eqn(1)
$$
\rho { d {\bf v} \over dt}=
-{\bf \nabla} p +\rho {\bf g}
+{1\over c} {\bf J \times B} +
\eta \nabla^2 {\bf v}~,
$$
$$
{\bf \nabla \times B} =
{4 \pi \over c} {\bf J}~,\quad
{\bf \nabla \times E } =
-\frac{1}{c}\frac{\partial {\bf B}}{\partial t}~,
$$
\begin{equation}
\label{l8}
{\bf J} =\sigma \left({\bf E}
+ {\bf v \times B}/c\right)~,
\end{equation}
where ${\bf v}$ is the flow velocity,
${\bf B}$ the magnetic field,
$p$  the plasma pressure,
$\sigma$  the plasma
conductivity, $\eta$ the dynamic
viscosity (with $\nu=\eta/\rho$
the kinematic viscosity), ${\bf J}$  the
current density, and ${\bf E}$
the electric field.
  These equations can be combined
to give the induction equation,
%%%%%%%% eqn(2)
$$
\frac{\partial {\bf B}}{\partial t}=
{\bf \nabla \times (v \times B) }-
{\bf \nabla \times}
(\eta_m {\nabla \times} {\bf B})~,
$$
\begin{equation}
\label{l9}
\approx {\bf \nabla \times (v \times B) }+
\eta_m{\bf \nabla}^2 {\bf B}~,
\end{equation}
where $\eta_m \equiv c^2/(4\pi \sigma)$ is the
magnetic diffusivity, and the
approximation involves neglecting
${\bf \nabla}\eta_m$.

  In equations (\ref{l8}) and (\ref{l9}) both the
viscosity $\nu$ and the magnetic
diffusivity $\eta_m$ have the same
units and both are assumed
to be due to turbulence
in the accretion flow.
  Thus, it is reasonable to express
both transport coefficients using
the ``alpha'' prescription of
Shakura (1972) and
Shakura and Sunyaev (1973),
%%%%%%%%% eqn(3)
\begin{equation}
\label{l10}
\nu = \alpha~ c_s ~\ell_t~,
\quad \quad \eta_m  = \alpha_m ~c_s~ \ell_t~,
\end{equation}
where $\ell_t$ is the outer scale of
the turbulence, and $c_s=\sqrt{p/\rho}$
is the isothermal sound speed.
   Bisnovatyi-Kogan
and Ruzmaikin (1976) introduced
$\alpha_m$ and proposed that
$\alpha_m \sim \alpha$.
   Note that $\alpha$ and $\alpha_m$
characterize a turbulent MHD flow
in which there is a Kolmogorov cascade of
energy from large scales ($\ell_t$)
to much smaller scales where the
actual (microscopic) dissipation
of energy occurs.

  We obtain stationary $\bf B$
field solutions for
accretion flows with time-averaged
flow
${\bf v}=\hat{\bf r} v_r
+\hat{\rvecphi~}v_\phi$ and with
finite diffusivity $\eta_m$
in the form of equations (\ref{l2}) and (\ref{l3}),
but with $k_r$ complex,
$k_r=k_r^R+ik_r^I$.
   Equation (\ref{l3}) still applies, but (\ref{l5})
does not.
   In its place we have from equation (\ref{l9}),
\begin{equation}
\label{l11}
0 \approx i{\bf k} \times ({\bf v \times  B})
- \eta_m {\bf k}^2 {\bf B}~,
\end{equation}
where the approximation involves neglect
of terms of order $1/(|{\bf k}|r) \ll 1$ or
smaller.
      Solution
of (\ref{l11}) gives
\begin{equation}
\label{l12}
k_r^R v_r +k_\phi v_\phi \approx 0~,\quad {\rm and}
\quad k_r^I \approx
- {\eta_m {\bf k}^2 \over |v_r|}~,
\end{equation}
where in this equation ${\bf k}^2 =
(k_r^R)^2 + k_\phi^2$.
   Consistent with the first part
of (12), we take $k_r^R = n_r/r$,
where $n_r=$ const.
   For the ADAF flows, we write
the second part of
(\ref{l10}) as $\eta_m=\alpha_m c_s r$.
  The second part of (\ref{l12}) then
gives
\begin{equation}
\label{l13}
k_r^I =-\left ({\alpha_m c_s \over |v_r|}\right)
{n_r^2 +m^2 \over r}~,
\end{equation}
where the quantity $\alpha_m c_s/|v_r|=$ const.

    The imaginary part of $k_r$ gives
an additional radial dependence of $B_r$ and
$B_\phi$ (and $rA_\theta$), multiplying the
earlier expressions (\ref{l2}),(\ref{l3}), and (\ref{l6}) by
a factor $r^\delta$, where
$\delta= (\alpha_m c_s/|v_r|)(n_r^2+m^2)$.
   For radial distances $r<r_{equi}$,
we have equipartition with magnetic
energy density ${E}_{m} \propto
r^{-{5/ 2}}$ so that $|{\bf B}_p| \propto
r^{-{5/ 4}}$.
   This requires  $\delta=3/4$
so that $|rA_\theta| \propto r^{3/4}$,
which corresponds to the flux within
a given tube decresing as $r$ decreases.
   The accretion speed is
\begin{equation}
\label{l14}
|v_r| ={4 \over 3}(n_r^2+m^2)\alpha_m c_s~.
\end{equation}
Note that for the present solution
  the accretion depends on
the magnetic diffusivity $\eta_m$ but  not
the viscosity $\nu$;  more general
solutions have $v_r$ dependent on
both viscosity and diffusivity.
  Validity of this solution
requires $|k_r^I|/|{\bf k}| =
(3/4)/\sqrt{n_r^2+m^2} \ll 1$.
  The field line pattern is similar
to that in Figure 2, but the number
of field lines in a tube $\propto
r^{3/4}$.

   The magnitude of the magnetic
field follows from the requirement
that the inflow of  angular
momentum carried by the
matter, $\dot{L}_{mech}=
4\pi r^3 <\sin\theta>\rho v_r v_\phi \approx
4 \pi r^3 \rho v_r v_\phi <0$,
equal the outflow
carried by the field,
$\dot{L}_{mag}=-r^3 B_r B_\phi>0$, which
gives $\rho v_r^2 \approx B_r^2/4\pi$.
   More generally,  turbulent viscosity
carries away part of the angular momentum
in which case the magnetic field
strength is reduced.
    For $r\leq r_{equi}$, we have
$B_r=B_0(r_{equi}/r)^{5/4}$, and
zero angular momentum flux
gives
\begin{equation}
\label{l15}
B_0= \left( {4 (n_r^2+m^2)\alpha_m (c_s/v_\phi)
\sqrt{GM} \dot{M}  \over
3 r_{equi}^{5/2} }\right)^{1/2}~.
\end{equation}
    This gives $B_0\approx 55$ G for
$r_{equi}=10^{15}$ cm, $M=10^6 M_\odot$,
$\dot{M}=0.1 M_\odot/yr$, $\alpha_m=0.1$,
$c_s/v_\phi=0.1$, and $n_r^2+m^2 =10^2$.

\subsection{Stationary picture of the      %%%%%%%%%%%%%%%%%%%%%
magnetic field distribution
in quasi-spherical accretion:
turbulence, equi\-parti\-tion and field
an\-ni\-hi\-la\-tion; two-tem\-pe\-ra\-tu\-re ad\-vec\-tive \\ discs}

In the optically thin accretion
discs at low mass fluxes the density
of the matter is low and
energy exchange between electrons and ions
due to binary collisions is slow.
   In this situation, due to different
mechanisms of heating and
cooling for electrons and ions, they may
have different temperatures.
First it was realized by Shapiro,
Lightman, Eardley (1976) where
advection was not included. It was
noticed by Narayan and Yu (1995) (see also Ichimaru, 1977),
that advection in this case is becoming
extremely important.
   It may carry the main energy flux into a black
hole, leaving rather low efficiency
of the accretion up to $10^{-3}\,-\,
10^{-4}$ (advective dominated
accretion flows - ADAF).
This conclusion is valid
only when the effects, connected
with magnetic field annihilation
and heating of matter due to it are
neglected.

  In the ADAF solution the ion
  temperature is about a virial one
$kT_i \sim GMm_i/r$, what means
that even at high initial angular
momentum the disk becomes very
thick, forming practically a quasi-spherical
accretion flow.
It is connected also with
an ``alpha'' prescription of
viscosity. At high ion temperatures,
connected with a strong
viscous heating, the ionic
pressure becomes high, making
the viscosity very effective.
So, due to suggestion of ``alpha''
viscosity in the situation,
when energy losses by ions are
very low, a kind of a ``thermo-viscous''
instability is developed,
because heating increases a viscosity, and
viscosity increases a heating.
Developement of this instability
leads to formation of ADAF.

A full account of the processes,
connected with a presence of magnetic
field in the flow, is changing
considerably the picture of ADAF. It was
shown by Schwarzman (1971),
that radial component of the magnetic field
increses so rapidly in the
spherical flow, that equipartition between
magnetic and kinetic energy
is reached in the flow far from the black hole
horizon. In the region where
the main enegry prodiction takes place, the
condition of equipartition
takes place. In presence of a high magnetic
field the efficiecy of a
radiation during accretion of an interstellar matter
into a black hole increase
enormously from $\sim 10^{-8}$ up to $\sim 0.1$
(Schwarzman, 1971), due to
efficiency of a magneto-bremstrahlung
radiation. So possibility of
ADAF regime for  a spherical accretion was
noticed long time ago.
To support the condition of equipartition a
continuous magnetic field reconnection
is necessary, leading to annihilation
of the magnetic flux and heating of
matter due to Ohmic heating.
It was
obtained by Bisnovatyi-Kogan and
Ruzmaikin (1974), that due to Ohmic heating
the efficiency of a radial
accretion into a black hole
may become as high as
$\sim 30\%$. The rate of the
Ohmic heating in the condition of
equipartition was obtained in the form

\begin{equation}
\label{ref3.5}
T\frac{dS}{dr} =
-\frac{3}{2}\frac{B^2}{8\pi \rho r}.
\end{equation}
In the supersonic flow of the
radial acrretion equipartition between
magnetic and kinetic energy
was suggested by Schwarzman (1971):

\begin{equation}
\label{ref3.6}
\frac{B^2}{8\pi} \approx
\frac{\rho v_r^2}{2}=\frac{\rho GM}{r}.
\end{equation}
For the disk accretion, where
there is more time for a field dissipation,
almost equipartition was suggested
(Shakura, 1972) between magnetic and
turbulent energy, what reduces
with account of "alpha" prescription
of viscosity to a relation

\begin{equation}
\label{ref3.7}
\frac{B^2}{8\pi}  \sim \frac{\rho v_t^2}{2} \approx
\alpha_m^2 P,
\end{equation}
where $\alpha_m$ characterises a
magnetic viscosity in a way similar to
the turbulent $\alpha$ viscosity.
It was suggested by Bisnovatyi-Kogan and
Ruzmaikin (1976) the similarity
between viscous and magnetic Reynolds
numbers, or between turbulent and
magnetic viscosity coefficients

\begin{equation}
\label{ref3.8}
Re=\frac{\rho v l}{\eta}, \quad
Re_m=\frac{\rho v l}{\eta_m},
\end{equation}
where the turbulent magnetic
viscosity $\eta_m$ is
connected with a turbulent
conductivity $\sigma$

\begin{equation}
\label{ref3.9}
\sigma=\frac{\rho c^2}{4\pi \eta_m}.
\end{equation}
Taking $\eta_m=\frac{\alpha_m}{\alpha}\eta$,
we get a turbulent conductivity

\begin{equation}
\label{ref3.10}
\sigma=\frac{ c^2}{4\pi \alpha_m h v_s},
\quad v_s^2=\frac{P_g}{\rho}
\end{equation}
in the optically thin discs.
For the radial accretion the turbulent
conductivity may contain mean
free path of a turbulent element $l_t$,
and turbulent viscocity
$v_t$ in (\ref{ref3.10}) instead of
$h$ and $v_s$.
In ADAF solutions, where the ion
temperature is of the order of the virial one
the two suggestions (\ref{ref3.6})
and (\ref{ref3.7}) almost coinside at
$\alpha_m \sim 1$.

The heating of the matter due
to an Ohmic dissipation
may be obtained from the Ohm's
law for a radial accretion in the form

\begin{equation}
\label{ref3.11}
T\frac{dS}{dr} =-\frac{ \sigma {\cal E}^2}{\rho v_r}
\approx -\sigma \frac{v_E^2 B^2}{\rho v_r c^2}
=-\frac{B^2 v_E^2}{4 \pi \rho \alpha_m v_r v_s l_t},
\end{equation}
what coinsides with (\ref{ref3.5})
when $ \alpha_m=\frac{4 r v_E}{3 v_r l_t}$, or $l_t
=\frac{4 r v_E^2}{3 v_r v_s \alpha_m}$.
Here a local electrical field strength
in a highly conducting plasma is of the
order of ${\cal E} \sim \frac{v_E B}{c}$,
$v_E\,\sim\,v_t\,\sim\,\alpha\,v_s$
for the radial accretion.

Equations for a radial temperature
dependence in the accretion disk,
separate for the ions and
electrons are written as

\begin{equation}
\label{ref3.12a}
{dE_i\over dt}-{P_i\over\rho^2}{d\rho\over dt}=
{\cal H}_{\eta i}+{\cal H}_{Bi}-Q_{ie}~,
\end{equation}
\begin{equation}
\label{ref3.12b}
{dE_e\over dt}-{P_e\over\rho^2}{d\rho\over dt}=
{\cal H}_{\eta e}+{\cal H}_{Be}+Q_{ie}
-{\cal C}_{brem}-{\cal C}_{cyc}~,
\end{equation}
Here $\frac{d}{dt}=\frac{\partial}{\partial t}
+v_r\frac{\partial}{\partial r}$.
A rate of a viscous heating
of ions ${\cal H}_{\eta i}$ is obtained from
(\ref{ref1.11}) as

\begin{equation}
\label{ref3.13}
{\cal H}_{\eta i}=
\frac{2\pi r}{\dot M}Q_+ \,=\,\frac{3}{2}
\alpha\frac{v_K v_s^2}{r}, \quad {\cal H}_{\eta e} \le
\sqrt{\frac{m_e}{m_i}} {\cal H}_{\eta i}.
\end{equation}
The equality in the last
equation is related to binary collisions.
Combining (\ref{ref1.3}),
(\ref{ref1.14a}),(\ref{ref1.9}),(\ref{ref1.13}),
we get
>\begin{equation}
\label{ref3.14}
v_r=\alpha\frac{v_s^2}{v_K {\cal J}},
\quad h=\sqrt{2}\frac{v_s}{v_K} r,\quad
\rho=\frac{\dot M}{4 \pi \alpha \sqrt{2}}
\frac{v_K^2 {\cal J}}{r^2v_s^3},
\end{equation}
where $v_K=r\Omega_K$, ${\cal J}=1-\frac{j_{in}}{j}$.
The rate of the energy
exchange between ions and
electrons due to the binary
collisions was obtained by Landau (1937),
Spitzer (1940) as

\begin{equation}
\label{ref3.15}
Q_{ie} \approx
{4(2\pi)^{1\over2}n e^4 \over m_im_e}
\left({T_e\over m_e}
+{T_i\over m_i}\right)^{-{3\over2}}
{\ell}n\Lambda(T_i-T_e),
\end{equation}
with $\ell n \Lambda={\cal O}(20)$
the Coulomb logarithm.
Thermodynamic functions must take
into account a possible relativistic
effects for electrons,
which energy may exceed $m_e c^2$.
    Neglecting
pair formation for an low density
accretion disk, we may write an exact
expression for a pressure

\begin{equation}
\label{ref3.16}
P_g=P_i+P_e=n_ikT_e+n_ekT_p = n_e k(T_e+T_i),
\end{equation}
and an approximate expression
for an energy, containing a smooth
interpolation between nonrelativistic
and relativistic electrons

\begin{equation}
\label{ref3.17}
E_g=E_e+E_i \approx \frac{\frac{3}{2}m_ec^2+3kT_e}{m_ec^2
+kT_e}\frac{kT_e}{m_p}+\frac{3}{2}\frac{kT_i}{m_p}.
\end{equation}
The electron bremstrahlung ${\cal C}_{brem}$
and magneto-bremstrahlung
${\cal C}_{cyc}$ cooling of
maxwellian semi-relativistic electrons, with
account of free-bound radiation
in nonrelativistic limit, may be written as
by inrerpolation of limiting
cases (Bisnovatyi-Kogan and Ruzmaikin, 1976)

\begin{equation}
\label{ref3.18}
{\cal C}_{brem} \approx
\frac{2\times 10^{30}m_ec^2+3\times
10^{32}(kT_e)^{3/2}}{m_ec^2+kT_e}(kT_e)^{1/2},
\end{equation}

\begin{equation}
\label{ref3.19}
{\cal C}_{cyc} \approx
\frac{1.7\times 10^{16}m_ec^2+1.7\times
10^{22}(kT_e)^2}{m_ec^2+kT_e}kT_e\,B^2.
\end{equation}
Effects of the cyclotron
self-absorption may be important for nonrelativistic
electrons, where radiation
itself is low. With increasing of the electron
temperature self-absorption
effects decrease, because the black-body
emission increase with a temperature
($\sim T^4$) more rapidly then
magneto-bremstrahlung volume losses.
Account of self-absorption was
investigated by Trubnikov (1973).

In the case of a disk accretion there
are several characteristic velocities,
$v_K$, $v_r$, $v_s$, and $v_t=\alpha v_s$,
 all of which may be used
 for determining "equipartition"
magnetic energy, and one
characteristic length $h$.
Consider three possible choices with
$v_B^2=v_K^2$, $v_r^2$, and
$v_t^2$  for scaling $B^2=4\pi\rho v_B^2$,
note that choice of cross products of different
velocities and of $v_s$, is also possible.
We get the following expressions for
magnetic field strengths in
different cases, with account of (\ref{ref3.14})

$$ {\bf a}.\quad\,\,\, v_B=v_K,\,\,\quad
\frac{B^2}{8\pi}=\frac{\rho v_K^2}{2}, \quad
B=\left(\frac{\dot M}{\alpha \sqrt 2}
\frac{v_K^4 {\cal J}}{r^2v_s^3}
\right)^{1/2},\,\,\,$$
\begin{equation}
\label{ref3.20}
 {\bf b}.\quad\,\,\,  v_B=v_r, \,\,\quad
 \frac{B^2}{8\pi}=\frac{\rho v_r^2}{2}, \quad \,\,\,
B=\left(\frac{\dot M \alpha}
{\sqrt 2{\cal J}} \frac{v_s }{r^2}
\right)^{1/2} \quad \,\,\,,
\end{equation}
$$ {\bf c}.\,\,\, v_B \sim v_t,\,\, \frac{B^2}{8\pi}=
\frac{\alpha_m^2}{\alpha^2}\frac{\rho v_t^2}{2}, \quad
B=\alpha_m\left(\frac{\dot M}
{\alpha \sqrt 2} \frac{v_K^2 {\cal J}}{r^2v_s}
\right)^{1/2}.$$
The expression for an Ohmic
heating in the turbulent accretion disk also
may be written in different ways,
using different velocities $v_E$ in the
expression for an effective electrical field
\begin{equation}
\label{ref3.21}
{\cal E}=\frac{v_E B}{c}.
\end{equation}
A self-consistency of the
model requires, that expressions
for a magnetic heating of the
matter ${\cal H}_B$, obtained from the
condition of stationarity of
the flow (\ref{ref3.5}), and from the Ohm's
law (\ref{ref3.11}), should be identical.
That gives some restrictions
for the choice of a characteristic
velocity in (\ref{ref3.21}).
Comparison
between (\ref{ref3.5}) and (\ref{ref3.11})
with account of
(\ref{ref3.10}),(\ref{ref3.14})
shows the identity of these two
expressions at

\begin{equation}
\label{ref3.22}
 \frac{\alpha}{{\cal J}\alpha_m}
\frac{v_E^2}{v_r^2}=
\frac{3\sqrt{2}}{4}.
\end{equation}
That implies $v_E\sim v_r \sim \frac{\alpha v_s^2}{v_K \sqrt{J}}
\simeq \frac{v_t v_s}{v_K \sqrt{J}} < v_t$
So, the model is becoming
self-consistent at the reasonable choice of
the parameters.
Note, that in the advective models
${\cal J}$ is substituted by
another function which is
not zero at the inner edge of the disk.
The heating due to magnetic
field reconnection ${\cal H}_B$
in the equations (\ref{ref3.12a}),
(\ref{ref3.12b}),
obtained from (\ref{ref3.5})
and (\ref{ref3.20}), may be written
with account of (\ref{ref3.13}) as

\begin{equation}
\label{ref3.23}
{\cal H}_B=\frac{3}{16 \pi}
\frac{B^2}{r \rho} v_r
    = \frac{1}{2{\cal J}}{\cal H}_{\eta i}
    \left(\frac{v_B}{v_K}\right)^2.
\end{equation}
So, at $v_B=v_K$ the expressions
for viscous and magnetic heating are
almost identical.
The distribution of the magnetic
heating between electrons and ions has
a critical influence on the model,
if we neglect the influence of a plasma
turbulence on the energy relaxation,
and take into account only
the energy exchange by binary
collisions from (\ref{ref3.15}).
Observations
of the magnetic field reconnection
in the solar flares show (Tsuneta, 1996),
that electronic heating prevails.

It follows from the physical picture of
the field reconnection,
that transformation of the magnetic energy into
a heat is connected with the
change of the magnetic flux, generation
of the vortex electrical field,
accelerating the particles.
This vortex
field has a scale of the
turbulent element and suffers rapid and
chaotic changes.
The accellerating forces on
electrons and protons in
this fields are identical,
but accelerations themselfs differ $\sim 2000$
times, so during a sufficiently
short time of the tubulent pulsation
the electron may gain much
larger energy, then the protons.
Additional particle
acceleration and heating
happens on the shock fronts, appearing around
turbulent cells, where
reconnection happens.
In this process
acceleration of the electrons is
also more effective than of the protons.
In the paper of Bisnovatyi-Kogan
and Lovelace (1997) the equations
(\ref{ref3.12a}), (\ref{ref3.12b})
have been solved in the
approximation of nonrelativistic
electrons, $v_B$=$v_K$, what
permitted to unite a viscous and magnetic
heating into a unique formila.
The combined heating of the electrons
and ions were taken as

\begin{equation}
\label{ref3.24}
{\cal H}_e=(2-g){\cal H}_{\eta i},
\quad {\cal H}_e =g{\cal H}_{\eta i}.
\end{equation}
In the expression for a
cyclotron emission self-absorption was taken into
account according to Trubnikov (1973).
The results of calculations for
$g=0.5 \div 1$ show that
almost all energy of the electrons is radiated,
so the relative efficiency of
the two-temperature, optically thin disk
accretion cannot become lower then 0.25.
Note again that accurate account
of a plasma turbulence for a
thermal relaxation and corresponding increase
of the term $Q_{ie}$ may restore
the relative efficiency to its unity value,
corresponding to the optically thick discs.

  Studies of the solar corona have
led to the general conclusion
that the energy build up in
the chaotic coronal magnetic field by
slow  photospheric
driving is released by
magnetic reconnection events or flares on
a wide range of scales and energies
(Benz 1997).
    The observations support the idea of
``fast reconnection,'' not
limited to a rate proportional
to an inverse power of the
magnetic Reynolds number (Parker 1979, 1990).
   The data clearly show the
rapid acceleration of electrons (e.g.,
hard X-ray flares, Tsuneta 1996)
and ions (e.g.,
gamma ray line events and
energetic ions, Reames {\it et al.} 1997), but
the detailed mechanisms of particle
acceleration are
not established.

   An essential aspect of the coronal
heating is the continual input
of energy to the chaotic coronal
magnetic field due to motion of
the photospheric plasma.
    As emphasized here, there is
also a continual input of energy
to the chaotic magnetic in a quasi-spherical
accretion flow due to compression of the
flow.
   The fact that the ratio of the plasma
to magnetic pressures is small in
the corona but of order unity in the
accretion flow may of course affect
the details of the plasma processes.

In magnetic field reconnection
  plasma flows into the neutral
layer from above and below with
a speed of the order of the turbulent
velocity $v_t$.  Magnetic flux is
destroyed in this layer as required
for the reasons discussed earlier.
  Consequently, there
is an electric field in the $z-$direction
${\cal E}_z =-(1/c)\partial A_z/\partial t
={\cal O}( v_t B_0/c)$, where
$B_0$ is the field well outside of the
neutral layer.
  This is the same as the
estimate of the electric field given
by Bisnovatyi-Kogan and Lovelace (1997).
   In the
vicinity of
the neutral layer ($|y| \ll \ell_t$)
this electric field
is typically
much larger than the Dreicer electric
field for electron runaway
$E_D =4\pi e^3(n_e/kT_e)\ln\Lambda$
(Parail \& Pogutse 1965), and
this  leads to streaming motion
of electrons in the $z-$direction.
   Note that it is much more
difficult to have ion runaway because
of the large gyro radii of ions
(Bisnovatyi-Kogan and Lovelace, 1997; Lesch, 1991).
  Of course, in the almost uniform magnetic
field assumed by Quataert (1998)
there is  no runaway of
particles.

    The streaming of the electrons
can give rise to a number of different
plasma instabilities the growth
of which gives an
anomalous resistivity.
   Galeev and
Sagdeev (1983) (see also Li {\it et al.} 1999) discuss
the quasi-linear theory
for such conditions and derive expressions
for the
rate of heating of electrons,
%%%%%%%%%% eqns(19) and (20)
\begin{equation}
{d T_e \over dt} = {1\over n}
\int {d^3k \over (2\pi)^3}~
\gamma_{\bf k}~W_{\bf k} ~{{\bf k \cdot u}_{e}
\over \omega_{\bf k}}~,
\end{equation}
and ions,
\begin{equation}
{d T_i \over dt} = {1\over n}
\int {d^3k \over (2\pi)^3}~
\gamma_{\bf k}~W_{\bf k} ~,
\end{equation}
where $W_{\bf k}$ is the wavenumber energy
spectrum of the turbulence,
$\gamma_{\bf k}$
is the linear growth rate of
the mode with real frequency
$\omega_{\bf k}$, $n$ is the
number density, and ${\bf u}_e$
is the electron drift velocity.
   If the quantity
${\bf k \cdot u}_e/\omega_{\bf k}$ is
assumed constant and taken out of the
integral sign in (19), then the ratio
of heating of electrons to that of
ions is
\begin{equation}
{dT_e \over dT_i} = { u_e \over (\omega /k)}~.
\end{equation}
This result is independent of the instability
type.
    Galeev and Sagdeev (1983)  point
out that for most instabilities, relation (21)
predicts   faster heating of electrons than
ions.
    Earlier, Lesch (1991) and Di Matteo (1998)
emphasized the role of reconnection
 in accelerating electrons
to relativisitic energies in accretion
flows of AGNs.

\section{Accretion on to Magnetized Stars}

\subsection{Disk Accretion on to a Rotating Star with an Aligned
Dipole B-Field}

\subsubsection{Introduction}

The problem of matter accretion on to
magnetized stars has been of continued
interest since the discovery of X-ray
pulsars in binary systems (Giacconi {\it et al.} 1971;
Schreier {\it et al.} 1972;
Tananbaum {\it et al.} 1972). The X-ray
pulsars were interpreted as magnetized
neutron stars, accreting matter from a
companion star (Pringle and Rees 1972;
Davidson and Ostriker 1973; Lamb,
Pethick, and Pines 1973; Rappaport and
Joss 1977). Of the more than 40 of known
X-ray pulsars, most are in high-mass
($M_{tot}>15
$) binary systems, while only few are in
low-mass ($M_{tot} < 3 $) binary systems
(see, for example, review by Nagase
1989).

 From the early observations, many X-ray
pulsars were found to show secular
decreases of pulse period (spin-up of
the neutron star  rotation) with
relatively large rates of change of
$-\dot P/P=10^{-2}-10^{-6}~{\rm yr}^{-1}$
(Gursky and Schreier 1975; Schreier and
Fabbiano 1976; Rappaport and Joss 1983).
The long-term monitoring of X-ray
pulsars during the last two decades have
shown a wide variety of pulse-period
evolution on different time-scales. Some
X-ray pulsars show a steady spin-up
evolution during many years which
rapidly changes to a steady spin-down
evolution (e.g., GX 1+4, SMC X-1, 4U
1626-67). Others show the wavy
variations of $P$ on time-scales of a
few years on a background of systematic
spin-up or spin-down (e.g., Cen X-3, Her
X-1, Vela X-1) (Nagase 1989; Sheffer {\it et al.} 1992;
Bildsten {\it et al.}, 1997).

    Short-term fluctuations of
pulse-period on time-scales of days to
months, including clear evidence of
spin-down episodes, were found for a
number of sources, for example, Her X-1
(Giacconi 1974), Cen X-3 (Fabbiano and
Schreier 1977), Vela X-1 (Nagase {\it et al.}
1984). Detailed observations of Vela X-1
with Hakucho and Tenma satellites, and
with HEAO-1 have revealed fluctuations
of large amplitude with time-scales about
3-10 days, which were estimated to be
random in both amplitude and sign
(Nagase 1981). BATSE observations have
shown that in some cases (Cen X-3, Her
X-1) the timescale of fluctuations is
less than one day with the change of
sign from spin-up to spin-down and
vice-versa (Bildsten {\it et al.} 1997). Note that
some X-ray pulsars (usually the
long-period ones) show no evidence of
regular spin-up or spin-down (for
example, 4U 0115+63, GX 301-2).

In some cases, like Her X-1, the
existence of an accretion disk is
indicated from various observations in
the X-ray and optical bands (for
example, Middleditch and Nelson 1976;
Bisnovatyi-Kogan {\it et al.} 1977;
Middleditch, Puetter, and Pennypacker
1985). In several other cases,  such as
Cen X-3, SMC X-1, GX 1+4 (Chakrabarty
{\it et al.} 1997; Cui
1997), and 4U 1626-67,
there are also different kinds
of evidence, that matter from companion
star forms a disk (for example, Tjemkes
{\it et al.} 1986; Nagase 1989). Most of these
are short-period pulsars with period
about several seconds or less (excluding
GX 1+4 with period 122 s.). However,
most of long-period pulsars, are supposed
to be powered by the capture of stellar
wind from companion. However if the
companion is the star of Be type with
slow rate of outflow, then it is
probable that once again the disk first
forms, opposite to the case when the
companion star is of O-B kind with high
speed of matter outflow (see review of
Nagase 1989).

Some aspects of matter accretion by
magnetized neutron stars were considered
for the first time by Bisnovatyi-Kogan
and Friedman (1969), and by Shvartsman
(1971). Ideas on the spinning-up of
pulsars by accretion were first proposed
in the case of disk-fed pulsars by
Pringle and Rees (1972), Lamb, Pethick
and Pines (1973), and Lynden-Bell and
Pringle (1974), and for wind-fed pulsars
by Davidson and Ostriker (1973), and
Illarionov and Sunyaev (1975) (see also
Lipunov 1993).  The spin-down of a
rapidly rotating magnetized star due to
``propellar'' action of the magnetosphere
was suggested by Shvartsman (1971) and
Illarinov and Sunyaev (1975), but a
detailed model has not been worked out.
Pringle and Rees (1972) considered
matter accretion from a viscous disk to
a magnetized rotating neutron star.
Their theory considered the idealized
case of an aligned rotator where the
magnetic moment of pulsar is aligned
parallel or anti-parallel with the
pulsar's spin-axis and the normal to the
plane of the accretion disk. They
supposed that accreting matter stops at
the point where the magnetic pressure of
the magnetosphere becomes equal to
pressure of matter. At this distance,
matter transfers its angular momentum to
the star through a thin boundary layer.
Another idea regarding the width of the
transition zone between the unperturbed
accretion disk and magnetosphere, was
considered in series of papers by Ghosh
and Lamb (1978, 1979a,b).
   They assumed that magnetic
field of the neutron star threads the
accretion disk at different radii in a
broad ``transition'' zone, in spite of
predominance of the disk stress compared
with magnetic stress in this region.
   Their work is based
on the idea of anomalous resistivity
connected with the supposed fast
reconnection of toroidal component of
the magnetic field lines across the
disk.

    Lovelace, Romanova,
and Bisnovatyi-Kogan (1995,
hereafter LRBK95;
Li and Wickramasinghe 1997)
proposed an essentially
different picture for disk accretion
onto a star with an aligned dipole
magnetic field.
    The idea is based on the
fact that when there is a large
difference between the angular velocity
of the star and that of the disk, the
magnetic field lines threading the star
and the disk undergo a rapid inflation
so that the field becomes open with
separate regions of field lines
extending outward from both the star and
the disk.
    As a result, the magnetosphere
consists of an open field line region
far from the star and a closed region
approximately corotating close to the
star.
    This is shown schematically in
Figure 9.
    The LRBK95 model does not assume an
anomalous resistivity, but rather a
turbulent magnetic diffusivity
proposed by Bisnovatyi-Kogan and Ruzmaikin
(1976), analogous
%of the
%disk comparable
to the turbulent
$\alpha$ viscosity of Shakura
(1972) and Shakura and Sunyaev (1973).
    This corresponds to a
resistivity about
$10^4$ times smaller than that assumed
by Ghosh and Lamb (1978,1979a,b).
Campbell (1992) has earlier discussed
closed field line models of disk
accretion onto an aligned dipole star
assuming a magnetic diffusivity
comparable to the turbulent
$\alpha$ viscosity.

%%%%%%%%%%%%%%%%%%%%%%%%%%%%%%%%%%%%%%%%%%%%%%%%%%%%%%%%%%%%%%%%%%%%%%%%%%%%%%%
%\begin{figure}
%\epsfysize=10cm % fix the y-dimension and scales x-dim. to y-dim.
%\centerline{\epsfbox{lrbk95_1.EPS}}%

%\label{LRBK1}
%\end{figure}
%%%%%%%%%%%%%%%%%%%%%%%%%%%%%%%%%%%%%%%%%%%%%%%%%%%%%%%%%%%%%%%%%%%%%%%%%%%%%%%

    The existence of an open
magnetic field region of the disk leads
to the possibility of magnetically
driven outflows.
   LRBK95 developed a model for
magnetohydrodynamic (MHD) outflows and
their back influence on the disk using
the work on MHD outflows and magnetized
disks by Blandford and
Payne (1982), Lovelace, Berk, and Contopoulos
(1991), Lovelace,
Romanova, and Contopoulos (1993),
and Lovelace, Romanova, and Newman
(1994).
     In \S 5.1.1 -5.1.8 we review the arguments
of LRBK95.
       In \S 5.2 we discuss the magnetic
propeller regime of accretion first considered
by Illarionov and Sunyaev (1975) and more
recently by Lovelace, Romanova and
Bisnovatyi-Kogan (1999; hereafter LRBK99).

\subsubsection{Theory}

The basic equations for an assumed
stationary plasma  configuration  are
$$
\nabla \cdot (\rho {{\bf  v}}) =
0~,~~{\bf{\nabla}}
\times {\bf B} = {4\pi \over c} {\bf
J}~,~~{\bf{\nabla}}
\cdot {{\bf B}} = 0~,~~{\bf{\nabla}}
\times {{\bf  E}} = 0~,
$$
\begin{equation}
\label{adi1}
{{\bf  J}} = \sigma_e
({\bf  E} + {\bf  v}
\times {\bf B}/c)~,~~\rho {\bf{(v
\cdot
\nabla)v}} =
  - \nabla
\rm p + \rho {\bf{g}} + {1\over c}
{{\bf  J}}
\times {{\bf B}} + {{\bf
F}}^{\rm{vis}}~.
\end{equation}
Here, ${{\bf  v}}$ is
the flow velocity, $\rho$ is the
density, ${\sigma_e}$ is the effective
electrical conductivity,
${\bf  F}^{\rm{vis}}$ is the viscous
force density, $p = \rho k_B T/m$ is the
gas pressure (with $k_B$ the Boltzmann
constant, $m$ the mean particle mass,
and with the radiation pressure assumed
negligible), and ${\bf{g}}$ is the
gravitational acceleration.  Outside of
the disk, dissipative effects are
considered to be negligible ($\sigma_e
\rightarrow
\infty,~{{\bf  F}}^{\rm{vis}} = 0$,
etc.).  We neglect the self-gravity of
the disk and relativistic effects so
that ${\bf{g}} = -{\bf{\nabla}}
\Phi_g$ with $\Phi_g = - GM/(r^2 +
z^2)^{1/2}$, where $M$ is the mass of
the central star.  Equations (\ref{adi1}) are
supplemented later by an equation for the
conservation of energy in the disk.

A general axisymmetric
${{\bf B}}$-field can be written as
${{\bf B}} =  {{\bf B}}_p + {\hat{
\phi}} B_\phi$,  where ${\bf B}_p =
\nabla
\times ({\hat{\bf \phi}}
\Psi/r)$ is the poloidal field,
$B_\phi$ is the toroidal field, and
$\Psi$ is the flux function (see, for
example, Mestel 1968, or Lovelace {\it et al.}
1986).
    We use a non-rotating cylindrical
coordinate system so that ${{\bf B}}_p =
(B_r, 0, B_z)$.
    Notice that $\Psi (r,z) =$ const
labels a poloidal field line,
$({\bf B}_p \cdot {\bf \nabla}) \Psi =
0$, or a flux-surface if the poloidal
field line is rotated about the z-axis.
For the present problem, the
${{\bf B}}_p$ field  can be represented
as
$\Psi = \Psi_* + \Psi'$, with
$\Psi_*$ the star's field-assumed to be
a dipole $\Psi_* = \mu r^2 (r^2 +
z^2)^{-{3\over 2}}$ and with
$\Psi'$ due to the non-stellar toroidal
currents.

\subsubsection{Inflation and Opening of
Coronal
${\bf B}$-Field}

Figure 10 shows  the poloidal
projections of two nearby field lines
connecting the star and the disk.
Because
$\partial /\partial t = 0$, the ${{\bf
E}}$ field is electrostatic and the
poloidal plane line integral of ${{\bf
E}}$ around the loop, $1
\rightarrow 2 \rightarrow 3
\rightarrow 4
\rightarrow 1$, is zero.  Because of
axisymmetry and the fact that
${{\bf  E}} + {{\bf  v}}
\times {{\bf B}}/c = 0$ outside of the
disk, the line integrals of ${{\bf  E}}$
along the two curved segments in Figure
10 vanish separately.  That is, the
electrostatic potential is a constant on
any given flux surface (see, for
example, Lovelace {\it et al.} 1986).  The
potential difference between the points
(1,2) on the star's surface is
$- {\bf{\delta r}}_{12} \cdot {{\bf
E}}_* = \omega_*
\delta \Psi/c$, with $\delta \Psi \equiv
\Psi_1 - \Psi_2$, where we have assumed
that the star is perfectly conducting
$[{{\bf  E}}_* = - ({{\bf  v}}
\times {{\bf B}})_*/c]$, and where
$\omega_*$ is the angular rotation rate
of the star.  Thus, we have
$\delta r_{34}(E_r)_d = - \omega_*
\delta \Psi /c =  - \omega_*
rB_z(r,0)\delta r_{34}/c$, where the
path element $\delta r_{34}$ is in the
mid plane of the disk, and the ``$d\,$''
subscript indicates that the quantity is
evaluated in the disk.  In turn, we have
$E_r\big|_d = (J_r/\sigma_e - v_\phi
B_z/c)_d$.  From Amp\`ere's law,
$J_r = -(c/4\pi) (\partial B_\phi/
\partial z)$, the fact that
$B_\phi$ is necessarily an odd function
of
$z$, and the approximation
$\partial B_\phi /\partial z =
(B_\phi)_h/h$, we have $J_r \alpha
\approx - (c/4\pi)(B_\phi)_h/h$, where
$h$ is the half-thickness of the disk,
and the notation $(\cdots)_h$ indicates
that the quantity is evaluated at
$z = h$.  Combining these results gives
for stationary conditions

\begin{equation}
\label{adi2}
(B_\phi)_h (r) = - {{hr}\over
\eta_t}[\omega_d (r) - \omega_*] B_z
(r)~,
\end{equation}
where $\eta_t = c^2/(4\pi
\sigma_e)$ is the magnetic diffusivity
of the disk, and $\omega_d$ is the
angular rotation rate of the disk.  The
fractional variation in $B_z$ between
the midplane and surface of the disk is
negligible
$[O(h/r)]$ if the disk is thin in the
sense that $h/r \ll 1$.  For the
conditions of interest here, it follows from
(\ref{ref1.3})
that $h/r \leq c_s/v_K$, where
$c_s \equiv (k_B T/m)^{1\over 2}$ the
Newtonian sound speed based on the
mid-plane temperature of the disk, and
$v_K = (GM/r)^{1\over 2}$ is the
Keplerian velocity at $r$. Campbell
(1992) has independently derived
equation (\ref{adi2}) but does not place any
limit on the ratio
$|B_{\phi}/B_z|$.

%%%%%%%%%%%%%%%%%%%%%%%%%%%%%%%%%%%%%%%%%%%%%%%%%%%%%%%%%%%%%%%%%%%%%%%%%%%%%%%
%\begin{figure}
%\epsfysize=7cm % fix the y-dimension and scales x-dim. to y-dim.
%\centerline{\epsfbox{lrbk95_2.EPS}} %

%\label{LRBK2}
%\end{figure}
%%%%%%%%%%%%%%%%%%%%%%%%%%%%%%%%%%%%%%%%%%%%%%%%%%%%%%%%%%%%%%%%%%%%%%%%%%%%%%%

In contrast with the work of Ghosh and
Lamb (1978, 1979a,b), we assume that the
magnetic diffusivity of the disk
$\eta_t$ is of the order of magnitude of
the disk's effective viscosity (see
Bisnovatyi-Kogan and Ruzmaikin 1976;
Parker 1979; Campbell 1992).
Additionally, we assume  that the disk
viscosity is the turbulent viscosity as
formulated by Shakura (1972) and Shakura
and Sunyaev (1973), $\nu_t = (2/3)
\alpha c_s h$, where $\alpha$ is a
dimensionless quantity less than unity.
In this work, we assume that
$\alpha$ is in the range $10^{-1}$ to
$10^{-2}$.  The possible contribution to
the turbulent momentum flux due to
small-scale magnetic field fluctuations
is assumed included in $\alpha$ (Eardley
and Lightman 1975;  Coroniti 1981;
Balbus and Hawley 1992; Kaisig, Tajima,
and Lovelace 1992;
Rast\"atter, L. \& Schindler, K. 1999).
We write $\eta_t =
D\nu_t$ where $D$ is dimensionless and of
order unity.

With $\eta_t$ the turbulent diffusivity,
equation (\ref{adi2}) gives $|(B_\phi)_h /B_z|$

\noindent
$= (3/2) r |\omega_d
(r) - \omega_*|/(\alpha D c_s)$. This ratio is
a measure of the twist of the ${{\bf B}}$
field between the star and the disk.
For the outer part of the disk
$\omega_d(r)$  is expected to be close
to the Keplerian value
$\omega_K = (GM/r^3)^{1\over 2}$.  Thus,
it would appear that the twist
$|(B_\phi)_h/ B_z|$ can be very much
larger than unity, say, $> 10$.  We
argue here that such large values of the
twist do not occur.

For a discussion of the limitation on
the twist $|(B_\phi)_h /B_z|$ we
consider that the plasma outside of the
disk is force-free, ${{\bf  J}}
\times {{\bf B}} = 0$, which is the
coronal plasma limit of Gold and Hoyle
(1960).  This limit is valid under
conditions where the kinetic energy
density of the plasma is much less than
${{\bf B}}^2/8\pi$.  The general
response of force-free coronal magnetic
field loops to stress (differential
twisting) applied to the loop foot
points (in the solar photosphere) has
been studied intensely by Aly (1984,
1991), Sturrock (1991), and Porter,
Klimchuk, and Sturrock (1992).  The
conclusion of these studies is that a
closed field loop with a small twist
evolves into an open field-line
configuration as the twist is
increased.  The related problem of the
twisting of a force free magnetic field
configuration with foot points at
different radii in a Keplerian disk has
been studied by Newman, Newman and
Lovelace (1992), and Lynden-Bell and
Boily (1994).  The twisting due to the
differential rotation of the disk acts
to increase the total magnetic energy of
the coronal field and this in turn acts
to ``inflate'' the field.

An alternative measure of the field
twist between the star and the disk is
simply the difference in the azimuthal
location,
$\Delta \phi$, of the stellar and disk
foot points of a given flux tube.  We
have $r d\phi/dl_p =  B_\phi
(r,z)/|{{\bf B}}_p(r,z)|$, where $dl_p$
is the length element along a poloidal
field line
$\Psi (r,z) =$ const.  Because $r
B_\phi (r,z)$ depends only on
$\Psi$ in a force-free plasma, we have

\begin{equation}
\label{adi3}
\Delta \phi = [r (B_\phi)_h]_{disk}
\int_{star}^{disk} {{dl_p}\over
{r|{\bf{\nabla}} \Psi|}}~.
\end{equation}

\noindent For small $\Delta \phi$ (say,
$<1$), the poloidal field is close to
that of a dipole, and the integral gives
$\Delta \phi = C_t [(B_\phi)_h /B_z]_d$
with
$C_t=8/15$.  For  increasing $\Delta
\phi$, the value of
$C_t$ increases because of the longer
path length and the smaller value of
$|{\bf{\nabla}} \Psi|$. The present
situation is analogous to a
current-carrying plasma column with a
longitudinal $B$ field.  For the plasma
column there is the well-known
Kruskal-Shafranov stability condition
(see, for example, Bateman 1980) on the
twist
$\Delta
\phi$ of the field (at the column's
surface) over the length of the column.
This condition, $\Delta \phi < 2\pi$, is
required for stability against symmetry
changing kink perturbations.
     A related,
non-axisymmetric instability may
accompany the above mentioned inflation
and opening of the star/disk field in
response to increasing $\Delta \phi$.

LRBK95 argued  that there is a definite
upper limit on the twist $\Delta \phi$
of any field lines connecting the star
and the disk.
    This limit implies a
corresponding limit on
$|(B_\phi)_h /B_z|_d$ in equation (\ref{adi2}).
If $\Delta
\phi$ is larger than this limit, then
the field is assumed to be open as
indicated in Figure 9.

\subsubsection{Equations for Disk}

The flow velocity in the disk is ${\bf v}(r)=-u(r)\hat
r+v_\phi(r)\hat\phi$, where $u$ is the accretion speed.
The  surface
density of the disk is $\sigma$.
The main  equations
are:

\medskip
\noindent 1.~~~Mass conservation:
\begin{equation}
  \dot M=2\pi r\sigma u,~=~{\rm const.}
\label{adi4}
\end{equation}
  We assume that the change of
$\dot M$ with $r$ due to possible MHD
outflows is small.

\noindent 2.~~~Radial force balance:
\begin{equation}
{\sigma v_\phi^2\over r}={GM\sigma\over r^2}-{(B_rB_z)_h\over
2\pi} +{d\over dr}\int\limits_{-h}^h dz\,p~.
\label{adi5}
\end{equation}

\noindent 3.~~~Angular momentum conservation:
\begin{equation}
  {d\over dr}(\dot M F) = -r^2 (B_\phi B_z)_h~, \quad \quad
F \equiv r^2 \omega + {{r^2 \nu_t}\over u} {{d\omega}\over dr}~,
\label{adi6}
\end{equation}
where $\omega = v_\phi/r$.  The term $-r^2 (B_\phi B_z)_h$
represents the outflow of angular momentum from the $\pm z$
surfaces of the disk (that is, a torque on the disk).  We assume
that the angular momentum carried by the matter of the outflow is
small compared with that carried by the field.  If the $B$ field
at $r$ is open as discussed in $\S$ 5.1.3, then the angular
momentum outflow from the disk is carried to infinity. On the
other hand, if the field is closed, then the angular momentum
outflow from (or inflow to) the disk is carried by the coronal $B$
field to (or from) the star.

\noindent 4.~~~Energy conservation:
\begin{equation}
\sigma\nu_t\left(r{d\omega\over dr}\right)^2 +{4\pi\over
{c^2}}\int\limits_{-h}^h dz ~{\eta_t{\bf J}^2} ={{4acT^4}\over
{3 \kappa \sigma}}\equiv 2 \sigma_B T_{eff}^4~.
\label{adi7}
\end{equation}
The first term is the viscous
dissipation, the second is the Ohmic
dissipation, while the third term
is the power per unit area carried
off by radiation (in the $\pm z$ directions)
from the disk which is
assumed optically thick.
Here, $\kappa$ is the
opacity assumed due to electron scattering,
$a$ and $\sigma_B = ac/4$ are the usual
radiation constants,  $T$ is the midplane
temperature of the disk, and $T_{eff}$
is disk's effective surface temperature.

\noindent 5.~~~Conservation of magnetic
flux for a general time-dependent
disk:
$$
{\partial \over \partial t}(rB_z)=
{\partial \over \partial r}\left[rB_zu-{\eta_t
r\over h}(B_r)_h\right]~,
$$
where the generally small
radial diffusion of the $B_z$ field
has been neglected (see LRN).
The first term inside the square brackets represents
the advection of the field while
the second term represents the
diffusive drift.
In a stationary state,
\begin{equation}
\beta_r (r) \equiv {{(B_r)_h}\over
B_z} ={ {uh}\over \eta_t}~.
\label{adi8}
\end{equation}

\noindent 6.~~~Vertical hydrostatic equilibrium:
The $z$-component of the Navier-Stokes
equation (\ref{adi1}) gives the condition
for vertical hydrostatic balance
which can be written as
\begin{equation}
\left({h\over r}\right)^2 + b \left({h\over r}\right)
- \left({c_s\over  v_K}\right)^2 = 0~,
\label{adi9}
\end{equation}
where $c_s$ is the Newtonian sound
speed based on the midplane temperature of the disk,
and $b \equiv r\left\lbrack \left(B_r\right)_h^2 +
\left(B_\phi\right)_h^2\right\rbrack / \left(4\pi\sigma v_K^2\right)$
(Wang, Lovelace, and Sulkanen 1990).
      Radiation pressure is
negligible for the conditions of interest.
     For $b \ll 2c_s/v_K$,
this equation gives the well-known relation $h/r = c_s/v_K $
(Shakura and Sunyaev 1973),  while for $b\gg 2c_s/v_K$ it gives
$h/r = b^{-1}(c_s/v_K)^2$ which is smaller than $c_s/v_K$ owing to
the compressive effect of the magnetic field external to the disk
(Wang et al. 1990) It is useful to write $b = \epsilon~ \beta^2$,
where $\epsilon \equiv r B_z^2(r,0)/(2\pi\sigma v_K^2)$ and
$\beta^2 \equiv [(B_r)_h^2+(B_\phi)_h^2]/(2B_z^2)$. In $\S$ 5.1.6,
we discuss that $\beta = O(1)$ in order to have outflows from the
disk, and $\beta \leq 1$ with no outflows. Thus, the Alfv{\'e}n
speed in the midplane of the disk is $v_A =
v_K\epsilon^{1\over2}(h/r)^{1\over2}$. We may term the magnetic
field as weak if $\epsilon < c_s/v_K \approx h/r$.  In this limit,
$v_A/c_s \approx (\epsilon r/h)^{1\over2} \leq 1$, and the
magnetic compression of the disk is always small. In the opposite,
strong field limit, $\epsilon > c_s/v_K$. If at the same time
$\beta = O(1)$, then the disk is magnetically compressed, and for
$\epsilon \gg c_s/v_K$, $v_A/c_s = 1/\beta$.

Note that with the magnetic field
terms neglected these equations give
the Shakura-Sunyaev (1973) solution for region ``$b$".

\subsubsection{Magnetically Driven Outflows}

Consider the outer region of the disk where the value of the field
twist $\Delta \phi$ given by equation (\ref{adi3}) [or
$|(B_\phi)_h/B_z|$ of equation (\ref{adi2})] is larger than a
critical value. In this region the $B$ field threading the disk is
open as discussed in $\S$ 5.1.3.
  The angular momentum flux
carried by the disk is $\dot M F$.
If the angular rotation of the disk
is approximately Keplerian,
$\omega (r) \approx \omega_K (r)$, then the
viscous transport contribution to $F$ is
$r^2 (\nu/u)(d\omega/dr) \approx -
(3/2) r^2 \omega [h/(\beta_r r)]$.
As discussed in detail below a necessary
condition for MHD outflows is that
$\beta_r = O(1)$.  Consequently, the
viscous contribution to $F$ in
equation (\ref{adi6}) is smaller than the bulk
transport term by
the small factor $h/r << 1$, and $F \approx \omega r^2$.

In equation (\ref{adi6}) we therefore
have $d(\dot M F)/dr \approx \dot M \omega_K
r/2$.  In equation (\ref{adi6}) we let
$(B_\phi)_h = -\beta_\phi B_z(r,0)$.  Studies of
MHD outflows (Blandford and Payne 1982, LBC, LRC)
indicate that $\beta_\phi = {\rm  const}= {\cal O}(1)$.
As a result, equation (\ref{adi6})
implies that
\begin{equation}
[B_z(r,0)]_w = k /r^{5/4},~ {\rm{with}}~ k =
  [\dot M (GM)^{1\over 2} /(2
\beta_\phi)]^{1\over 2}~,
\label{adi10}
\end{equation}
\noindent where the $w$ subscript indicates the wind region of the disk.
Of course at a sufficiently small radius, denoted $r_{wi}\;$, the outflow
ceases, and the dependence of $B_z(r,0)$ reverts to approximately the
stellar dipole field, $\mu/r^3$.  As a consistency condition we must
have
\begin{equation}
{\cal E} = (k /r_{wi}^{5/4}){\Big /}(\mu /r_{wi}^3) =
\bigg [ {{\dot M
(GM)^{1/2}r_{wi}^{7/2}}\over
{\mu^2 (2\beta_\phi)}}\bigg ]^{1\over 2} < 1~.
\label{adi11}
\end{equation}

\noindent The transition between the dipole and outflow field dependences is
handled by letting $B_z (r,0) = (\mu /r^3) g(r) + (k /r^{5/4}) [1 -
g(r)]$, where, for example, $g(r) = \{ 1 + \exp [(r - r_{wi}) / \Delta r ]
\}^{-1}$ and $\Delta r/r_{wi} \ll 1$.

Equations (\ref{adi4}) -- (\ref{adi9}) can be used to obtain the
disk parameters in the region of outflow. As shown below the
viscous dis\-sipation is much larger than the Oh\-mic, and then
Eq. (\ref{adi7}) gives $T = [81 \dot M^2 (GM)^{1/2}\kappa \big
/(128\pi^2 \alpha D \beta_r^2 ac)]^{1/ 4}$ $r^{-{7/ 8}}$.
  For illustrative values, $M = 1M_\odot$, $\dot M = 10^{17}$g/s $\approx
1.6 \times 10^{-9} M_\odot$/yr, $\alpha = 0.1$, $\beta_r = 0.58$, and $D
= 1$, we find at the representative distance $r = 10^8$cm that $T \approx
0.44 \times 10^6K$, $v_K = 1.15 \times 10^9$cm/s, $c_s = 7.6 \times
10^6$cm/s, $u = 0.88 \times 10^6$cm/s, $\sigma = 180$g/cm$^2$, and $h/r =
6.6 \times 10^{-3}$.  The ratio of the ohmic dissipation to that due to
viscosity is: $(4/27) \alpha D \beta_r^3 \ll 1$.  The ratio of the
viscous dissipation to the power output of the outflows (the $\pm\, z$
Poynting fluxes)   (per unit area of the disk) is: $(9/2)(h/r)(\beta_r
D)^{-1}
\ll 1$.  Thus, most of the accretion power for $r > r_{wi}$
goes into the outflows.  However, the fraction of the total accretion
power in outflows is small if $r_{wi} \gg r_*$, where $r_*$ is the
star's radius ($\sim 10^6$ cm for a neutron star).  Note that $\dot M =
10^{17}$g/s and $M = 1 M_\odot$ correspond to a total accretion power or
luminosity $L_0 = GM\dot M/r_* \approx 1.33 \times
10^{37}$(erg/s)($10^6$cm/$r_*$).

Conservation of the poloidal flux threading the disk implies that
there is an outer radius, denoted $r_{wo}$, of the outflow from
the disk.  That is, we have $\int_{r_{wi}}^{r_{wo}} rdr (k
/r^{5/4}) = \int_{r_{wi}}^{r_{wo}} rdr (\mu /r^3)$. So, the outer
radius is $r_{wo}$ $\approx r_{wi} (f {\cal E})^{-{4/3}}$, where
$f \approx 1$ for $1-{\cal E} \ll 1$ and $f =3/4$ for ${\cal E}
\ll 1$. For $r > r_{wo}$, $B_z(r,0)$ rapidly approaches zero.

Equation (\ref{adi9}) gives the disk thickness.
 From equation (\ref{adi10}),
we have $\epsilon = rB_z^2/(2\pi\sigma v_K^2)$
$ = (u/c_s)(c_s/v_K)/(2\beta_\phi)$.
The magnetic compression of the disk
is small if $\epsilon < c_s/v_K$ or
equivalently if $u/c_s < 1$.
Our numerical solutions have $u/c_s < 1$ in the
region of outflow.

\subsubsection{Necessary Condition for MHD Outflows}

A necessary condition on $\beta_r$
for MHD outflows from the disk can be
obtained by considering
the net force ${\cal F}_p$ on a fluid particle in the
direction of its poloidal motion above the disk.
Notice that for stationary flows, the poloidal
flow velocity ${\bf v}_p$
is parallel to ${\bf B}_p$ owing to the assumed axisymmetry and perfect
conductivity.
However, in general ${\bf v} \times {\bf B} \not= 0$ so that
the fluid particles do not move like ``beads on a wire".
Above, but close
to the disk ($h \leq z \ll r$), we assume that the poloidal field lines
are approximately straight.  Thus, the poloidal position of a fluid
particle above the disk is ${\bf r}=(r_o+S\sin(\theta)){\hat {\bf r}} +
S \cos(\theta){\hat {\bf z}}$,
where $S$ is the distance along the path from the starting value $S_o =
h/cos(\theta)$ at a radius $r_o$, $\tan(\theta)=(B_r)_h/B_z=\beta_r$, and
$z=Scos(\theta)$.  The effective potential is
$U(S)=-(1/2)\omega_o^2(r_o+S\sin(\theta))^2-GM/|\bf r|$~(LRN),
and the force in the direction of the particle's poloidal motion
is ${\cal F}_p =-\partial U/\partial S$.  For distances $z$ much less
than the distance to the Alfv{\'e}n point, $\omega_o \approx \omega_K(r_o)$.
In this way we find
\begin{equation}
{\cal F}_p = - (\omega_K^2 - \omega^2)r \biggr ({{B_r} \over
|{\bf B}_p|}\biggr ) +
\omega_K^2 z\; (3 \beta_r^2 - 1)
\biggr ({{B_z} \over |{\bf B}_p|}\biggr )~,
\label{adi12}
\end{equation}
\noindent for $h \leq z \ll r$, where we
assume $B_z > 0$.

The slow
magnetosonic point of the outflow
occurs where ${\cal F}_p = 0$ (LRC) at
the distance
\begin{equation}
z_s = r \biggr ( 1 - {\omega^2 \over \omega_K^2} \biggr )
{\beta_r \over
{3\beta_r^2 -1}}~.
\label{adi13}
\end{equation}

\noindent The factor $1 - (\omega/\omega_K)^2$
can be obtained from the radial force
balance equation (\ref{adi5}).
The radial pressure force is small compared with
the magnetic force for the conditions we consider, $\alpha
\beta_r^2/(3\beta_\phi) > h/r \ll 1$.  Thus,
\begin{equation}
{{z_s} \over h} = \Big({{2 \alpha DB_z^2 r}
\over {3 \dot M \omega_K} }\Big)
{\beta_r^3\over {3 \beta_r^2 - 1}}~.
\label{adi14}
\end{equation}
\noindent For $r_{wi} < r < r_{wo}$,
the quantity in brackets is simply $\alpha
/ (3 \beta_\phi)$ owing to equation (\ref{adi10}).
Note that the minimum of
$\beta_r^3/(3\beta_r^2 - 1)$ is $1/2$ at $\beta_r = 1$.
If the gas
near the surface of the disk ($h \leq z \leq z_s$)
is assumed isothermal
with temperature $T_1 \ll T$ and sound
speed $c_{s1} \equiv (k_B T_1/m)^{1\over2}$,
then the density at the slow magnetosonic
point $\rho_s$ can be obtained
using the MHD form of Bernoulli's
equation.
     At the slow magnetosonic
point, the poloidal flow speed is
$v_p = c_{s1}|{\bf B}_p|/|{\bf B}|
\equiv v_{sm}$, the slow magnetosonic speed,
and $\rho_s = \rho_h
exp\left\{-1/2 -[U(z_s) -U(h)] /c_{s1}^2\right \}$,
where U is now
regarded as a function of $z$.
In turn, the mass flux density from the
$+z$ surface of the disk is
$\rho_s v_{sm}  \cos(\theta) = \rho_s c_{s1}
B_z/|{\bf B}|$.

The consistency of the magnetically
driven outflow solutions requires
$z_s \geq h$.
If this were not the case, the outflows would be matter
rather than field dominated near the disk.
There are two possible
regimes having $z_s \geq h$:
one is with $\alpha D/(3\beta_\phi) < 1$ [or
$\alpha D/(3 \beta_\phi) \ll 1$]
and $\beta_r \geq 1/\sqrt 3$; another is with
$\alpha D/(6 \beta_\phi) > 1$ and $\beta_r > 1$.
In the present work, we
consider the first regime which is
consistent in the following respect:
The magnetic pinching force on the
inner part of the outflow $(r <
r_{wo})$ increases as the mass flux
in the outflow increases (LBC).
In
turn, an increase in the pinching force
acts to decrease $\beta_r =
(B_r)_h/B_z$ thereby increasing $z_s/h$
and decreasing the mass flux in
the outflow.

In addition to the condition $z_s \geq h$, the slow magnetosonic
point must be at a distance $z_s$ {\it{not}} much larger than $h$,
say $2h$, in order for the outflow to have a non-zero mass flux.
At the inner radius of the outflow, $r_{wi}\;$, the $B_z(r,0)$
field in the disk is $\mu/(2 r^3_{wi})$ as discussed in $\S$
5.1.5. We then have
\begin{equation}
r_{wi} \geq \bigg [ {{\alpha D \mu^2}\over
{24 \dot M(GM)^{1\over 2}}}\bigg
]^{2\over 7}
\label{adi15}
\end{equation}
$$
\approx 0.75 \times 10^8{\rm{cm}} \bigg [ \Big( {\alpha D
\over 0.1}\Big)
\Big ( {\mu \over {10^{30} {\rm{Gcm}}^3} } \Big )^2 \Big
({{10^{17}{\rm{g/s}}}\over \dot M}\Big)
\Big( {M_\odot \over M}\Big )^{1\over
2}\bigg ]^{2\over 7}~,
$$
from equation (\ref{adi14}).
Note that from equation (\ref{adi11}), we get
${\cal E}
\geq [\alpha D/(48 \beta_\phi)]^{1\over 2}$
or ${\cal E} \geq 0.046$ and
$r_{wo}
\leq 90 r_{wi}\;$ for $\alpha = 0.1$,
$D = 1$ and $\beta_\phi = 1$.

\subsubsection{MHD Outflows and Spin-Up ($r_{to} < r_{cr}$)}

We have numerically integrated
equations (\ref{adi4}) -- (\ref{adi9}) assuming
magnetically driven outflows
from the disk.
We integrate the equations inward starting from a large radial
distance $r < r_{wo}\;$.
The inner radius of the region of outflow is
determined as discussed above.
For $r < r_{wi}\;$, $d F/dr = 0$,
and the solutions in this range of $r$ all exhibit a ``turn-over
radius", $r_{to}\;$, where $d\omega /dr =0$.
   This ``turn over" results
from the radially outward magnetic
force in equation (\ref{adi5}) which becomes
stronger as $r$ decreases.
    In the region close to the turnover, the
magnetic compression of the disk becomes
strong in that $\epsilon = O(1)$.
 From a least squares fitting of many integrations, we find
\begin{equation}
r_{to} \approx 0.91 \times 10^8{\rm{cm}} \Big ( {\alpha D \over 0.1}\Big
)^{0.3} \Big ( {\mu \over {10^{30}{\rm{G cm}}^3} }\Big)^{0.57} \Big (
{{10^{17}{\rm{g/s}}}\over \dot M}\Big)^{0.3} \Big ( {M_\odot \over
M}\Big)^{0.15}~.
\label{adi16}
\end{equation}
   The turn-over radius is less than but in
all cases close to $r_{wi}$.  Thus,
equation (\ref{adi16}) is compatible with  (\ref{adi15}).
Our radius $r_{to}$ has
a role similar to that of the Alfv\'en
radius $r_A$ of Ghosh and Lamb
(1978, 1979a,b).
    The dependences we find of $r_{to}$ on $\mu$, $\dot
M$, and $M$ are close to those of $r_A$.
   However, our analysis shows an
important dependence on $\alpha D$
which is proportional to the magnetic
diffusivity of the disk.
Furthermore, our $r_{to}$ is smaller
than $r_A$ by a factor of about
$(\alpha D/12)^{0.3}$, which is $\approx
0.24$ for $\alpha D = 0.1$.

The turn-over radius is important in
the respect that the inward angular
momentum flux carried by the disk
$(\dot M F)$ is $\dot M \omega_{to}
r_{to}^2$ because $d\omega/dr = 0$ at $r = r_{to}$.
[Note that we do not
consider here the possibility of
significant poloidal current flow along
the open field lines extending
from the polar caps of the star ($B_\phi
> 0$ for $z > 0$ in figure 9)
which would act to remove angular
momentum from the star to infinity.]
 From numerical integrations, we find to
a good approximation that $\omega_{to}
= (GM/r^3_{to})^{1\over 2}$, so that the
influx of angular momentum to the
star is $\dot M F_{to} = \dot M (GM r_{to})^{1\over 2}$.
Thus, the rate
of increase of the star's angular momentum $J$ is:
\begin{equation}
{{dJ}\over {dt}} = \dot M (GM r_{to})^{1\over 2}~.
\label{adi17}
\end{equation}
\noindent With $I$ the moment of
inertia of the star, $J = I \omega_*$,
and $dJ/dt = (dI/dM)\dot M\omega_* + I (d\omega_*/dt)$.
For the situation
of interest, the term proportional
to $dI/dM$ is negligible because
$r_{to}$ is much larger than the star's radius.
Thus, we have ``spin-up" of the star,
$$
{{d\omega_*} \over {dt}} = {{\dot M (GM r_{to})^{1\over 2}}\over I}~,
$$
\noindent or
\begin{equation}
{1\over P} {dP \over dt} \approx -5.8 \times 10^{-5} {1\over {\rm{yr}}}
\Big( {P\over 1s} \Big)
\Big( { \dot M\over {10^{17}}{\rm{g/s}} }\Big) \Big(
{ {10^{45}{\rm{g cm}}^2}\over I } \Big)
\Big( {M\over M_\odot} \Big)^{1\over
2} \Big( { r_{to} \over {10^8{\rm{cm}}} }
\Big)^{1\over 2}~,
\label{adi18}
\end{equation}
where $P= 2\pi/\omega_*$ is the pulsar period.
Because the total accretion luminosity
$L=GM{\dot M}/r_*$ with $r_*$ the star's
radius, ${\dot P}\propto -P^2 L^{0.85}
\mu^{0.285}$ for constant $\alpha D, r_*,$ and $ M$.

Our outflow solutions are consistent
only in the case where $\omega_{to} =
(GM/r_{to}^3)^{1\over 2}$
$> \omega_*$.
In this case, with $r$ decreasing from
$r_{to}$, the rotation rate of the
disk $\omega(r)$ decreases and
approaches the rotation rate of the
magnetosphere $(\approx \omega_*)$ from
above where it matches onto the
magnetospheric rotation through a
radially thin turbulent boundary layer.
We find no consistent stationary
solutions when $\omega(r)$ decreases below $\omega_*$.
The condition
$\omega_{to} > \omega_*$ is the same as
\begin{equation}
r_{to} < r_{cr} \equiv
\bigg ( {{GM}\over \omega_*^2}\bigg )^{1\over 3} \approx 1.5
\times 10^8{\rm{cm}}\bigg ({M \over M_\odot}\bigg)^{1\over 3}
\bigg ({P \over
{1s}}\bigg)^{2\over 3}~,
\label{adi19}
\end{equation}
where $r_{cr}$ is commonly
referred to as the ``co-rotation
radius".  Note that $r_{cr}$ decreases during spin-up.
Thus,
$r_{cr}$ approaches $r_{to}$ if $r_{to}$
does not decrease rapidly.

\subsubsection{Magnetic Braking of Star by Disk; Spin-Up or Spin-Down
($r_{to} > r_{cr}$)}}

For $r_{to} > r_{cr}$, the $B$ field in the outer part of the disk $(r
\gg r_{cr})$ is open as discussed above, but there are no
consistent stationary  MHD
outflows, $(B_\phi)_h = 0,~\beta_r = h u/\eta_t \ll 1$, and the Ohmic
dissipation is small compared with that due to viscosity.  Thus, the
outer part of the disk obeys essentially the equations of Shakura and
Sunyaev (1973) for a Keplerian disk with $F = {\rm const.} \equiv
F_\infty$ in equation (\ref{adi6}) undetermined.  We show below that $F_\infty$
can be determined self-consistently by considering the region near
$r_{cr}$ where the Keplerian accretion flow is
brought into co-rotation with the star.  Because there are no
outflows, the angular momentum influx to the star is $\dot M F_\infty$
which is equal to $d J_*/dt$.

In the region of the disk where $\omega (r)$ is close to $\omega_*$, that is,
where $r \sim r_{cr}$, $|(B_\phi)_h/B_z|$
given by equation (\ref{adi2}) is
{\it{not}}  much larger than unity.  In this region, closed but
twisted field lines link the star and the disk.  The twist of the field
acts to remove (or add) angular momentum from (to) the disk if $B_\phi
B_z > 0$ (or $< 0$).
The angular momentum removed from the disk is
deposited on the star via the $B$ field.  In this region of the disk the
key equations are (\ref{adi2}) and (\ref{adi6}) which we rewrite as
\begin{equation}
{d\omega \over dr} =
{u \over {r^2\nu_t}} (F - \omega
r^2)~, \quad
{dF \over dr} = {hr\over D\nu_t}(\omega -
\omega_*)\Big( {{r^2 B_z^2}\over \dot
M}\Big) H \Big( {\tau\over\tau_{max}}\Big)~.
\label{adi20}
\end{equation}
Here, $H(x)$ is a Heaviside function
such that $H(x) = 1$ for $|x|
< 1$ and $H(x) = 0$ for $|x| > 1$;
$$
\tau(r) \equiv { {(B_\phi)_h (r)}
\over {B_z(r,0)} } = -{hr \over {D
\nu_t}} (\omega -\omega_*)
$$
is a measure of the field twist;  $\tau_{max} =$ const is
the maximum value of the twist;  and
$$
F(r) \equiv r^2 \omega + {{r^2 \nu_t}\over u}{{d \omega}\over
dr}~.$$
We consider $\tau_{max}$ to be a universal
constant.  For $|\tau| > \tau_{max}$
the field configuration becomes
open.  In equations (\ref{adi18}), $B_z$ is assumed to be the star's
dipole field.
The other equations of \S 4.1.4 are still needed.  Note
that both the viscous and Ohmic heating must be retained in the energy
equation (\ref{adi7}).
  The viscous dissipation is dominant for $r > r_{cr}$,
while the Ohmic dissipation dominates for $r < r_{cr}$.
  Equations (\ref{adi5})
and (\ref{adi8}) can be combined to
give $u$ as a function of $r, T,$ and
$\omega$.
In turn, this expression for $u$ can be combined with the energy
conservation equation (\ref{adi7}) to
give both $u$ and $T$ as functions of $r,
\omega,$ and $\partial
\omega/\partial r$ if viscous heating
dominates, or as functions of $r$ and
$\omega$ if Ohmic heating dominates.
Furthermore, $\nu_t = (2/3)\alpha c_s h$
in equation (\ref{adi20}) can be derived
from $T$ and $h/r$ from equation (\ref{adi9})
using
$\beta_r$ from equation (\ref{adi8}),
$\beta_\phi = - \tau$, and $\epsilon$ from
$B_z$ and $u$.

We have solved the equations (\ref{adi5},
\ref{adi7}-\ref{adi9}, \& \ref{adi20})
by numerical integration
starting from the outside, at a distance where $\tau >
\tau_{max}$ and $F = $const$ = F_\infty$,
and integrating inward through $r_{cr}$.
The fact that the value of $F_\infty$ is not known a priori
points to the use of a ``shooting method"
for its determination.
   Using this approach, we find that there
is in general a unique value of
$F_\infty$, denoted $F_\infty^0$,
such that the solution for $\omega(r)$
smoothly approaches $\omega_*$ for $r$
decreasing below $r_{cr}$.  If
$F_\infty$ is smaller than $F_\infty^0$, then $\omega(r)$ follows the
Keplerian law as $r$ decreases below $r_{cr}$.  On the other hand, if
$F_\infty$ is larger than $F_\infty^0$, then $\omega(r)$ goes through a
maximum and decreases rapidly to values much less than $\omega_*$.  The
only physical solution is that with $F_\infty = F_\infty^0$.
Hereafter, the zero superscript on $F_\infty$ is implicit.

The behavior of equations (\ref{adi20}) can be
understood qualitatively by noting that if $\omega(r)$ were to decrease
linearly through $r_{cr}$, then the radial width of the region of
braking, $dF/dr < 0$ where $\omega < \omega_*$, would equal that the width of
the region of acceleration, $dF/dr > 0$ where $\omega > \omega_*$.
Consequently, $F$ would change by only a small fractional amount.
However, if within the braking region $F(r)$ attains a value equal to
$\omega r^2$, then $d\omega/dr = 0$ at this point and $d\omega /dr$
remains small
and $\omega \approx \omega_*$ for smaller $r$. A rough estimation gives the
radial width of the braking region as $\Delta r \sim \tau_{max} Dc_s/\omega_K$,
where $c_s$ is based on the temperature just outside of this region,
and the jump in $F$ as $\Delta F \sim \tau_{max} \Delta r (r^2B_z^2/\dot M)$.

Figure 11 shows the dependence
of $F_\infty$ on $\mu$ and on $\dot M$ for sequences of cases where
$r_{cr}$ is assumed comparable to $r_{to}$.  This is clearly not a
necessary choice; that is, $r_{to}$ could differ significantly from
$r_{cr}$.  However, with $r_{cr} = r_{to}$, Figure 11
allows a direct comparison of the angular momentum influx to the star
($\dot M F_K$) for the case of outflows $(r_{to} \leq r_{cr})$ with the
influx $(\dot M F_\infty)$ in the case of no outflows and magnetic
braking $(r_{to} > r_{cr})$.
   Figure 11 suggests that a bimodal behavior
can occur where the star switches between spin-up and spin-down, owing
to some extraneous perturbation.

%%%%%%%%%%%%%%%%%%%%%%%%%%%%%%%%%%%%%%%%%%%%%%%%%%%%%%%%%%%%%%%%%%%%%%%%%%%%%%%
%\begin{figure}
%\epsfysize=6cm % fix the y-dimension and scales x-dim. to y-dim.
%\centerline{\epsfbox{GRFIG9B.EPS}} %

%\label{LRBK4}
%\end{figure}
%%%%%%%%%%%%%%%%%%%%%%%%%%%%%%%%%%%%%%%%%%%%%%%%%%%%%%%%%%%%%%%%%%%%%%%%%%%%%%%

\medskip

\subsection{Magnetic Propeller Regime}

\subsubsection{Introduction}

Observations of some
X-ray pulsars show remarkable `jumps' between
states where the pulsar is spin-ning-down
to one where it is spinning-up.
  Examples include the
objects Cen X-3 (Bildsten {\it et al.} 1997) and GX 1+4
(Chakrabarty {\it et al.} 1997; Cui 1997).
  The theoretical problem of disk accretion on
to a rotating magnetized star has been
discussed in many works over a long
period (Pringle \& Rees 1972; Lynden-Bell
\& Pringle 1974; Ghosh \& Lamb 1979a,b;
Wang 1979;
Lipunov 1993;
Lovelace, Romanova, \&
Bisnovatyi-Kogan 1995 (referred to as LRBK95);
Li \& Wickramasinghe 1997).
   However, except for the
work by Li \& Wickramasinghe
(1997), the studies do not specifically
address the ``propeller'' regime (Illarionov
\& Sunyaev 1975) where the
rapid rotation of the star's magnetosphere
acts to expell most of the accreting matter
and where the star spins-down.
    Recent computer simulation studies
of disk accretion on to a rotating star
with an aligned dipole magnetic field
(Hayashi, Shibata, \& Matsumoto 1996; Goodson,
Winglee, \& B\"ohm 1997; Miller \& Stone 1997)
provide evidence of time-dependent outflows
but do not give definite evidence for
a ``propeller'' regime with spin-down of the star.
    Recent computer simulation studies
of disk accretion on to a rotating star
with an aligned dipole magnetic field
(Hayashi, Shibata, \& Matsumoto 1996; Goodson,
Winglee, \& B\"ohm 1997; Miller \& Stone 1997)
provide evidence of time-dependent outflows
but do not give definite evidence for
a ``propeller'' regime with spin-down of the star.

     Lovelace, Romanova,
and Bisnovatyi-Kogan (1999,
hereafter LRBK99) developed
a model for magnetic
``propeller''-driven outflows
which were first considered by
Illarionov and Sunyaev (1975).
    Such outflows
can  cause a rapidly rotating
magnetized star accreting from a
disk to spin-down.
Energy and angular
momentum lost by the star goes
into expelling most of the accreting
disk matter.
     The envisioned geometry is shown
in Figure 9.
    The theory gives
an expression for the effective Alfv\'en
radius $R_A$ (where the inflowing
matter is effectively stopped)
which depends on the mass
accretion rate, the star's mass and
magnetic moment, {\it and} the star's
rotation rate.
  The model points to a mechanism
for `jumps' between spin-down and
spin-up evolution and for the reverse
transition, which are changes between
two possible equilibrium configurations
of the system.
  In for example the transistion from
spin-down to spin-up states
the Alfv\'en radius $R_A$ decreases
from a value larger than
the corotation radius to one which
is smaller.  In this transistion
the ``propeller'' goes from being ``on''
to being ``off.''
  The ratio of the spin-down
to spin-up torque (or the ratio
for the reverse change) in a jump
is shown to be of order unity.

%%%%%%%%%%%%%%%%%%%%%%%%%%%%%%%%%%%%%%%%%%%%%%%%%%%%%%%%%%%%%%%%%%%%%%%%%%%%%%%
%\begin{figure}
%\epsfysize=9cm % fix the y-dimension and scales x-dim. to y-dim.
%\centerline{\epsfbox{lrbk99.EPS}} %

%\label{LRK1}
%\end{figure}
%%%%%%%%%%%%%%%%%%%%%%%%%%%%%%%%%%%%%%%%%%%%%%%%%%%%%%%%%%%%%%%%%%%%%%%%%%%%%%%

\subsubsection{Theory}

    Consider disk accretion on to a
rapidly rotating
star so that the disk-star
  configuration is as
sketched in Figure 12.
   Consider the flow of mass, angular
momentum, and energy into and out
of the annular region indicated
by the box $A'ACC'$ in this figure,
where $A'A$ is at radius $r_1$ and
$CC'$ is at $r_2$.   Notice that
at this point the values of $r_1$
and $r_2$ are {\it unknown}.   They
are determined by the physical
considerations discussed here.

Consider first the outer surface $CC'$ through the disk.  The
influx of mass into the considered region is
\begin{equation}
\dot{M}_{accr}=2\pi r_2\int_{-h}^{h}
dz~[\rho~ (-v_r)]_{2}~>0~,
\end{equation}
where $h$ is the half-thickness of the
disk, and the $2$ subscript
indicates evaluation at $r=r_2$.
We assume that mass accretion
rate for $r \geq r_2$ is approximately
constant equal to $\dot M_{accr}$.
That is, we consider that outflow
from the disk is negligible for $r\geq r_2$.
The influx of angular momentum into
the considered region is
\begin{equation}
\dot{{L}}_{accr} = \dot{M}_{accr}(rv_\phi)_2
-T_2~, \quad \quad
  T_2 = 2\pi r_2^2
\int_{-h}^h dz~ (T_{r\phi }^{vis})_2~,
\label{rmg2}
\end{equation}
where $T^{vis}_{r\phi }$ is the viscous
contribution to the stress tensor
which includes {\it both} the turbulent
hydrodynamic and turbulent magnetic stresses.
The influx of energy into the considered region
is
\begin{equation}
\dot{E}_{accr} = \dot{M}_{accr}
\left(-{ GM \over 2 r_2}+w_2\right)
- \omega_2 T_2 ~,
\label{rmg3}
\end{equation}
where $\omega \equiv v_\phi/r
\approx v_K/r$ with $v_K\equiv
\sqrt{GM/r}$ the Keplerian speed and $M$
the mass of the star, and where $w_2$ is
the enthalpy.  For conditions of interest here the
disk at $r_2$ is geometrically thin
so that $w \sim c_s^2 \ll v_K^2$, where
$c_s$ is the sound speed.

Consider next the fluxes of mass, angular momentum, and energy
across the surface $AA'$ in Figure 12.
  For the physical regime considered,
where $v_K(r_1)/r_1$
$\ll \omega_*$ ($=$ the angular
rotation rate of the star), the
mass accretion across the $AA'$ surface
is assumed to be small
compared with $\dot{M}_{accr}$.
  The reason for this is that any plasma
which crosses the $AA'$ surface will
be `spun-up' to an angular velocity
$\omega_*$ (by the magnetic force)
which is substantially larger than
the Keplerian value, and thus it
will be thrown outwards.
Thus the efflux of angular
momentum across this surface
from the considered region is
$\dot{L}_1= -T_1~.
$
The efflux of energy across this surface is
$\dot{E}_1 = -\omega_*T_1,
$
where $\omega_*$ is the angular rotation rate of the star and the
inner magnetosphere as shown in Figure 12.  For the conditions
considered, the star slows down and loses rotational energy so
that $T_1 >0$.

  We have
${\dot E_1 / \dot{L}_1} = \omega_*$.
Because the interaction of the star with the
accretion flow is by assumption entirely
across the surface $AA'$,
this is consistent with the spin-down of
a star with constant moment of inertia $I$;
that is,
$ {\Delta E_* /\Delta {L}_*}
={I_* \omega_* \Delta \omega_*/
(I_* \Delta \omega_*)} = \omega_*$.

Next we consider the mass, angular momentum, and energy fluxes
across the surfaces $A'C'$ and $AC$ in Figure 12. As mentioned,
accretion on to the star is small for $r_{cr} \ll r_1$ where
$r_{cr}$ is the corotation radius as indicated in Figure 12. Thus,
the mass accretion goes mainly into outflows, $\dot{M}_{out}
\approx \dot{M}_{accr},
$
where ``$out$'' stands for outflows.
   The angular momentum outflow
across the surfaces $A'C'$ and
$AC$,
$\dot{L}_{out}$,
must be the difference between the angular momentum
lost by the star and the incoming angular momentum
of the accretion flow.   The angular momentum
carried by radiation from the disk is negligible
because $(v_K/c)^2 \ll 1$, where $c$ is the speed
of light.
That is,
$
\dot{L}_{out} =\dot{L}_{accr}
-\dot{L}_1.
$
The energy outflow across the $A'C'$ and $AC$ surfaces
is
$\dot  E_{out} +\dot{E}_{rad}=
\dot{E}_{accr}-\dot{E}_1,
$
where $\dot{E}_{rad}$ is the radiation energy loss
rate from the disk surfaces between $r=r_1$ and $r_2$,
and $\dot{E}_{out}$ is the rate of energy loss
carried by the outflows.

  Angular momentum conservation gives
\begin{equation}
\dot{L}_{out} =\dot{M}_{accr} (r v_\phi)_2 -
T_2 +T_1~.
\label{rmg4}
\end{equation}
We have
\begin{equation}
\dot  E_{out} +\dot{E}_{rad}=
-{GM \dot{M}_{accr} \over 2 r_2}
-\omega_2 T_2 +\omega_* T_1~.
\label{rmg5}
\end{equation}
We can solve equation (\ref{rmg4}) for $T_2$
and thereby eliminate this quantity from the energy
equation (\ref{rmg5}).  This gives
\begin{equation}
\dot  E_{out} +\dot{E}_{rad}=
-{3G M \dot{M}_{accr} \over 2 r_2} + \omega_2 \dot
{L}_{out} +(\omega_* -\omega_2)T_1~.
\label{rmg6}
\end{equation}

The preceeding equations are independent
of the nature of the outflows from the disk.
At this point we consider the case of
magnetically driven outflows as treated
by Lovelace, Berk, and Contopoulos (1991,
hereafter LBC).
   In the LBC model the outflows
come predominantly from an annular
inner region of the disk of radius $\sim R_A$
where the disk rotation rate is
$\omega_0 = \sqrt{GM/R_A^3}$.
   Thus we assume that the outflows
come from a region of the disk which
is approximately in Keplerian rotation.
  For the present situation, shown in
Figure 1, it is clear that we must have
$
r_1 <R_A <r_2~.
$
Further, we will assume $R_A$ is close in
value to $r_1$ with $(R_A -r_1)/R_A
{\buildrel < \over \sim 1}$.
For conditions where the outflow from the
disk is relatively low temperature (sound
speed much less than Keplerian speed),
equations (16) and (18) of LBC imply
the general relation
$
\dot{E}_{out} = \omega_0~\dot{L}_{out}-
{3GM\dot M_{out}/( 2 R_A)}.
$
This equation can be used to
eliminate $\dot{L}_{out}$
in favor of $\dot{E}_{out}$ in equation (\ref{rmg6}).
Recalling that $\dot{M}_{out} =\dot{M}_{accr}$
we have
\begin{equation}
(1-\delta^{3\over2})
\dot{E}_{out} +\dot{E}_{rad}=
(\omega_*-\omega_2) T_1
-{3GM\dot{M}_{accr} \over 2 r_2}
(1-\delta^{1\over2})~,
\label{ref7}
\end{equation}
where $\delta \equiv R_A/r_2 <1$.

The energy dissipation in the
region of the disk $r=r_1$ to $r_2$
heats the disk and this heat energy
is transformed into outgoing radiation
$\dot{E}_{rad}$.
  Thus we have
\begin{equation}
\dot{E}_{rad} ={3\over 2}\int_{r_1}^{r_2} dr~
{G M\dot{M}_{disk}(r)
\over r^2}~,
\label{rmg8}
\end{equation}
(Shakura 1973; Shakura \& Sunyaev 1973),
where $\dot{M}_{disk}(r_2) =\dot{M}_{accr}$
and $\dot{M}_{disk}$ $(r_1) \approx 0$.
The essential change in $\dot{M}_{disk}(r)$ occurs
in the vicinity of $R_A$ so that
$
\dot{E}_{rad} \approx (3 G M \dot{M}_{accr}/
2)({R_A^{-1}} - {r_2^{-1}} ).
$
Thus equation (\ref{ref7}) becomes
\begin{equation}
(1-\delta^{3\over2}) \dot{E}_{out}=
\left(\omega_*-\omega_2 \right)T_1
-{3GM\dot{M}_{accr} \over 2 R_A}(1-\delta^{3\over
2})~.
\label{rmg9}
\end{equation}
Thus the power from the spin-down of the star
$\omega_*T_1$ must be larger than a certain
value in order to drive the outflow.

   For magnetically driven outflows, the
value of $\dot{E}_{out}$ can be written as
\begin{equation}
\dot{E}_{out} = {3\over 2} ~{\cal F}_0^{2\over3}~
\omega_0^{4\over 3}~ R_A^{8\over 3} ~B_0^{4\over 3}~
\dot{M}^{1\over 3}_{accr}~,
\label{rmg10}
\end{equation}
(equation 34 of LBC),
where ${\cal F}_0$ is a dimensionless numerical
constant $\buildrel >\over \sim 0.234$, and $B_0$ is
the poloidal magnetic field at the base of the outflow
at $r=R_A$.   We take the simple estimate
$
B_0= {\mu / R_A^3}~,
$
which omits corrections for example for compression of the
star's field by the inflowing plasma.

    Next we consider the torque on the
star $T_1$.
    Because most of the matter
inflowing in the accretion disk at $r=r_2$ is driven off
in outflows at distances $r>r_1$, the  stress
is necessarily due to the magnetic field.
   The magnetic field in the vicinity of $r_1$
has an essential time-dependence owing to the
continual processes of stellar flux leaking
outward into the disk, the resulting field
loops being inflated by the differential
rotation (LRBK95), and the reconnection between
the open disk field and the closed stellar
field loops.
   The time scale of these
processes is
$t_1 ~{\buildrel < \over \sim}~ 2\pi r_1/v_K(r_1)$.
We make
the estimate of the torque
$
T_1 = -2\pi r_1^2
(2 \Delta z )
{\langle B_{r} B_{\phi}\rangle_1 /( 4 \pi)},
$
where $\Delta z$ is the vertical half-thickness of
the region where the magnetic stress is significant,
and where the angular brackets denote a time average
of the field quantites at $r \sim r_1$.
    The magnetic field components, $B_r,~B_\phi$, with
$B_\phi \propto - B_r$ necessarily, must be of
magnitude less than or of the
order of the dipole field
$B_1=\mu/r_1^3$ at $r=r_1$.
   The fact that $B_\phi$ has the opposite sign
to that of $B_r$ is due to the fact that
the $B_\phi$ field arises differential
rotation between the region $r<r_1$, which
rotates at rate $\omega_*$, and the region
$r>r_1$, which rotates at
rate $\omega_K(r_1) <\omega_*$.
Also,  it is reasonable to
assume $\Delta z~ {\buildrel < \over \sim}~ r_1$.
Therefore,
\begin{equation}
T_1\approx  \bar{\alpha}_1 r_1^3 B_1^2 \equiv
\bar{\alpha}~{\mu^2 \over
R_A^3}~,
\label{rmg11}
\end{equation}
where $\bar{\alpha} \leq 1$ (the time average
of $\alpha(t)$) is a dimensionless
constant analogous to the $\alpha-$parameter
of Shakura (1973) and Shakura \& Sunyaev (1973).
(Note that because $r_1~ {\buildrel < \over \sim}~ R_A$,
$\bar{\alpha}$ and
$\bar{\alpha}_1 = (r_1/R_A)^3
\bar{\alpha}$ are of the same order of magnitude.)

Substituting equations (\ref{rmg10}) and (\ref{rmg11}) into
equation (\ref{rmg9}) gives
\begin{equation}
{3\over 2}~(1-\delta^{3\over 2})~{\cal F}^{2\over 3}_o \left({r_A
\over R_A}\right)^{7\over 3} =
{\bar{\alpha}~r_A^{7\over 2} \over R_A^2~ r_{cr}^{3\over2}}
\left[1-\left({r_{cr}\over R_A}\right)^{3\over 2}
\delta^{3\over 2}\right] -{3\over 2}(1-\delta^{3\over 2})~.
\label{rmg12}
\end{equation}
Here, we have introduced  two
characteristic radii -- the corotation radius,
$$
  r_{cr} \equiv \left( {GM \over \omega_*^2}\right)^{1 / 3}
\approx 1.5 \times 10^8 {\rm cm}~M_1^{1/3}
P_1^{2/3}~,
$$
with $P_1 \equiv (2\pi/\omega_*)/1{\rm s}$
the pulsar period and $M_1 \equiv M/M\odot$.
[For a young stellar object, $r_{cr} \approx
1.36\times 10^{12} M_1^{1/3} P_{10d}^{2/3}$,
where $P_{10d}$ is the period in units of
$10$ days.]
   The second is
the {\it nominal} Alfv\'en radius

$$ R_A \equiv \left[{ \mu^2 \over \dot{M}_{accr}
\sqrt{GM}}\right]^{2/ 7} \approx 3.6\times 10^8{\rm cm}~{
\mu_{30}^{4/7} \over \dot{M}_{17}^{2/7} M_1^{1/7}}~, \eqno(330a)
$$ where $\dot{M}_{17}$ $ \equiv \dot{M}_{accr}/(10^{17} {\rm
g/s})$ is the accretion rate
  with $10^{17}{\rm g/s}$
$\approx 1.6 \times 10^{-9} M_\odot/yr$ and $\mu_{30}
\equiv \mu/(10^{30} {\rm Gcm}^2)$ with
the magnetic field at the star's equatorial surface
$\mu/r^3 = 10^{12} G (10^6 {\rm cm}/r)^3$.
  [For a young stellar object $r_A \approx
1.81 \times 10^{12}{\rm cm}
  \mu_{36.5}^{4/7}$ $/(\dot{M}^{2/7}M_1^{1/7})$,
where the normalization corresponds to a stellar
radius of $10^{11}$ cm, a surface magnetic
field of $3\times 10^3$ G, and an accretion
rate of $1.6 \times 10^{-8} M_\odot/{\rm yr}$.]
The corotation radius is the distance
from the star where the centrifugual force
on a particle corotating with the star ($\omega_*^2r$)
balances the gravitational attraction ($GM/r^2$).
The Alfv\'en radius $r_A$ is the
distance from a {\it non-rotating}
star where the free-fall of a quasi-spherical
accretion flow is stopped, which occurs
(approximately) where the kinetic
energy-density
of the flow equals the
energy-density of the star's
dipole field.
  Note that the assumptions leading
to equation (330a) require $R_A > r_{cr}$.

Notice that $R_A$ (or $r_1 \sim R_A$)
is the {\it effective}
Alfv\'en radius for a {\it rotating} star.
It depends on {\it both} $r_A$ and $r_{cr}$
in contrast with the common notion that the
Alfv\'en radius is given by $r_A$
even for a rotating star.
 From equation (\ref{rmg11}), the {\it
spin-down} rate of the star is
$
I(d\omega_* / dt) =
- ~\bar{\alpha}~{\mu^2 /R_A^3}
$,
where $I$ is the moment of inertia of
the star (assumed constant).
Thus the spin-down rate
depends on both $r_A$ and $r_{cr}$.

  Figure 13 shows the dependence of $R_A$ on
$r_A$ and $r_{cr}$ for a sample case.
  For conditions
of a newly formed disk around a young pulsar,
the initial system point would be on the upper
left-hand part of the curve.
  Due to
the pulsar slowing down (assuming $\mu$ and
$\dot{M}$ constant), the system
point would move downward and to the right as indicated
by the arrow.
  In this region of the diagram, $R_A \approx
\sqrt{\bar{\alpha}}~r_A^{7/4}/r_{cr}^{3/4}
\propto \sqrt{\omega}$
(for $\delta \ll 1$), so that the
torque on the star is $T_1 \approx (GM\dot{M})^{3/2}
/(\mu \sqrt{\bar{\alpha}}~\omega^{3/2})$.  Thus
the braking index is $n=-3/2$, where
$n$ is defined by the relation
$\dot{\omega}_* = -{\rm const}~\omega_*^n$.
Numerically,
\begin{equation}
{1\over P}{dP \over dt} =
{1.55 \times 10^{-5}\over {\rm yr }}~
{P_1^{5/2} (M_1~\dot{M}_{17})^{3/2} \over
\sqrt{\bar{\alpha}} ~\mu_{30}~ I_{45} }~,
\label{rmg13}
\end{equation}
where $I_{45}\equiv I/(10^{45} {\rm g~cm}^2)$.
     For $R_A \gg r_{cr}$, the mass accretion rate
to the star is  small
compared with $\dot{M}_{accr}$, but some
accretion may occur due
to `leakage' of relatively
low angular momentum plasma across field
lines near $r_1$ (Arons \& Lea 1976).

As $R_A$ decreases, the spin-down torque
on the star increases.  Over a long
interval, $R_A$ will decrease to a value
larger than $r_{cr}$ but not much larger.
In this limit, mass
accretion on to the star $\dot{M}_1$ may
  become significant.
     Our treatment
can be extended to this limit
by noting that $\dot{M}_1 =\dot{M}_2
-\dot{M}_{out}$, $\dot{L}_1=\dot{M}_1\omega_1r_1^2
-T_1$, and $\dot{E}_1=\dot{M}_1(-GM/2r_1)-\omega_*T_1$,
where $T_1$ is given by equation (\ref{rmg11}).
In this limit the accretion luminosity is
$GM\dot{M}_1/r_*$, where $r_*$
is the star's radius.   Figure 13 is not
changed appreciably for
$\dot{M}_1 < \dot{M}_{accr}$.

  Further spin-down of the star
will cause the system point in Figure 13
to approach the right-most part of the
curve.
  Further spin-down of the
star is impossible.
  At this point of the evolution,
the only possibility is a transition
to the spin-up regime.
  In this regime the effective
Alfv\'en radius is the `turnover radius'
$r_{to}$ of the disk
rotation curve calculated by
LRBK99, the star spins-up at
the rate $I d \omega_*/dt = \dot{M}_{accr}
(GMr_{to})^{1/2}$, and most of the disk
accretion $\dot{M}_{accr}$ falls onto the
star.
  The dashed
horizontal line in Figure 13 indicates $r_{to}$
which is necessarily less than $r_{cr}$.

  The location of the turnover line $r_{to}$ in
Figure 13 suggests the possible evolution shown
by the sequence of points $a \rightarrow
b \rightarrow c\rightarrow d \rightarrow a$.
    The system can jump
down
from point $a$ where the star spins-down
to point $b$ where it spins-up.
  The spin-down torque at $a$ is
$-\bar{\alpha}\mu^2/R_{Aa}^3$
(where $R_{Aa}$ is the effective Alfv\'en
radius at point $a$), whereas
the spin-up torque at $b$ is
$\dot{M}_{accr} (GMr_{to})^{1/2}$.
  The magnitude
of the ratio of these torques is
\begin{equation}
{{\rm spin\!-\!down} \over {\rm spin\!-\!up}} =
\bar{\alpha}\left({r_A \over R_{Aa}}\right)^{7/2}
\left({R_{Aa}\over r_{to}}\right)^{1/2}~.
\label{rmg14}
\end{equation}
With the system on the $r_{to}$ line, it
evolves to the left.
   Because $r_{to} \leq r_{cr}$,
there must be
an upward jump from point $c$ to point $d$.
    For this case the torque ratio is given
by equation (\ref{rmg14}) with $R_{Aa} \rightarrow R_{Ad}$.
   From point $d$ the system evolves to the right.
For the example shown in Figure 2, the torque
ratio is $\approx 3.45$ for $a \rightarrow b$
whereas it is $\approx 0.87$ for $c \rightarrow d$.
    The vertical line $c\rightarrow d$ is at
the left-most position allowed
for the considered conditions, but
this transistion could also occur if the
line is shift to the right.
    The line $a \rightarrow b$  can be displaced
slightly to the right or it can be displaced
to the left to be coincident with the
$c \rightarrow d$ line.
    In the latter case the torque ratio for the
spin-down to spin-up jump is
approximately equal to the torque ratio for
the spin-up to spin-down jump and is $\approx 0.87$.
   The smaller the horizontal separation of the
$a \rightarrow b$ and the $c \rightarrow d$
lines, the shorter is the time interval
between jumps.

   Summarizing, we can say that the
horizontal locations of the transistions, $a \rightarrow b$,
and $c \rightarrow d$,
are indeterminate within a definite range.
   The locations of the jumps in the $(R_A,~r_{cr})$
plane may in fact be a
stochastic or chaotic in nature and give
  rise to chaotic hysteresis in the
of the spin-down/spin-up
behavior of the pulsar.
The jumps could be triggered by small variations in
the accretion flow ($\dot{M}$ for example)
and magnetic field configuration (the time-dependence
of $\alpha$ in the torque $T_1$).
   Analysis of  the accreting
neutron star system Her X-1
(Voges, Atmanspacher, \& Scheingraber 1987;
Morfill {\it et al.} 1989)
suggests that the intensity variations
are described by a low dimensional deterministic
chaotic model.
    The transistions between
spin-down and spin-up and the reverse transistions
may be described by an analogous model.

  The allowed values in Figure 13 have
$r_{cr}/r_A \leq {\rm const} \equiv k$,
where
$k \approx \bar{\alpha}^{2/3}$.
This
corresponds to pulsar periods
\begin{equation}
P \leq 5.2{\rm s}
\left({\bar{\alpha}\over M_1^{5/7}}\right)
\left({\mu_{30}^2 \over \dot M_{17}}\right)^{3/7}~.
\label{rmg15}
\end{equation}
For some long period pulsars such as GX 1+4
this inequality  points to  magnetic
moment values $\mu_{30}$ appreciably larger
than unity.
   Periods much longer than allowed by
(\ref{rmg15}) can result for pulsars which accrete
from a stellar wind (Bisnovatyi-Kogan 1991).
   [For a young stellar object, equation (\ref{rmg15})
gives
$P \leq 8{\rm d}(\bar{\alpha}/M_1^{5/7})
(\mu_{36.5}^2/\dot{M}_{18})^{3/7}$.]

%%%%%%%%%%%%%%%%%%%%%%%%%%%%%%%%%%%%%%%%%%%%%%%%%%%%%%%%%%%%%%%%%%%%%%%%%%%%%%%
%\begin{figure}
%\epsfysize=7cm % fix the y-dimension and scales x-dim. to y-dim.
%\centerline{\epsfbox{GRFIG10B.EPS}}

%\label{LRK2}
%\end{figure}
%%%%%%%%%%%%%%%%%%%%%%%%%%%%%%%%%%%%%%%%%%%%%%%%%%%%%%%%%%%%%%%%%%%%%%%%%%%%%%%

\subsection{ Cyclotron emission from magnetic
poles in X-ray pulsars}

Accretion flow into a strongly magnetized neutron star is canalyzed by the
magnetic field and is falling into magnetic poles. Observational spectra
of X-ray pulsars are formed by combination of black body and comptonization in
the hot coronal layer. In presence of a strong magnetic fields
$\sim 10^{10}\,\,-\,\, 10^{13}$ Gs the cyclotron or synchrotron emission
of electrons also falls into an X-ray band.
There are a lot of observations of the spectral feature, interpreted as a
cyclotron line in the X-ray spectrum of the pulsar Her X-1
%\cite{a,b,c,d,e,f,o,p}
(Tr\" umper et al., 1978;
Tueller et al., 1984;
Ubertini et al., 1980;
Voges et al., 1982;
Gruber et al., 1980;
McCray at al., 1982;
Mihara et al., 1990;
Sheepmaker et al. 1981)
at 39-58 KeV (see \tablename \,\,\ref{tcyc1}).
Magnetic field intensity was usually calculated
from the non-relativistic formula

\begin{equation}
\label{cyc1}
H=\frac{m c \omega}{e},
\end{equation}
where $\omega$ is the cyclic frequency of the photons, $m$ is the mass of
the electron. From observations of the line frequency in the Table
the magnetic field strengths are
of the order of $(3-5)\cdot 10^{12}$~Gs.
But as large as this value comes into conflict with some theoretical
reasonings like interpretation of the observations of pulsar spin
acceleration
(Bisnovatyi-Kogan and Komberg, 1973),
condition for the transparency for the outgoing of the
directed radiation
(Bisnovatyi-Kogan, 1973, 1974),
consideration of the interrelation between radio and X-ray pulsars
(Bisnovatyi-Kogan and Komberg, 1974),
interpretation of beam variability during the 35-day cycle
(Sheffer et.al., 1992; Deeter et al., 1998; Scott et al., 2000),
see also Lipunov (1987).
It seems likely that the reason of this conflict is an unsuitability of the
non-relativistic formula in this case. According to
%\cite{k}
Bisnovatyi-Kogan and Fridman (1969), the temperature
of the electrons emitting a cyclotron line could be $\sim 10^{11}K$,
and therefore they are ultrarelativistic.
By this means the mean energy of the cyclotron line is broadened and shifted
relativistically by a factor of $\gamma \simeq \frac{kT}{mc^2}$. In the
paper of
%\cite{bbk}
Baushev and Bisnovatyi-Kogan (1999)
the spectral profile of the cyclotron line had been calculated for
various electron distributions. Furthermore, the model of the hot spot
 on the pulsar magnetic pole was
considered and it was shown that the overall observed
X-ray spectrum (from 0.2 to 120 KeV) could arise
in presence of the surface magnetic field
($\sim 5\cdot 10^{10}$~Gs) which is well below
 then those, obtaind from (\ref{cyc1}).

\subsubsection{The spectrum of the cyclotron radiation of the anisotropic
relativistic electrons}

According to
%\cite{l,i}
Bisnovatyi-Kogan (1973),
Gnedin and Sunyaev (1973),
in the magnetic field near the pulsar the
component of a momentum perpendicular to the magnetic field emits rapidly,
while the parallel component of a velocity remains constant.
Hence the momentum distribution of the electrons is
anisotropic

\begin{equation}
\label{cyc2}
p_\perp^2\ll p_\parallel^2,
\end{equation}
 where $p_\perp\ll mc$, $p_\parallel\gg mc$.
Assume for simplicity that the transverse electron
distribution is two-dimensional Maxwellian

\begin{equation}
\label{cyc3}
  dn=\frac{N}{T_1}{\exp \left( -\frac{m u^{2}}{2T_1} \right)}\,
  d\frac{m u^{2}}{2},
\end{equation}
where  $T_1 \ll mc^2$.
Let us calculate the cyclotron emission of N such particles that move
at a rate  $V$ along the magnetic field.

For a single particle, having the transverse velocity $u$, we find
%\cite{m}:
(Ginzburg, 1975):

\begin{equation}
\label{cyc4}
j(\theta)=\frac{e^4 H^2 u^2 (1-\frac{V^2}{c^2})^2
[(1+\cos\theta)(1+\frac{V^2}{c^2})-4\frac{V}{c} \cos\theta]}
{8 \pi c^5 m^2 (1-\frac{V}{c} \cos\theta)^5},
\end{equation}
where $\theta$ is an observational angle in a laboratory frame of reference.
Integrating over the distribution (\ref{cyc3}), we obtain for $N$ particles:

\begin{equation}
\label{cyc5}
J(\theta)=\int j(\theta) dn=N \frac{e^4 H^2 T_1 (1-\frac{V^2}{c^2})^2
[(1+\cos\theta)(1+\frac{V^2}{c^2})-4\frac{V}{c} \cos\theta]}
{4 \pi c^5 m^3 (1-\frac{V}{c} \cos\theta)^5}.
\end{equation}
For the spectrum we find:

\begin{equation}
\label{cyc6}
\omega (\theta)=\omega_H \frac{\sqrt {1-\frac{V^2}{c^2}}}
{1-\frac{V}{c}\cos\theta},\quad
\omega_H=\frac{eH}{m c}.
\end{equation}
When $V\simeq c$ the cyclotron radiation is highly directed and diagram
has a pencil beam along $V$ ($\theta \simeq 0$). Under these conditions
 $(\theta=0,\,\,\, V\simeq c)$ we obtain from (\ref{cyc5}),(\ref{cyc6}):

\begin{equation}
\label{cyc7}
J(0)=\frac{2 N e^4 H^2 T_1}{\pi c^5 m^3 (1-\frac{V}{c})},
\end{equation}
\begin{equation}
\label{cyc8}
\omega(0)=\omega_H \sqrt \frac{1+\frac{V}{c}}{1-\frac{V}{c}}
\approx 2 \omega_H\frac{E_\parallel}{m_e c^2},
\end{equation}
what gives
\begin{equation}
\label{cyc9}
1-\frac{V}{c}=\frac{2 \omega_H^2}{\omega^2}.
\end{equation}
Let us consider the parallel momentum distribution of the electrons as:

\begin{equation}
\label{cyc10}
dn=f(p_\parallel)\,dp_\parallel.
\end{equation}
Substituting of $dn$ for $N$ and using

\begin{equation}
\label{cyc11}
p_\parallel=\frac{mc}{2}\frac{\omega}{\omega_H}\, ;\quad
 1-\frac{V}{c}=\frac{2 \omega_H^2}{\omega^2},
\end{equation}
we obtain for the spectral density:

\begin{equation}
\label{cyc12}
J_\omega=\frac{e^2 T}{2 \pi c^2 \omega_H} \omega^2
 f \left( -\frac{m c }{2}\frac{\omega}{\omega_H}\right )\,d\omega.
\end{equation}
Let us consider two important cases. When $f$ is a relativistic Maxwell:

\begin{equation}
\label{cyc13}
f=\frac{n_0 c}{T_2} \exp\left( -\frac{p_\parallel c}{T_2}\right )\,  , \quad
T_2\gg m c^2\gg T_1,
\end{equation}
where $n_0$ is a number of emitting electrons. Then the spectrum is:

\begin{equation}
\label{cyc14}
J_\omega=\frac{n_0 e^2}{2 \pi c \omega_H}\frac{T}{T_2} \omega^2
 \exp\left( -\frac{m c^2 \omega}{2 \omega_H T_2}\right )
\,d\omega.
\end{equation}
This spectrum has a single maximum at

\begin{equation}
\label{cyc15}
\frac{\omega}{\omega_H}=\frac{4 T_2}{m c^2}.
\end{equation}
In the second case consider the function $f$ as

\begin{equation}
\label{cyc16}
f=\frac{n_0 }{\sqrt {\pi} \sigma}
\exp\left( -\frac{{(p_\parallel-a)}^2}{\sigma^2}\right ).
\end{equation}
The spectrum of radiation is

\begin{equation}
\label{cyc17}
J_\omega=\frac{n_0 e^2}{2 \pi c^2 \omega_H} \omega^2
 \exp\left( -\frac{(\frac{m c}{2}\frac{\omega}{\omega_H }-a)^2}
 {\sigma^2}\right )
\,d\omega.
\end{equation}
When $\sigma \ll a$ this spectrum has a single maximum at

\begin{equation}
\label{cyc18}
\omega \simeq \frac{2 a}{m c} \omega_H.
\end{equation}
Notice that in all cases the maximum is shifted to

\begin{equation}
\label{cyc19}
\frac{\omega}{\omega_H} \sim \frac{\bar E_e}{m c^2}.
\end{equation}
It is a common property of relativistic cyclotron line, that is independent
of the particular form of $f$.
We had approximated experimental spectrum taken from
%\cite{o}
Mihara et al. (1990) (solid line)
by (\ref{cyc14}),(\ref{cyc17}), representing the second variant
with better coinsidence by the
dot line in Figure 14. The spectrum from
%\cite{p}
McCray et al. (1982) was approximated
by (\ref{cyc17}) only (see Figure 15).

%\begin{figure}
%\epsfysize=6cm % fix the y-dimension and scales x-dim. to y-dim.
%\hspace{3.5cm}\epsfbox{fcyc1.eps} %for centering: act on hspace argument

%\label{fcyc1}
%\end{figure}

%\begin{figure}
%\epsfysize=6cm % fix the y-dimension and scales x-dim. to y-dim.
%\hspace{3.5cm}\epsfbox{fcyc2.eps} %for centering: act on hspace argument

%\label{fcyc2}
%\end{figure}
Setting in accordance with
%\cite{k})
Bisnovatyi-Kogan and Fridman (1969) the
longitude electron temperature as $\sim 2 \cdot 10^{11}$~K, that is $T_2=
2\cdot 10^{11}$~K and $a=7 \cdot 10^{-4}$~${\rm\frac{eV \cdot s }{cm}}$, we
obtain for the magnetic field strength $B=4\cdot 10^{10}$~Gs, $8\cdot
10^{10}$~Gs, and $4\cdot 10^{10}$~Gs respectively.  Here we estimate the
spectral form of the cyclotron line averaged over the pulsar period,
supposing
a uniform distribution of $f(p\parallel)$ over the polar cap.
In this model
the beam of the cyclotron feature is determined by the number distribution of
the emitting relativistic electrons, moving predominantly along the magnetic
field, over the polar cap.

\subsubsection{Model of the X-ray spectrum of Her X-1.}

In order to obtain the whole experimental spectrum of the Her X-1
the following model of the hot spot is considered (see Figure 16).
A collisionless shock wave is generated in the accretion flow
near the surface on the magnetic pole of a neutron star. In it`s front the
ultrarelativistic electrons are generated having
small pitch-angle values, so the condition (\ref{cyc2}) is fullfilled
automatically. Behind the shock
there is a hot turbulent zone with a temperature $T_e$,
and optical depth $\tau_e$, and under this zone a heated spot on the surface
of the neutron star with a smaller temperature is situated.

The whole X-ray spectrum of pulsar Her X-1 taken from
%\cite{p}.
McCray et al. (1982) is represented
in Figure 15 by the solid line, and in
Figure 16
together with the schematic
picture of the emition region of Her X-1.

%\begin{figure}
%\epsfysize=8cm % fix the y-dimension and scales x-dim. to y-dim.
%\hspace{3.5cm}\epsfbox{fcyc3.eps} %for centering: act on hspace argument

%\label{fcyc3}
%\end{figure}

There are three main regions in the spectrum:
a quasi-Plankian spectrum between 0,3 and 0,6 KeV, that is generated
near the magnetosphere of the X-ray pulsar; power-law spectrum $(0.6\div
20)$~KeV with a rapid decrease at 20 KeV, and the cyclotron feature.

The power-law spectrum area appears as follows.
A surface emits the black-body spectrum with
a temperature $T_s$. Travelling through the turbulent zone this radiation is
comptonized. This comptonized spectrum has been calculated according to
%\cite{s}.
Sunyaev and Titarchuk (1980).
Setting the neutron star radius equal to 10 km, distance from the
X-ray pulsar 6 Kps, hot spot area $S=2 \cdot 10^{12}$~cm$^2$,
we have found the best
approximation conditions at $T_s=1$~KeV, $T_e=8$~KeV, $\tau_e=14$.
The best approximation of the X-ray spectrum of the pulsar Her X-1 is
represented in Figure 15
by the dashed line. It agrees nicely with the experimental
curve.

 Observation of the variability of the cyclotron line
is reported in
%\cite{n}.
Mihara et al. (1997).
Ginga detected the changes of the cyclotron energies
from 4 pulsars. The change is as much as 40 \% in the case of 4U 0115+63.
Larger luminosity of the source corresponds to smaller average energy of the
cyclotron feature. These changes might be easily explained in our model.
The velocity of the accretion flow decreases with increasing of the
pulsar`s
 luminosity   because localy the luminosity is close to the Eddington limit.
As a result the shock wave intensity drops as well as the energy of the
ultrarelativistic electrons in it`s front. The cyclotron energy decreases in
accordance with (\ref{cyc19}).

Observations on Beppo-Sax and RXTE satellites
(Santangelo et al., 1999; Heindl et al., 1999)
%\cite{bep,rx}
revealed
4 cyclotron-like spectral features (harmonics)
in the spectum of the X-ray pulsar
4U 0115+63. Interpretation of these features
is still not completed.

\begin{table}
\caption{Observations of the cyclotron line in Her X-1}
\medskip
%\begin{tabular}{|l|c|c|c|}
\begin{tabular}{|l|c|c|c|}
\hline
%\hline

Date&References&$\omega_{\rm max}$ (KeV)&Line width (KeV)\\
\hline

May 1978&Tr\"umper et al. (1978)&58&$11^{+26}_{-11}$\\
%\hline

September 1977&Voges et al. (1982)&51&$21^{+9}_{-7}$\\
%\hline

February 1978&Gruber et al. (1980)&48&$28\pm 7$\\
%\hline

April 1980&Gruber et al. (1980)&54&$11^{+14}_{-11}$\\
%\hline

May 1980&Ubertini et al. (1980)&49.5&$18^{+6}_{-3}$\\
%\hline

September 1980&Tuller et al. (1984)&39&$27^{+21}_{-20}$\\
\hline
\end{tabular}
\label{tcyc1}
\end{table}

\section{Magnetohydrodynamic\,  Origin\, of\, Jets\,
from\,\,\qquad\qquad\qquad Accretion\,\, Disks}

\subsection{Introduction}

    Powerful,
highly-collimated, oppositely directed jets are observed in
ac\-ti\-ve ga\-la\-xi\-es and quasars (see for example Bridle \&
Eilek 1984), and in the ``microquasars'' -- old compact stars in
binaries (Mirabel \& Rodriguez 1994; Eiken\-berry {\it et al.}
1998).
   Further, highly collimated
emission line jets are
seen in young stellar
objects
(Mundt 1985;  B\"uhrke, Mundt,
\& Ray 1988).
    Different
ideas and models have been put forward to explain astrophysical
jets (see reviews by Begelman, Blandford, \& Rees 1984;
Bisnovatyi-Kogan 1993b;
   Lovelace et al. 1999b).
 Recent observational and
theoretical work favors
models where twisting of an
ordered magnetic field
threading an accretion
disk acts to magnetically
accelerate the jets as first
proposed by Lovelace (1976)
and Blandford (1976).
   The nature of the ordered magnetic
field threading an accretion disk
envisioned by Bisnovatyi-Kogan and
Ruzmaikin (1976) is shown in Figure 17.
   Figure 18 shows the outflows from
a disk with such an ordered field
from Lovelace (1976).
    Two main regimes have
been considered in theoretical
models, the
{\it hydromagnetic regime} where the
energy and angular
momentum is carried by both
the electromagnetic field and
the kinetic flux of matter, and
the {\it Poynting flux regime}
where the energy and angular
momentum outflow from the
disk is carried predominantly by the
electromagnetic field.

%\begin{figure}
%\epsfysize=6cm % fix the y-dimension and scales x-dim. to y-dim.
%\hspace{3.5cm}\epsfbox{bkr.eps} %for centering: act on hspace argument

%\label{fjet1}
%\end{figure}

%\begin{figure}
%\epsfysize=6cm % fix the y-dimension and scales x-dim. to y-dim.
%\hspace{3.5cm}\epsfbox{rvel.eps} %for centering: act on hspace argument

%\label{fjet2}
%\end{figure}

     The theory of the origin of
hydromagnetic outflows has been
developed by  many authors
 (Blandford \& Payne 1982;
Pudritz \& Norman 1986; Sakurai 1987;
Koupelis \& Van Horn 1989; Lovelace, Berk \&
 Contopoulos 1991;  Pelletier \& Pudritz 1992;
K\"onigl \& Ruden 1993;  Lovelace,
Romanova, \& Contopoulos 1993;
Cao \& Spruit 1994;
   Contopoulos \& Lovelace 1994;
Contopoulos 1995;
Ostriker 1997).
   Physical understanding of the
hydromagnetic
outflows from disks has developed
from MHD simulations (Uchida \& Shibta 1985;
Shibata \& Uchida 1986; Bell 1994;
Ustyugova {\it et al.} 1995, 1999;
Koldoba {\it et al.} 1995;
Romanova {\it et al.} 1997, 1998;
Meier {\it et al.} 1997;
Ouyed \& Pudritz 1997;
    Krasnopolsky, Li, \& Blandford 1999;
Koide {\it et al.} 2000).

   Recent work on
hydromagnetic outflows
assumes
the geometry sketched in
Figure 19.
   A strong case for
hydromagnetic
jets as an explanation of
jets in protostellar
systems emerges because the
temperature of the inner
regions of these systems is
insufficient to permit
driving by thermal or
radiation pressure
(K\"onigl \& Ruden 1993).
   Part of the
investigations have
been analytical or
semi-analytical and
outgrowths of the
self-similar solution of
Blandford \& Payne (1982)
(Pudritz \&
Norman 1986; K\"onigl 1989;
Pelletier \& Pudritz 1992;
Contopoulos \& Lovelace 1994).
    The outflows in this
model are often referred to
as ``centrifugally driven''
owing to the driving force
close to the disk:
   Close to the disk, MHD fluid particles move
along the magnetic field lines (sometimes
described as ``beads on a wire'').
   If the poloidal magnetic field lines
threading the disk diverge away from the $z-$axis
as shown in Figure 17
(making an angle with the $z$-axis of more than
$30^\circ$), then the sum of the gravitational and
centrifugal forces is outward, away from the disk.
              The self-similar models are
unsatisfactory in the respect
that they must be cutoff at
small cylindrical radii,
$r \leq r_{min}$.
  This is the most important region of the
jet flow.
   Observations of optical stellar
jets (Mundt 1985) reveal jet velocities
$\sim 200 - 400 $ km/s, which are
comparable to the Keplerian disk velocity
close to the star's surface.
    This suggests that the jets originate from
the inner region of the disk close to the
star (Shu et al. 1988;
Pringle 1989).

%\begin{figure}
%\epsfysize=6cm % fix the y-dimension and scales x-dim. to y-dim.
%\hspace{3.5cm}\epsfbox{gbkl1.eps} %for centering: act on hspace argument

%\label{fjet3}
%\end{figure}

    The limitations of analytical
models
has motivated
efforts to study jet formation using
MHD simulations.
    Simulation
studies of hydromagnetic jet formation
have addressed two main regions:
  The  jet formation region where the
matter enters with  sub
slow-magnetosonic speed and exits
with super fast-magnetosonic speed.
   The second region includes the disk and the
problem of the Velikhov (1959) - Chandrasekhar (1981)
instability (
Balbus-Hawley 1998)
and the resulting 3D MHD
turbulence.
   A number of studies have addressed
the coupled
problem of the disk and near jet
regions (Uchida \& Shibata 1985;
Shibata \& Uchida 1986;
Stone \& Norman 1994;
Bell and Lucek 1995;).
    MHD simulations of the
near jet region have been
carried out by several
groups (Bell 1994; Ustyugova et
al. 1995, 1999;
Koldoba et al. 1995; Romanova et al. 1997, 1998;
Meier et al. 1997; Ouyed \& Pudritz 1997).

   The powerful jets observed from active
galaxies and quasars
are probably {\it not} hydromagnetic
outflows but rather Poynting flux
dominated jets.
   The  motion of these
jets measured by very long
baseline interferometry correspond
to bulk Lorentz factors of $\Gamma={\cal O}(10)$
which is much larger than the Keplerian
disk velocity predicted for hydromagnetic
outflows.
   Furthermore, the low Faraday rotation
measures observed
for these jets at distances $<$kpc
from the central object
implies a very low plasma densities.
   Similar arguments indicate that
the jets of microquasars are {\it not}
hydromagnetic outflows but rather
Poynting jets.
   Poynting jets have been proposed
to be the driving mechanism for gamma
ray burst sources (Katz 1997)
    Theoretical studies
have developed
models for Poynting jets
from accretion disks
(Lovelace, Wang, \& Sulkanen 1987;
Colgate \& Li 1998;
   Lynden-Bell 1996;
Romanova \& Lovelace 1997;
Levinson 1998; Romanova 1999;
Lovelace {\it et al.} 2001;
 and
       Li {\it et al.} 2001).
Stationary
Poynting flux dominated
outflows were found by
Romanova {\it et al.} (1998) and
Ustyugova {\it et al.} (2000)
in axisymmetric MHD simulations of
the opening of  magnetic loops
threading a Keplerian disk.

   Here, we first review recent results
on  MHD simulations of hydromagnetic
jet formation and later discuss the
Poynting flux regime.
  Section 4.3
discusses general considerations of
hydromagnetic outflows.
   Sections 4.4
discusses MHD simulations  which
give non-stationary and stationary
hydromagnetic outflows.
     Sections 4.5 describes the
Poynting flux regime which
remains to be fully explored by
simulations.
   Section 4.6
gives the conclusions.
\medskip

\subsection{Basics of MHD Outflows
from Disks}

   The main forces which
drive a hydromagnetic outflow from a
disk threaded by a magnetic field
are the centrifugal
force and the magnetic
pressure gradient force.
  If disk has
a hot corona, the pressure gradient may
also be important.
   We neglect the radiative force
but in this regard see Phinney (1987).
   Accreting matter
of the disk carries magnetic field
inward thus generating a $B_r$
component of the magnetic field as
sketched in Figure 19.
   On the
other hand, rotation of
the disk acts to generate a
toroidal component of the field
$B_{\phi}$ ($<0$ if $B_z >0$).

    For a sufficiently inclined
magnetic field
($\theta$ in Figure 19 sufficiently large),
outflows can result from the centrifugal
force (Blandford \& Payne 1982) and/or
the magnetic pressure gradient force
($-{\nabla_z}~B^2_{\phi}/(8\pi)$)
(Lovelace et al. 1989, 1991;
Koupelis \& Van Horn 1989).
 This depends on
the ratio of energy densities at the
base of the outflow at the inner
radius of the disk denoted $r_i$.
Thus, the main parameters are
$
\varepsilon_{th} = (c_s/v_K)_i^2$ and
$\varepsilon_B=(v_{A}/v_K)_i^2$,
where $v_K$ is Keplerian velocity,
$c_s$ the sound speed, $v_A$
the Alfv{\'e}n speed, and the $i$
subscript indicates evaluation on
the surface of the disk at its
inner radius, $r=r_i$.
   For $\varepsilon_B \sim 1$ the
outflow is magneto-centrifugally driven, whereas for
$\varepsilon_{th}\sim 1$, the flow is thermally driven.

   Processes in
the disk are of course coupled to the outflows
(Lovelace et al. 1994, 1997;
Falcke, Malkan, \& Biermann 1995).
    However, it is difficult to simultaneously
simulate the
disk and the outflow regions
because the
time scales of the accretion
and outflow are in general very different.
    The radial accretion speed of the disk is
typically $|v_r| \ll c_s \ll v_K$, where
$c_s$ is the sound speed in the disk, and
$v_K$ is the Keplerian velocity.
    On the other hand the velocity of
magnetic outflows from disks is
of the order of $v_K$.   Clearly, it
is ``easier'' (less computation time
is required) to simulate the outflows.

 Furthermore, processes
in the disk may involve the small scale
MHD instability of Chandrasekhar-Velikhov
instablity emphasized by Balbus, and Hawley,
and therefore
require high spatial resolution.
 Stone \& Norman (1994) attempted
to simultaneously simulate the internal
MHD dynamics of a disk and MHD
dynamics of outflows.
   This proved impractical because
essentially all of the spatial
resolution was needed for treating the
unstable dynamics of the disk.
  Also, there was the problem that
the initial configuration was far
from equilibium.
   Simulation of the internal MHD
disk dynamics has led several groups
to the problem of simulating 3D MHD
turbulence in a sheared flow of a
patch of a disk
(for example, Hawley et al. 1995;
Brandenburg et al. 1995).
   This is a much larger project than that
of understanding MHD outflows.
  At the same time it is widely thought, and
observations of cataclysmic variables
support the view, that the disk
turbulence - including MHD
turbulence - can be modeled approximately using the
 Shakura (1973), Shakura \&
Sunyaev (1973) ``alpha''
viscosity model (Eardley \& Lightman 1975;
Coroniti 1981; see also section 1.5).
      In contrast with the
internal disk dynamics, there is
theoretical and simulation
evidence that the outflows can
be treated using axisymmetric (2D) MHD
(Blandford \& Payne 1982;
Lovelace et al. 1991; Ustyugova et al. 1995).
  Here, we consider outflows from a disk
represented as a boundary
condition.
    This approach has subsequently
been adopted by other groups
(Meier et al. 1997;
Ouyed \& Pudritz 1997).
   This treatment of the
disk is justified for outflows
from a disk where the accretion speed is small
compared with the Keplerian speed
(Ustyugova et al. 1995).

\subsection{Theory of Stationary MHD Flows}

  The theory of stationary,
axisymmetric, ideal MHD flows
was developed by
Chandrasekhar (1956),
Woltjer (1959),
Mestel (1968),
Kulikovskyi \& Lyubimov (1962),
and others.
   Under these conditions the
MHD equations can be reduced to
a single equation for the ``flux function''
$\Psi(r,z)$ in cylindrical $(r,\phi,z)$
coordinates (Heinemann \& Olbert 1978; Lovelace
et al. 1986).
   The flux function
$\Psi$ labels flux surfaces so that
$\Psi(r,z)=$const represents the poloidal
projection of a field line.
  The equation for $\Psi$ is commonly referred to as
the Grad-Shafranov equation (Lovelace et al. 1986).
   The discussion of this section follows
that of Ustyugova {\it et al.} (1999).

\subsubsection{Integrals of Motion}

 For axisymmetric conditions the flow field
can be written as ${\bf v}={\bf v}_p
+ v_\phi {\bf e}_\phi$ where ${\bf v}_p$
is the poloidal $(r,z)$ component, $v_\phi =
\omega r >0$ is the toroidal component, and
${\bf e}_\phi$ is the unit toroidal vector.
Similarly, the magnetic field can be
written as ${\bf B} =
{\bf B}_p + B_\phi {\bf e}_\phi$.
The ideal MHD equations then imply
that certain quantities are constants
on any given flux surface $\Psi(r,z)=$const
or equivalently they are constants
along any given stream line or
a given magnetic field line.
These integrals are  functions of
$\Psi$
(see for example Lovelace et al. 1986),

\begin{equation}
\label{j1}
{\bf v}_p  =
\frac{K(\Psi)}{4 \pi \rho} {\bf B}_p
\end{equation}
\begin{equation}
\label{j2}
 \omega r^2 - \frac{r B_\phi}{K} =
\Lambda(\Psi)
\end{equation}
\begin{equation}
\label{j3}
\omega - \frac{K B_\phi}{4 \pi
\rho r} = \Omega(\Psi)
\end{equation}
\begin{equation}
\label{j4}
S = S(\Psi)
\end{equation}
\begin{equation}
\label{j5}
 w + \Phi - \frac{\Omega^2 r^2}{2} +
\frac{{\bf v}_p^2}{2}  + \frac{1}{2}
( \omega - \Omega )^2 r^2 = E(\Psi)
\end{equation}
Here, $S$ is the entropy,
$w$ is the enthalpy, and
$\Phi$ is the gravitational
potential. The quantity $K$
corresponds to the conservation
of mass along a streamline,
$\Lambda$ to the conservation
of angular momentum, $\Omega$
to the conservation of helicity, $S$
to the conservation of entropy,
and $E$ (Bernoulli's constant) to
the conservation of energy.

  The remaining MHD equation (which cannot
be written in the integral form)
is the Euler force equation across
the poloidal magnetic
 field line (Bogovalov 1997),

\begin{equation}
\label{j6}
({\bf v}_p^2 - v_{Ap}^2) \frac{\partial\theta}{\partial s}
 - \frac{\cos \theta}{r}
(v^2_\phi - v^2_{A\phi}) +
+ \frac{1}{\rho}
\frac{\partial}{\partial n}
\left( p + \frac{{\bf B}^2}{8 \pi} \right)
+ \frac{\partial \Phi}{\partial n}  =0~,
\end{equation}
which is equivalent to the Grad-Shafranov equation.
Here,
${\partial}/{\partial n}$ is the
derivative in the direction perpendicular
to magnetic field lines
and directed outward from the axis,
$\theta$ is the angle of
inclination of the poloidal magnetic
field line away from the $z-$axis,
$s$  is the distance from the disk
along a magnetic field
line, $v_{Ap}\equiv
|{\bf B}_p|/\sqrt{4\pi\rho}$ and
$v_{A\phi}\equiv |B_\phi|/\sqrt{4\pi \rho}$
 are the poloidal and azimuthal
Alfv\'en velocities.
The quantity
${\partial\theta}/{\partial s}$ is the curvature
of magnetic field line.
The first two terms in equation (6) are
determined by the non-diagonal
(tension) part of the stress tensor,
 $\rho v_i v_k  +
\left( p + {{\bf B}^2}/{8 \pi} \right) \delta_{ik}
- {B_i B_k}/{4 \pi}$.
  The third term
is determined by the total
(matter plus magnetic) pressure
$p + {{\bf B}^2}/{8 \pi}$ and the gravity
force $\partial \Phi/\partial n$.

   To clarify the physical sense
of the integrals of motion,
it is useful to derive
the fluxes of mass,
angular momentum (about the $z-$axis), and energy.
   The corresponding conservation
laws for stationary conditions are

\begin{equation}
\label{j7}
{\bf\nabla}\cdot (\rho {\bf v}_p)  = 0~,
\end{equation}
\begin{equation}
\label{j8}
{\bf\nabla}\cdot \left (r\rho {\bf v}_p v_\phi   -
{{\bf B}_pr B_\phi  \over 4\pi}\right ) = 0 ~,
\end{equation}
\begin{equation}
\label{j9}
{ \bf\nabla}\cdot \left [ \rho {\bf v}_p
\left ( \frac{{\bf v}^2}{2} +
\frac{{\bf B}^2}{4 \pi \rho} + w + \Phi \right )
-{{\bf B}_p({\bf v}\cdot {\bf B})\over 4\pi}
\right ] = 0 ~.
\end{equation}
  Because ${\bf v}_p \parallel {\bf B}_p$,
the vector flux densities
are directed along the field lines.

   Consider the fluxes through an annular
region with surface
area element
$d{\bf S}$.
   The matter flux through the
axisymmetric surface $\bf S$
extending out from the $z-$axis is
\begin{equation}
\label{j10}
{\cal F}_M  = \int \limits_{\bf S}
 d{\bf S}\cdot\rho {\bf
v}_p  = {1\over 4\pi}\int \limits_{\bf S}
d{\bf S}\cdot {\bf B}_p {K(\Psi) } ~,
\end{equation}
where we  took into account
the integral (\ref{j1}).
   $d{\bf S}\cdot{\bf B}_p$ is  the magnetic flux
through the annular region bounded by flux surfaces
$\Psi$ and $\Psi + d\Psi$.
   Thus we can change from space integration
to integration over $\Psi$.
   Because
 $B_r=-(1/r){\partial\Psi/\partial z}$
and
 $B_z=(1/r){\partial \Psi/\partial r}$, we
have
\begin{equation}
\label{j11}
{\cal F}_M(\Psi) = \frac{1}{2}
\int \limits_0^\Psi d \Psi^\prime K(\Psi^\prime) ~,
\end{equation}
where $\Psi=0$ corresponds to the $z-$axis.
Similarly,
$$
{\cal F}_L(\Psi) =\int d {\bf S} \cdot
\left (r \rho {\bf v}_p
v_\phi  -
\frac{{\bf B}_p rB_\phi}{4 \pi}  \right )
$$
\begin{equation}
\label{j12}
= \frac{1}{2}
\int \limits_0^\Psi d \Psi^\prime
\Lambda (\Psi^\prime) K(\Psi^\prime)~,
\end{equation}
$$
{\cal F}_E(\Psi) \!=\!\! \int \!\! d{\bf S}
\cdot \rho {\bf v}_p
\left [ \left (
\frac{{\bf v}^2}{2} +
\frac{{\bf B}^2}{4 \pi \rho} + w +
\Phi \right )
 -
 \frac{ {\bf B}_p{\bf v} \cdot {\bf B}}
{4 \pi} \right ]
$$
\begin{equation}
\label{j13}
=
\frac{1}{2} \int \limits_0^\Psi
d\Psi^\prime (E + \Lambda
\Omega) K~.
\end{equation}
Thus,
$Kd\Psi/2 $  is the matter
flux between the flux
surfaces separated by $d\Psi$,
$\Lambda Kd\Psi/2 $ is the angular
momentum flux,
and $(E + \Lambda \Omega) Kd\Psi/2$  is the
energy flux.
     Note that
$\Lambda(\Psi)$ is the specific
angular momentum carried
along the magnetic field line
$\Psi={\rm const}$,
$E(\Psi) + \Lambda \Omega (\Psi)$ is the specific energy,
and $\Omega(\Psi)$ is the angular velocity
of the disk at the point where
the magnetic field line  or
flux surface $\Psi ={\rm const}$
intersects the disk
(for $|{\bf v}_p| \rightarrow 0$ at
the disk).

\subsubsection{Alfv\'en Surface}

   Conditions at the Alfv\'en surface
are known to be important for the
global properties of MHD flows (Weber \&
Davis 1967).
  Equations (2) and
(3) constitute a  linear
system of equations
for $\omega$ and $B_\phi$.
The determinant of this system is zero if
$K^2=4 \pi \rho$.  Under this condition
  a solution exists  if
$\Lambda= r^2 \Omega$ (Weber \& Davis 1967)
which corresponds to
${\bf v}_p={\bf B}_p/{\sqrt{4 \pi \rho}}=
{\bf v}_{Ap}$.   This is the condition
which defines the Alfv\'en surface.
  Figure 20 shows the
Alfv\'en surface for a sample case.
  The radius at which this field line intersects the
disk is $r_d(\Psi)$. The radius at which it crosses
the Alfv\'en surface is $r_A(\Psi)$.
 The density
at this point on the Alfv\'en surface
is $\rho_A (\Psi)$.
Thus,
\begin{equation}
\label{j14}
\rho_A (\Psi)= K^2 (\Psi)/ 4 \pi ~, \quad
r^2_A (\Psi)= \Lambda (\Psi) / \Omega (\Psi)~.
\end{equation}
Equations (\ref{j2}) and (\ref{j3}) give
\begin{equation}
\label{j15}
\omega = \Omega~
\frac{1-\rho_A r^2_A / {\rho r^2}}{1 -
\rho_A /\rho}~,
\end{equation}
\begin{equation}
\label{j16}
B_\phi =
r\Omega  \sqrt{4\pi \rho_A}~ \frac{1 -
 r^2_A /r^2}{1-\rho_A /\rho}.
\end{equation}
Taking into account equations (\ref{j14}) - (\ref{j16}),
one can express the
fluxes of mass, angular momentum and energy,
using only the values of
physical quantities on the
Alfv\`en surface:
\begin{equation}
\label{j17}
{\cal F}_M(\Psi) = \int \limits_0^\Psi d \Psi^\prime
\sqrt{\pi \rho_A}~,
\end{equation}
\begin{equation}
\label{j18}
{\cal F}_L(\Psi) = \int \limits_0^\Psi d \Psi^\prime
 \sqrt{\pi \rho_A}~\Omega ~r^2_A ~,
\end{equation}
\begin{equation}
\label{j19}
{\cal F}_E(\Psi) = \int \limits_0^\Psi
d \Psi^\prime \sqrt{\pi\rho_A}~
 \left( E+ \Omega^2 r^2_A \right)~ .
\end{equation}

\subsubsection{Forces}

   For understanding
the plasma acceleration, we
project the different forces
onto the poloidal magnetic
field lines.
   As mentioned,  in a stationary
state,  matter flows
along the poloidal magnetic field lines.
   The acceleration in the  poloidal
$(r,z)$ plane is
\begin{equation}
\label{j20}
({\bf v}_p \cdot {\bf \nabla}) {\bf v}_p +
v_\phi ({\bf e}_\phi \cdot {\bf \nabla} )
(v_\phi {\bf e}_\phi)~.
\end{equation}
  The last term represents the centrifugal
acceleration
$ - (v^2_\phi /r ){\bf e}_r=-r\omega^2  {\bf
 e}_r$.
   To get the force per unit
mass along a magnetic field line,
we multiply the Euler equation by a
 unit vector  $\hat{\bf b}$ parallel to
${\bf B}_p$.
  This gives
\begin{equation}
\label{j21}
f = \omega^2 r \sin \theta -
\frac{1}{\rho} \frac{ \partial p}{ \partial s}
- \frac{\partial \Phi}{\partial s}
+ \frac{1}{4 \pi \rho}
{\hat{\bf b}} \cdot [({\bf \nabla}
\times {\bf B})\times {\bf B}]~,
\end{equation}
where $\theta$ is the inclination angle
of the field line to the $z-$axis.
The final
term of (\ref{j21}) is the projection
of the magnetic force in the
direction of $\hat{\bf b}$,
which can be transformed to
$$
f_M=\frac{1}{4 \pi \rho}
{\hat{\bf b}} \cdot [({\bf \nabla}
\times {\bf B})\times {\bf B}]=
-{1\over{8 \pi \rho r^2}}
{\partial(rB_\phi)^2\over\partial s}~,
$$
which is useful for
understanding our results.

  When magnetic field lines
are inclined outward, away from the
symmetry axis, the gravitational force
$f_G=\partial \Phi/\partial s$
opposes the outflow of matter
from the disk.
    If the matter is relatively cold then
the pressure gradient force
$f_P=-({1}/{\rho}) ({ \partial p}/{ \partial s})$
is unimportant.
   Then matter is accelerated outward
 by the centrifugal force
$f_C=\omega^2 r {\rm sin} \theta$ and/or
the magnetic force $f_M$.
 This determines
the driving mechanisms of the outflow, centrifugal
and/or magnetic.
   The centrifugal force always acts to
accelerates matter outward if the distance between
magnetic field line and the axis increases.
  However, under the same
condition, the magnetic force can act
to accelerate or decelerate the flow
depending on the direction of
$-{\bf \nabla}(rB_\phi)^2$ (see discussion
in Ustyugova {\it et al.} 1999).

  Thus, magnetic and centrifugal forces may
both accelerate matter, but this
depends on the
configuration
of magnetic field and current-density lines.

\subsubsection{Collimation}

   Consider now the collimation of
the flow.
   From equation (\ref{j6}), taking into
account that
${\rm cos} \theta = {\partial r}/{\partial n}$,
we have
\begin{equation}
\label{j22}
({\bf v}^2_p - {\ v}^2_{Ap})
\frac{\partial \theta}{\partial s} =
- \frac{1}{8 \pi \rho r^2}
\frac{\partial}{\partial n}
\left( r B_\phi \right)^2 +
\frac{\cos \theta v^2_\phi}{r}
- \frac{1}{\rho}
\frac{\partial}{\partial n} \left(p +
 \frac{{\bf B}^2_p}{8 \pi}
\right) - \frac{\partial \Phi}{\partial n}~.
\end{equation}
   At large distances from the Alfv\'en surface
$r\gg r_A$, the density $\rho \ll \rho_A$,
but values $\rho r^2$ and ${\bf v}_p^2$ remain
finite (Heyvaerts and Norman 1989).
   Then $v^2_{Ap}=\left(\rho/\rho_A\right)^2
{\bf v}_p^2 \ll {\bf v}_p^2$,
so that the second term on the left-hand side
of (\ref{j22})
is negligible.
    On the right-hand side,
only the first term is important
for $r \gg r_A$.
    Then, equation (\ref{j22}) simplifies to
\begin{equation}
\label{j23}
v^2_p ~\frac{\partial \theta}{\partial s} =
- \frac{1}{8 \pi \rho r^2}
\frac{\partial}{\partial n} \left( r B_\phi \right)^2~ .
\end{equation}
   Thus, the curvature of magnetic field lines
in the region $r \gg r_A$ is
determined by the gradient
$(rB_\phi)^2$.
   This force can act either to ``collimate'' the flow
or ``anticollimate'' it depending on the orientation of
the $rB_\phi(r,z)=$const and $\Psi(r,z)=$const surfaces
(Ustyugova {\it et al.} 1999).
   The simulation study of Ustyugova
{\it et al.} (1999) in fact finds
only very small collimation of
the hydromagnetic outflows over
distances of the order of $100$
times the inner disk radius.
   Recent work by Ustyugova
{\it et al.} (2001) extents this
conclusion by showing only small
collimation out to distances of
$>1000$ times the inner disk radius.
   The collimation observed
in protostellar jets at small
distances, for example, in HH 30
observed with the space telescope,
may result from the pressure of
the external medium (Lovelace {\it et al.} 1991).

   The conclusion by Heyvaerts
and Norman (1989) that the
magnetic collimation of MHD jets is
a mathematical rather than physical result.
   The conclusion is based on
the premise that the
entire half-space above the disk is
described by a stationary solution $\Psi(r,z)$
of the Grad-Shafranov equation.
    In reality, MHD jets result from
initial conditions at the
disk (as in the simulation
studies) which result in the outward
propagation of the jet
into a highly conducting ionized
external (interstellar or
intergalactic) medium.
    At a given time, the region inside
the surface $r={\cal R}(z,t)$ is filled by
the jet flow and the region outside
it is filled by external plasma.
   The net axial current  carried
by the jet $2\pi\int_0^{\cal R} r dr J_z$
(for say $z>0$)
is necessarily zero for any $(z,t)$.
   Therefore, the toroidal magnetic field is zero
on the surface ${\cal R}$, and there is
{\it no} magnetic confinement of the flow.
    Of course, it is possible that a
fraction of the flux $\Psi$ is collimated
along the $z-$axis while  the remaining flux
is ``anti-collimated'' (that is, pushed
away) from the $z-$axis.
   Exactly such a case is discussed in
\S 1.5 below for the case of Poynting
jets.
     This case also has the physical
condition that the jet carries no
net magnetic flux, $2\pi\int_0^{\cal R} r dr B_z =0.$
     As in Landau's (1946) treatment
of the damping of plasma waves, the physically
correct treatment of the collimation of MHD
jets requires  consideration of an
initial value problem.

\subsection{Numerical Simulations of MHD
Outflows}

   In order to test and more
fully understand
analytical models
of stationary outflows,
MHD simulations of flows from a disk treated
as a boundary condition have
been carried out by a number
of groups.

\subsubsection{Non-Stationary Outflows}

   The initial magnetic field
configuration was chosen
so that the magnetic field
was significantly inclined to the disk
($\theta>30^o$) over most of
the disk surface (Ustyugova et al. 1995;
Koldoba et al. 1995).
    The simulations involve solving the
complete system of ideal non-relativistic
MHD equations using a Godunov-type
code assuming axisymmetry but
taking into account all three
components of velocity and magnetic
field.
     Matter of the
corona was initially in thermal
equilibrium with the gravitating center.
  At $t=0$, the disk is set into
rotation with Keplerian velocity
and at the same time matter is pushed
from the disk with a small poloidal velocity
equal to a fraction of the slow magnetosonic
velocity ( $v_p = \alpha v_{sm}$,
with $\alpha=0.1-0.9$).
  A relatively high
temperature and small magnetic field
was considered.
    We found that at the
maximum of the outflow,
matter is accelerated to
speeds in excess of the escape
speed and in excess of the
fast magnetosonic speed within
the simulation region ($\sim 100 r_i$).
   The acceleration is
due to both thermal and magnetic pressure
gradients.
   The outflow
collimates within the simulation
region due to strong amplification,
`wrapping up' of the toroidal
magnetic field and the associated
pinching force.

  However, the outflows are {\it not stationary}.
   The matter flux grows to a peak
and then decreases to relatively small
values.
     The strong collimation of the outflow
reduces the divergence of the field away
from the $z-$axis ($\theta < 30^o$) and this
``turns off'' the outflow of matter and
leads to flow velocities less than the
escape speed.
   Thus, this simulation is
an example of a temporary outburst
of matter to a jet.
    Unfortunately, this type of
flow has a significant dependence
on the initial conditions.

\subsubsection{Stationary Outflows}

 More recently, stationary magnetohydrodynamic
outflows from a rotating
accretion disk have been obtained
by time-dependent axisymmetric
simulations by Romanova et al. (1997)
and systematically analyzed by
Ustyugova et al. (1999).
 The initial magnetic field in the
latter work was taken to
be a split-monopole poloidal
field configuration (Sakurai 1987)
frozen into the
disk.
   The disk was treated as a perfectly
conducting, time-independent density
 boundary [$\rho(r)$]
in Keplerian rotation which is different
from the  earlier specification of a
small velocity outflow (\S 4.3.1, Ustyugova
et al. 1995).
  The outflow velocity
from the disk is
determined self-consistently from the MHD
equations.
   The temperature of
the matter outflowing from
the disk is  small in the
region  where the magnetic
field is inclined away from
the symmetry axis
($c_s^2 \ll v_K^2$), but
relatively high
($c_s^2 ~{\buildrel < \over \sim}~v_K^2$)
at very small radii in the disk
where the magnetic field is not inclined
away from the axis.
   We have found a large class of stationary
MHD winds.

   Figure 20 shows the nature of the stationary
MHD outflows found by Ustyugova {\it et al.} (1999).
   The outflows are approximately spherical, with
only small collimation within the simulation region.
   The {\it collimation distance} over which
the flow becomes collimated (with divergence less
than, say, $10^o$) is much larger than the size of
our simulation region.
      Because the flow is  super
fast-mag\-ne\-to\-so\-nic (see below), the flow is not
expected to magnetically
collimate at larger distances.
   Notice for example that the solar wind is
also super fast-magnetosonic and essentially
uncollimated.
 Collimation may however result from the pressure of
external plasma (Lovelace {\it et al.} 1991).

%\begin{figure}
%\epsfysize=5.5cm % fix the y-dimension and scales x-dim. to y-dim.
%\hspace{3.5cm}\epsfbox{gbkl2.eps} %for centering: act on hspace argument

%\label{fjet4}
%\end{figure}

   Figure 21 shows the variation of the flow
speed and other characteristic speeds along
a sample streamline.
  The outflow
accelerates from thermal velocity ($\sim c_s$)
to a much larger asymptotic poloidal flow
velocity of the order of one-half $\sqrt{GM/r_i}$
where $M$ is the mass of the central object
and $r_i$ is the inner radius of the disk.
   This asymptotic velocity is much larger than
the local escape speed and is larger than
fast magnetosonic speed by a factor of $\sim 1.75$.
  The {\it acceleration distance} for the outflow, over
which the flow accelerates from $\sim 0$
to, say, $90\%$ of the asymptotic speed, occurs
at a flow distance of about $80 r_i$.
   Close to the disk the outflow is driven by
the centrifugal force while at all larger
distances the flow is driven by the
magnetic force which is proportional to $-{\bf \nabla}
(rB_\phi)^2$, where $B_\phi$ is the toroidal field.

%\begin{figure}
%\epsfysize=6cm % fix the y-dimension and scales x-dim. to y-dim.
%\hspace{3.5cm}\epsfbox{gbkl3.eps} %for centering: act on hspace argument

%\label{fjet5}
%\end{figure}

  The stationary MHD flow solutions allow us
(\ref{j1}) to compare the results  with MHD theory
of stationary flows,
(\ref{j2}) to investigate the influence of different
outer boundary conditions
on the flows, and (\ref{j3}) to investigate the influence
of the shape of the simulation region
 on the flows.
   The ideal MHD integrals of motion (constants
on flux surfaces discussed by Lovelace et
al. 1986) were
calculated along magnetic field
lines and were shown
to be constants with accuracy $5-15 \%$.
   Other characteristics of the
numerical solutions were compared with the theory,
including conditions at the
Alfv\'en surface.

    Different outer boundary conditions
on the  toroidal component
of the magnetic field can significantly
influence the calculated flows.
    The commonly used ``free'' boundary condition on
the toroidal field leads
to  artificial magnetic
forces on the outer boundaries,
which can give spurious collimation
of the flows.
   New outer boundary conditions
which do not give artificial forces have
been proposed and
investigated by Ustyugova et al. (1999).

   The simulated flows
may also depend  on the shape
of the simulation region.
   Namely, if the simulation region
is elongated in the $z-$direction,
then Mach cones on the outer
cylindrical boundaries may be
partially directed inside the
simulation region.
   Because of this, the boundary can have
an artificial influence on the calculated
flow.
   This effect is reduced if the
computational region is approximately square
or if it is spherical as in Figure 20.
  Simulations of MHD outflows with an
elongated computational
region can lead to
{\it artificial} collimation
of the flow.

   Recent simulation studies have treated
MHD outflows from disks with
more general initial ${\bf B}$
field configurations, for example, that
where the poloidal field has different
polarities as a function of radius
(Hayashi, Shibata, \& Matsumoto 1996;
Goodson, Winglee,
\& B\"ohm 1997;
Romanova et al. 1998).
   The differential rotation
of the foot-points of
${\bf B}$ field loops at different
radii on the
disk surface causes
twisting of the coronal magnetic
field, an increase in the
coronal magnetic energy, and an
opening of the loops in
the region where the magnetic pressure
is larger than the matter pressure
($\beta {\buildrel < \over \sim} 1$)
(Romanova et al. 1998).
   In the region where
$\beta {\buildrel >\over \sim}1$,
the loops may be only
partially opened.
  Current layers form in the
narrow regions
which separate oppositely
directed magnetic field.
    Reconnection occurs in these
layers as a result of the
small numerical magnetic
diffusivity.
    In contrast with the case of the
solar corona,
there can be a steady outflow of energy and
matter from the disk surface.
 We find that the power output
mainly in the form of a
Poynting flux.
    Opening of magnetic field loops
and subsequent closing
can give
reconnection events which may be responsible
for X-ray flares in disks
around both stellar mass objects
and massive black holes (Hayashi et al. 1996;
Goodson,  Winglee,
\& B\"ohm 1997;
 Romanova et al. 1998).

\subsection{Poynting  Jets}

   In a very different regime from
the hydromagnetic flows discussed
in \S 2 - 4,  a Poynting outflow or jet
transports energy and angular
momentum mainly by the
electromagnetic fields with
only a small contribution due to the
matter
     (Lovelace {\it et al.} 1987;
Colgate \& Li 1998; Romanova {\it et al.}
1998;
Ustyugova {\it et al.} 2000;
Li {\it et al.} 2001;
Lovelace {\it et al.} 2001).
  A steady Poynting jet can be
characterized in the lab frame
by its asymptotic (large distance)
magnetic field $B_\phi = -B_0 [r_o/r_j(z)]$
at the jet's edge, $r=r_j(z)$, where
$r_0$ is the jet's radius at $z=0$
and $B_0$ is the poloidal field strength
at this location.
   The electric field in the
jet is ${\bf E} = -{\bf v}\times {\bf B}/c$
and consequently the energy flux
or luminosity of the jet is
$L_j = v B_0^2r_0^2/8 \sim 4\times 10^{45}
{\rm erg/s}(v/c)(r_0/10^{14}{\rm cm})^2
(B_0/10^4 {\rm G})^2$.
  Propagating disturbances in such field
dominated
jets provide a simple, but
self-consistent physical
model for the gamma ray flares observed
in Blazars (Romanova \& Lovelace 1997;
Levinson 1998;
Romanova 1999).
   Owing to pair production close to
the black hole, the main constituent of
a Poynting flux jet may be electron-positron
pairs.

\subsubsection{Theory of Poynting Jets}

    Consider
the  coronal magnetic field
of a differentially rotating Keplerian
accretion disk for a given poloidal field
threading the disk.
  That is, the disk is perfectly conducting
with a very small accretion speed.
    Further, consider ``coronal
or force-free'' magnetic fields in
the non-relativistic limit.
   We use cylindrical $(r,\phi,z)$
coordinates as in \S 3 and consider
axisymmetric field configurations.
   Thus the magnetic field has
the form
$ {\bf B}~ = {\bf B}_p +
B_\phi \hat{\rvecphi~}~,
$
with
$
{\bf B}_p = B_r{\hat{\bf r}}+
B_z \hat{\bf z}~.
$
Because ${\bf \nabla \cdot B}=0$,
${\bf B} = {\bf \nabla \times A}$ with
${\bf A}$ the vector potential.
  Consequently,
$
B_r =
-(1 / r)(\partial \Psi
/ \partial z)~,
B_z =(1 / r)
(\partial \Psi / \partial r)~,
$
where $\Psi(r,z) \equiv r A_\phi(r,z)$.
  The $\Psi(r,z)=$ const
lines label the poloidal field lines;
that is,
$({\bf B}_p{\bf \cdot \nabla}) \Psi
=0={\bf B \cdot \nabla}\Psi=0 $.
Note that
$2\pi \Psi(r,z)$ is the magnetic flux
through a horizontal, coaxial
circular disk of radius $r$.

   The magnetic field threading
the disk at $z=0$ is assumed to
evolve slowly so that it can be
considered approximately
time-independent,  $\Psi(r,z=0)
=\Psi_0(r)$.
    However, the magnetic field above
the disk will in general
 be time-dependent,
$\Psi=\Psi(r,z,t)$,
due to the differential rotation
of the disk.
   Figure 22 shows a possible vacuum
poloidal magnetic field threading
the disk.  This field may be regarded
as the ``initial'' field for the
considered case where the space
exterior to the disk is filled
with a low density plasma.

%\begin{figure}
%\epsfysize=6cm % fix the y-dimension and scales x-dim. to y-dim.
%\hspace{3.5cm}\epsfbox{gbkl4.eps} %for centering: act on hspace argument

%\label{fjet6}
%\end{figure}

     The non-relativistic equation of
plasma motion in the corona of
an accretion disk is

\begin{equation}
\label{j24}
\rho { d{\bf v} / dt}=
-{\bf \nabla}p +\rho {\bf g} +
{\bf J \times B}/c,
\end{equation}
where ${\bf v}$ is the flow velocity,
$p$ is the pressure, and ${\bf g}$
is the gravitational acceleration.
  The equation for the
${\bf B}$ field is
$
{\bf \nabla \times B} ={4\pi \over c}{\bf J},
$
because the displacement current is
negligible in the non-relativistic
limit.
   In the ``coronal or
force-free plasma limit,''
the magnetic energy density ${\bf B}^2/(8\pi)$
is much larger than the kinetic or thermal
energy densities; that is,
sub-Alfv\'enic flow speeds ${\bf v}$,
${\bf v}^2 \ll {v}_A^2 = {\bf B}^2/4\pi \rho$,
where $v_A$ is the Alfv\'en velocity.
Equation (\ref{j24}) then simplifies
to
$
{ 0} \approx {\bf J \times B}$;
   therefore, ${\bf J} = \lambda {\bf B}$
(Gold \& Hoyle 1960).
  Because ${\bf \nabla \cdot J}=0$,
$({\bf B \cdot \nabla}) \lambda =0$
and consequently $\lambda = \lambda(\Psi)$,
as  well-known.
   Thus
Amp\`ere's law becomes
$
{\bf \nabla \times B} = 4 \pi \lambda(\Psi)
 {\bf B}/c$.

   The $r$ and $z$ components of
this equation  imply
$
 rB_\phi = H(\Psi),$ and
$dH(\Psi) / d\Psi$
$={4\pi}
\lambda(\Psi)/c$,
where $H(\Psi)$ is another function of
$\Psi$.
   Thus, $H(\Psi)=$ const are lines of
constant poloidal current density;
${\bf J}_p=(c / 4\pi)
(dH / d\Psi){\bf B}_p$ so that
$({\bf J}_p \cdot{\bf \nabla}) H=0$.
     The toroidal component of
Amp\`ere's law gives
\begin{equation}
\label{j29}
\Delta^\star \Psi = -
H(\Psi) {d H(\Psi) \over d\Psi}~,
\end{equation}
which is the Grad-Shafranov equation for
$\Psi$ (see for
example Batemann 1980;
 Lovelace {\it et al.} 1986).
   Here,
$ \Delta^\star \equiv
{\partial^2 / \partial r^2}
-(1/r)(\partial / \partial r)
+{\partial^2 / \partial z^2}
$
is the adjoint Laplacian operator.
   Note that $\Delta^\star \Psi =
r(\nabla^2 -1/r^2) A_\phi$ and that
$H(dH/d\Psi) = 4\pi r J_\phi/c$.
   From Amp\`ere's law,
$\oint d{\bf l}\cdot {\bf B}
=(4\pi/c)\int d{\bf S}\cdot {\bf J}$,
so that
$rB_\phi(r,z)=H(\Psi)$ is $(2/c)$ times
the current flowing through a circular
area of radius $r$ (with normal $\hat{\bf z}$)
labeled by $\Psi(r,z)$= const.
  Equivalently, $-H[\Psi(r,0)]$
is  $(2/c)\times$
the current flowing into the
area of the disk $\leq r$.
   For all cases studied here,
$-H(\Psi)$ has a maximum so that
the total current flowing into
the disk for $r\leq r_m$ is
$
 I_{tot} = {2 \over c}(-H)_{max},
$
where $r_m$ is such that
$-H[\Psi(r_m,0)]=(-H)_{max}$
so that $r_m$
is less than
the radius of the $O-$point, $r_0$.
  The same total current $I_{tot}$
flows out of the region of the disk
$r=r_m$ to $r_0$.

   The function $H(\Psi)$ must be
determined before the GS
equation can be solved.
Lynden-Bell and Boily (1994)
determined the form of $H(\Psi)$
by requiring that
the solution $\Psi({\bf r})$
be a self-similar
function of $|{\bf r}|$.
  Their stationary solution
has {\it no} energy or angular
momentum outflow from the disk.
   However, in general the
magnetic field threading
a disk will have definite
length-scales and the
field solution will not
be self-similar.

     For definiteness we consider
the {\it initial value problem} where the
disk at $t=0$ is threaded by a poloidal field
such as shown in Figure 22.
   Such a field can arise from
a disk-dynamo driven by star-disk collisions
as discussed by
Pariev, Finn, \& Colgate (2001).
We assume for simplicity
that the field threading the disk
has the symmetry of a dipole although
  it is predicted in the reference above
that the quadrupole field grows
somewhat faster.

     The form of $H(\Psi)$
is then determined
by the differential rotation of the
disk:
   The azimuthal {\it twist} of a given field
line going from an inner footpoint
at $r_1$ to an outer footpoint at $r_2$
is fixed by the differential rotation
of the disk.
  The field line slippage speed through
the disk due to the disk's finite
magnetic diffusivity is estimated
to be negligible compared with the
Keplerian velocity.
  For a given field line
we have $rd\phi/B_\phi = ds_p/B_p$,
where $ds_p \equiv \sqrt{dr^2+dz^2}$ is the
poloidal arc length along the field
line, and
$B_p \equiv \sqrt{B_r^2+B_z^2}$.
   The total twist of a field line
loop is
\begin{equation}
\label{j31}
\Delta \phi(\Psi) =
\int_1^2 ds_p ~{-B_\phi \over r B_p}
=-H(\Psi) \int_1^2 {ds_p \over r^2 B_p}~,
\end{equation}
with the sign included to give
$\Delta \phi >0$.
  For a Keplerian disk around
an object of mass $M$ the angular
rotation is $\omega_K = \sqrt{G M/r^3}$
so that
the field line twist after a time $t$ is
\begin{equation}
\label{j32}
\Delta \phi(\Psi)
=   \omega_0 ~t
\left[\left({r_0\over r_1}\right)^{3/2} -
\left({r_0\over r_2}\right)^{3/2}\right]
= (\omega_0~t) ~{\cal F}(\Psi/\Psi_0)
\end{equation}
where $r_0$ is the radius of the
$O-$point,
$\omega_0\equiv\sqrt{GM/r_0^3}$,
and ${\cal F}$ is a dimensionless
function (the quantity in the
square brackets).   At sufficiently small
$r_1$ one  reaches the inner
radius of the disk $r_i~(\ll r_0)$
where we assume $\omega_K$ saturates
at the value $\omega_{Ki}=\sqrt{GM/r_i^3}$.
    For the dipole-like field
of Figure 22, ${\cal F}\approx 3^{9/8}
(\Psi_0/\Psi)^{3/4}$
for $\Psi/\Psi_0 \ll 1$, while
${\cal F}
\approx  3.64(1-\Psi/\Psi_0)^{1/2}$
for $1-\Psi/\Psi_0 \ll 1$.

   The quantity
$T \equiv  \omega_0 t$
(in radians)
we refer to as the ``twist'' of the
disk.
   For the dashed field lines
of Figure 22, $ T$ is
the twist $\Delta \phi(\Psi_1)$
 of the field line with
$\Psi_1/\Psi_0 \approx 0.932$
which has footpoints at $r_1\approx 0.73r_0$
and $r_2 \approx 1.4r_0$.
  For $\Psi/\Psi_0=0.1$,
where the footpoints are at
$r_1 \approx 0.14r_0$
and $r_2 \approx 7.8r_0$,
the
field line twist is $\Delta \phi \approx
18.3$ rad. for $T=1$.

   For example, for an accretion
disk around a massive
 $M=M_810^8M_\odot$ black hole in
the nucleus of galaxy,
the twist is
$ T= (t/3.17{\rm d})(r_0/10^{15}{\rm cm})^{3/2}
/\sqrt{M_8}$.  The inner radius of
the disk is $r_i\approx M_8 9 \times 10^{13}$ cm
for a Schwarzschild black hole.

  The Grad-Shafranov equation (\ref{j29})
is an elliptic equation and
 can be readily solved by the method
of Successive Over-Relaxation (SOR)
(see for example, Potter 1973).
   Figure 23 shows a sample high
twist solution
exhibiting a Poynting jet (Lovelace
{\it et al.} 2001).
  These high-twist solutions consist
of a region near the axis which
is {\it magnetically collimated} by
the toroidal $B_\phi$ field and
a region far from the axis, on
the outer radial boundary, which
is {\it anti-collimated} in the sense
that it is pushed against the outer
boundary.
  The field lines returning
to the disk at $r>r_0$ are
anti-collimated by the pressure
of the toroidal magnetic field.

    Figure 24 shows a three dimensional
view of two sample field lines for the
same case as Figure 23.

%\begin{figure}
%\epsfysize=6cm % fix the y-dimension and scales x-dim. to y-dim.
%\hspace{3.5cm}\epsfbox{gbkl5.eps} %for centering: act on hspace argument

%\label{fjet7}
%\end{figure}

%\begin{figure}
%\epsfysize=6cm % fix the y-dimension and scales x-dim. to y-dim.
%\hspace{3.5cm}\epsfbox{gbkl6.eps} %for centering: act on hspace argument

%\label{fjet8}
%\end{figure}

\subsubsection{Analytic Model of Poynting Jets}

    Figure 24 shows that most of the twist
$\Delta \phi$ of a field line occurs
along the jet from $z=0$ to $Z_{max}$.
Because $-r^2 d\phi/H(\Psi) = dz/B_z$,
we have
\begin{equation}
\label{j34}
{\Delta \phi(\Psi) \over -H(\Psi)}=
{(\omega_0 t) {\cal F}(\Psi/\Psi_0)
\over -H(\Psi)} \approx
{Z_{max} \over r^2 B_z}~,
\end{equation}
where $r^2 B_z(r,z)$ is evaluated
on the straight part of the jet
at $r=r(\Psi)$.
   In the core of the jet
$\Psi \ll \Psi_0$, we have
${\cal F} \approx 3^{9/8}
(\Psi_0/\Psi)^{3/4}$, and in
this region we can take
\begin{equation}
\label{j35}
\Psi=C \Psi_0 \left({r \over r_0}\right)^q~,~~~
{\rm and} ~~~ H=-{\cal K}
\left({\Psi_0 \over r_0}\right)
\left({\Psi\over \Psi_0}\right)^s~,
\end{equation}
where $C,~q,{\cal K},$ and $s$ are
dimensionless constants.
   The GS equation for the straight
part of the jet, $\Psi_{rr}
-\Psi_r/r= - H H^\prime$ implies
$q=1/(1-s)$ and $C^{2(1-s)}=
s(1-s)^2 {\cal K}^2/(1-2s)$.
   In turn, equation (\ref{j35})
implies $s=1/4$ so that $q=4/3$,
$C =[9/32]^{2/3}{\cal K}^{4/3}$,
and
$
{\cal K}=3^{1/8} 4({r_0 \omega_0  t
/ Z_{max}})$.

   In order to have
a Poynting
jet, we must have ${\cal K}$ larger than
 $\approx 0.5$.
    For the case of uniform expansion
of the top boundary, $Z_{max}=V_z t$,
this condition is the same as
$V_z <9.2 (r_0\omega_0)$.
   For the case of Figures 23 and 24,
${\cal K} \approx 0.844$.

  The field components in the
straight part of the jet
are
\begin{equation}
\label{j37}
B_\phi =-\sqrt{2}B_z=-\sqrt{2}
\left({3 \over 16}\right)^{1/3} {\cal K}^{4/3}
\left({\Psi_0 \over r_0^2}\right)
\left({r_0 \over r}\right)^{2/3}~.
\end{equation}
These field components of
course satisfy the radial
force balance equation $~~~~~~~$
$d(B_z^2)/dr
+(1/r^2)d(r^2B_\phi^2)/dr =0$
which is equivalent
to the GS equation.
   These dependences agree approximately
with those found independently in the numerical
simulations of Poynting jets by
Ustyugova {\it et al.} (2000).
  On the disk,
$\Psi \approx 3^{3/2}\Psi_0(r/r_0)^2$
for $r < r_0/3^{3/4}$.
  Using this and equation (16)
for $\Psi(r)$
gives the relation between the
radius of a field line in the
disk, denoted $r_d$, and its
radius in the jet,
$
{r / r_0} =2^{5/2}3^{1/8}
({r_d / r_0})^{3/2} {\cal K}^{-1}$.
Thus the power laws apply for
$r_1 < r <  r_2$, where
$r_1 = 3^{1/8}2^{5/2}
r_0(r_i/r_0)^{3/2}/{\cal K}$ and
$r_2=(2^{5/2}/3)r_0/{\cal K}$,
with $r_i$  the inner radius
of the disk.
  The outer edge of the Poynting
jet has a transition layer where
the axial field changes from $B_z(r_2)$
to zero while (minus) the toroidal field
increases from $-B_\phi(r_2)$
to $(-H)_{max}/r_2$.
    Using equations (19), which are
only approximate at $r_2$, gives
$(-H)_{max}\approx (3/2)^{1/2}{\cal K} \Psi_0/r_0$.
   This expression agrees approximately
with our numerical results.

\subsubsection{MHD Simulations of Poynting Jets}

Ustyugova {\it et al.} (2000)
did full, axisymmetric MHD
simulations of a Keplerian
disk initially threaded by
a dipole-like magnetic field
as shown in Figure 22.
   For these simulations the outer
boundaries at $r=R_{max}$ and
$z=Z_{max}$ were treated as
free boundaries following the
methods of Ustyugova {\it et al.} (1999).
   These simulations  established
that a quasi-stationary collimated
Poynting jet arises from the inner
part of the disk while a steady
uncollimated hydromagnetic outflow
arises in the outer part of the disk.
Figure 25 shows the evolution towards
a stationary Poynting jet.

%\begin{figure}
%\epsfysize=6cm % fix the y-dimension and scales x-dim. to y-dim.
%\hspace{3.5cm}\epsfbox{gbkl7.eps} %for centering: act on hspace argument

%\label{fjet9}
%\end{figure}

 Quasi-stationary Poynting jets
 from the two sides of the disk within $r_0$
give an energy outflow
per unit radius of the disk
$d\dot{E}_B/dr=
r v_K(-B_\phi B_z)_h$, where the $h$
subscript indicates evaluation at
the top surface of the disk.
   This outflow
is $\sim r_0d\dot{E}_B/dr \sim
v_K(r_0)(\Psi_{max}/r_0)^2 \sim  10^{45}({\rm erg/s})
r_{015}^{3/2}\sqrt{M_8}[B_z(0)/6{\rm kG}]^2
$ where  $r_{015}$
is in units of $10^{15}$cm,
and $M_8$ in units of
$10^8{\rm M}_\odot$.
  This formula  agrees
approximately with the values
derived from the simulations.
   The jets also give an
outflow of angular momentum from
the disk which
causes disk accretion
(without viscosity) at the rate
$\dot{M}_B(r) \equiv - 2\pi \Sigma v_r=
-2(r^2/v_K)(B_\phi B_z)_h \sim
2 \Psi_{max}^2/[r_0^2 v_K(r_0) ]$, where
$\Sigma$ is the disk's surface mass density.
(Lovelace {\it et al.} 1997).
   The Poynting jet has a net axial
momentum flux $\dot{P}_z= (1/4)\int rdr
(B_\phi^2 -B_z^2) \sim
0.5 (\Psi_{max}/r_0)^2$, which acts to
drive the jet outward through an external
medium.
  Further, the Poynting jet
generates toroidal magnetic flux at the rate
$\dot{\Phi}_t \sim
- 12 [v_K(r_0)/r_0]\Psi_{max}$.

  For long time-scales,
 the Poynting jet is of course time-dependent
due to the angular momentum it extracts
from the inner  disk ($r<r_0$).
  This loss of angular momentum leads to
a ``global magnetic instability'' and collapse
 of the inner disk (Lovelace {\it et al.} 1997).
   An approximate model of this collapse
can be made if the
inner disk mass $M_d$ is concentrated
near the $O-$point radius $r_0(t)$,
if the
field line slippage through the
disk is negligible (Lovelace {\it et al.} 1997),
$\Psi_{max}=$const, and if
$(-rB_\phi)_{max}\sim \Psi_{max}/r_0(t)$ (as
found here).  Then,
$
M_d {d r_0 / dt} =  {-2 \Psi_{max}^2
(GMr_0})^{-1/2}.$
   If  $t_i$ denotes the time at which
$r_0(t_i)=r_i$ (the inner radius of the disk), then
$
r_0(t)=r_i [1 -(t-t_i)/
t_{coll} ]^{2/3}~,
$
for $t \leq t_i$,
where $t_{coll} =\sqrt{GM}~M_d r_i^{3/2}
/(3 \Psi_{max}^2)$
is the time-scale for
the  collapse of the inner disk.
(Note that the time-scale for $r_0$ to decrease
by a factor of $2$ is $\sim t_i(r_0/r_i)^{3/2}\gg t_i$
for $r_0 \gg r_i$.)
  The power output to the Poynting jets is
$
\dot{E}(t)=(2/ 3)(\Delta E_{tot}/
 t_{coll})[1-(t-t_i)/
 t_{coll}]^{-5/3},$
where $\Delta E_{tot} = G M M_d /2 r_i$ is
the total energy of the outburst.
  Roughly, $t_{coll} \sim 2{\rm day}
M_8^2(M_d/M_\odot)(6\times 10^{32}{\rm
G cm}^2/\Psi_{max})^2$ for
a Schwarzschild black hole, where validity of the
analysis requires
$t_{coll} \gg t_i$.

\subsection{Summary Regarding Jets from
Accretion Disks}

    MHD simulations carried
out by a number of groups
over the last several years
support the idea that an
ordered magnetic field of
an accretion disk can give
powerful outflows or jets of matter,
energy, and angular momentum.
   Most of the studies  have
been in the hydromagnetic
regime and find
asymptotic flow speeds of the
order of the maximum Keplerian
velocity of the disk, $v_{Ki}$.
   These flows are clearly relevant
to the jets from protostellar systems
which flow speeds  of the order of $v_{Ki}$.
    In contrast,
observed VLBI jets in quasars and
active galaxies point to bulk
Lorentz factors $\Gamma \sim 10$ - much
larger than the disk Lorentz factor.
   The large Lorentz factors
as well as the small Faraday rotation
measures point to the fact that these
jets are in the Poynting flux
regime.
   These jets  may involve
energy extraction
from a rotating black hole (Blandford
and Znajek 1977;  Livio et al. 1999).

\section{Discussion}

Observational evidence for
massive black holes at the center
of our Galaxy
and in the nuclei of other galaxies
(Cherepashchuk, 1996; Ho, 1998; Haswell, 1999; Richstone
{\it et al.} 1998; Gebhardt {\it et al.} 2000;
Merritt \& Ferrarese 2001)
make it necessary to revise
theoretical models of the disk accretion.
    A large part of the high energy
radiation originates
close to the black hole,
where the standard accretion
disk model is not
appropriate.
   Improvements of  the accretion models
have been made by taking into
account the advective terms and including
the important consequences of
ordered and disordered magnetic fields.
      The conclusions obtained
from ADAF solutions - for optically thin
accretion flows at low mass accretion rates -
are open to question because of omission
of all magnetic field effects.
    Heating due to the annihilation
of the magnetic
field in the accretion flow
is predicted to be important.
    Further, the energy exchange between electrons
and ions may be increased substantially by
the turbulent electric fields in the plasma.
    Account of the magnetic field leads
to the conclusion that the radiative
efficiency of the accretion is {\it not}
smaller than about $0.25$ the efficiency for
  standard disk accretion (Bisnovatyi-Kogan \& Lovelace,
1997, 2000)
It is expected that more accurate
treatment of the relaxation
connected with the plasma turbulence will
increase the efficiency further,
making it close to unity (see also
Fabian and Rees, 1995).

Some observational data which
were interpreted as an evidence for the
existence of the ADAF regime have
disappered after additional accumulation
of data.
     The most interesting example
     of this sort is connected with the
claim of the proof of the
existence of event horizon of the black holes
due to manifestation of the
ADAF regime of accretion (Narayan {\it et al.}, 1997).
     Analysis of a more complete
set of  observational data
(Chen {\it et al.}, 1997) had shown the
disappearance of the statistical effect
claimed as an evidence for ADAF.
This example shows how dangerous is to
base a proof of the theoretical
model on the preliminary observational data.
It is even more dangerous,
when the model is physically
not fully consistent.
Then even a reliable set of the observational
data cannot serve as a proof of the model.
   The classical example from
astrophysics of this kind gives
the theory of the origin of the elements
presented in the famous book
of Gamov (1952), where the model of
the hot universe was developed.
    In addition to rich advantages of this model,
the author also wanted to explain
the origin of heavy elements in the primordial
explosion, neglecting the
problems connected with an absence of the
stable elements with the number
of barions equal to $5$ and $8$.
    Gamov
considered the good coincidence
of his calculations, where the mentioned
problem was neglected, and the
observational curve, as a  proof of
his theory of the origin of the elements.
    Further developments
have shown that his outstanding
theory explains many things, but not
the origin of the heavy elements
which are produced due to stellar
evolution.

    It appears difficult explain
under-luminous AGNs with ADAF
solutions where the radiative
efficiency is much smaller than
for standard accretion disk models.
     Possible explanations include
the fact that
the mass accretion rates to
the black holes may have been
  overestimated.
    More importantly, other mechanisms
of  energy loss may be involved
in the accretion flows, for example,
in the form of
accelerated particles, as in
the radio-pulsars, where the energy
carried by the relativistic wind
far exceeds the electromagnetic
radiation (e.g., the Crab pulsar).
     Winds and/or jets are likely to happen in
the presence of a large scale
magnetic field which can have a key
role in their
the formation.
     To extend this,  we may  suggest that
under-luminous AGN loose significant
part of their energy to the formation of winds
and/or jets.
    A search for a correlation
between the occurrence of outflows or jets and low
luminosity could be very informative.
     [Note that the ADAF model of Blandford and
Begelman (1999) takes into account
outflows, but  it requires the same assumption as
earlier ADAF models that there is no
magnetic field and no
heating of electrons by field anihilation.  As
we have argued, these assumptions are implausible.]

    The behavior of accretion disks in the
presence of ordered and disordered magnetic
fields was discussed.
     The  interaction of an accretion
disk with a rotating magnetized star
was reviewed including the formation of
magnetohydrodynamic (MHD) outflows,
magnetic braking of the star's rotation, and
the propeller effect.
   The discussed problems of the MHD origin of jets
in both in the hydromagnetic
and Poynting flux regimes are currently under
active study.
   Both regimes are being studied
by different groups using axisymmetric and
more recently 3D MHD simulations.
    Many simulation papers have been published
on MHD outflows and jets,
but the results often represent the time
evolution for only a short interval (typically
the period of rotation at the inner edge of
the disk). These results are therefore strongly
dependent on the initial conditions which are
unknown.
    Furthermore, the boundary conditions
used in the simulation codes can give false
results, for example, collimation of an MHD
flow when there actually is no collimation
(see e.g. Ustyugova {\it et al.}, 1999).
    The difficulty of obtaining  generally valid
results from MHD simulations is even more
challenging in the case of relativistic MHD
(Komissarov, 1999) including
the Kerr  metric near a rapidly rotating
black hole (Koide {\it et  al.} 2000).
   Energy losses connected with
magnetohydrodynamic (MHD) outflows and jets
from the accretion disk may be very important
and represent an alternative explanation of
under-luminous AGN black holes.

\newpage

\section*{Figure captions}

Fig.1

\smallskip
\noindent
Geometry of a thin accretion disk from Shapiro and Teukolsky
(1983). Here, $\dot M$ is the accretion rate, $j$ is the total
angular momentum flux, and
$f_{\phi}=(1/r^2)\partial(r^2t_{r\phi})$ is the viscous force with
$t_{r\phi}=\eta \partial \Omega/\partial r$ the main component of
the viscous stress tensor and $\eta$ the dynamic viscosity
coefficient.

\bigskip\noindent
Fig.2

\smallskip
\noindent
The dependences of the optical
 depth $\tau_0$ on
radius, $r_*=r/r_{g}$, for the case $M_{BH}=10^8\;M_\odot$,
$\alpha=1.0$ and different values of $\dot m$. The thin solid,
dot-triple dash, long dashed, heavy solid, short dashed, dotted
and dot-dashed curves correspond to $\dot m=1.0, 3.0, 8.0, 9.35,
10.0, 11.0, 15.0$, respectively. The upper curves correspond to
the optically thick family, lower curves correspond to the
optically thin family.

\bigskip\noindent
Fig.3

\smallskip
\noindent
The specific angular momentum
$j_{\rm in} \equiv {\ell}_{\rm in}$
as a function of the mass accretion rate $\dot{M}$
for different viscosity parameters
$\alpha=0.01$ (squares), $0.1$ (circles) and $0.5$ (triangles),
corresponding to viscosity prescription (\ref{ref1.7}).
The solid dots represent models with the saddle-type
inner singular points, whereas the empty dots
correspond to the nodal-type ones,
from Artemova et al. (2001).

\bigskip\noindent
Fig.4

\smallskip
\noindent
The specific angular momentum
$j_{\rm in}\equiv {\ell}_{\rm in} $
as a function of the mass accretion rate $\dot{M}$
for different viscosity parameters
$\alpha=0.01$ (squares), $0.1$ (circles) and $0.5$ (triangles),
corresponding to viscosity prescription (\ref{ad2}).
The solid dots represent models with the saddle-type
inner singular points,
whereas the empty dots correspond to the nodal-type ones,
from Artemova et al. (2001).

\bigskip\noindent
Fig.5

\smallskip
\noindent
The position of the inner singular points
as a function of the mass accretion rate $\dot{M}$
for different viscosity parameters
$\alpha=0.01$ (squares), $0.1$ (circles) and $0.5$ (triangles),
corresponding to viscosity prescription (\ref{ref1.7}).
The solid dots represent models with the saddle-type
inner singular points, whereas
the empty dots correspond to the nodal-type ones,
from Artemova et al. (2001).

\bigskip\noindent
Fig.6

\smallskip
\noindent
The position of the inner singular points
as a function of the mass accretion rate $\dot{M}$
for different viscosity parameters
$\alpha=0.01$ (squares), $0.1$ (circles) and $0.5$ (triangles),
corresponding to viscosity prescription (\ref{ad2}).
The solid dots represent models with the saddle-type
inner singular points, whereas
the empty dots correspond to the nodal-type ones,
from Artemova et al. (2001).

\bigskip\noindent
Fig.7

\smallskip
\noindent
Poloidal projection
of the instantaneous
magnetic field in an accretion flow.

\bigskip\noindent
Fig.8

\smallskip
\noindent
Equatorial projection
of the magnetic field given by
equation (\ref{l6}) for an infinitely
conducting accretion flow.

\bigskip\noindent
Fig.9

\smallskip
\noindent
 Schematic drawing of the magnetic field configuration
considered by LRBK95.
    The magnetosphere consists of an
inner part, where the magnetic
field lines are closed and an outer part
where the field lines are open.
     The outer region where open field lines
thread the disk gives rise to a MHD
wind from the disk.

\bigskip\noindent
Fig.10

\smallskip
\noindent
This shows two poloidal field lines used in the
derivation of equation (\ref{adi2}).

\bigskip\noindent
Fig.11

\smallskip
\noindent
The figure shows the
dependence of $F_\infty$ and $F_K = (GMr_{cr})^{1/2}$
on the pulsar period $P$
assuming $r_{cr} = r_{to}$ for ${\dot M}= 10^{17}$g/s
and $4\times 10^{17}$
g/s.  The other parameters have been taken to be
$\alpha = 0.1$, $ D = 1$, $\tau_{max} = 5$, and $M =M_\odot$.
Thus, each ${\dot M}=const.$ curve corresponds to different magnetic
moments as $\mu \propto P^{1.17}$.
$F_\infty$
and $F_K$ are measured in units of $(GMr_o)^{1/2}$ with $r_o =
10^8$cm.
This plot allows a comparison of the angular momentum
influx to the star $(\dot M F_K)$ in the case of outflows and spin-up
$(r_{to} < r_{cr})$ with the case of magnetic braking where the
influx is $\dot M F_\infty (r_{to} > r_{rc})$.

\bigskip\noindent
Fig.12

\smallskip
\noindent
Geometry of disk accretion on
to a rapidly rotating star with an aligned
dipole magnetic field considered by
LRBK99.
   Here, $r_1$
and $r_2$ are the boundaries of the
region; $R_A$
is the effective radius where the
outflow begins;  $\omega_*$
is the star's rotation rate;  $\omega_K$
is the Keplerian rotation rate of the
accretion disk; and $r_{cr} \equiv
(GM/\omega_*^2)^{1/ 3}$
is the corotation radius.
   The magnetic field in the vicinity of $r_1$
has an essential time-dependence owing to the
continual processes of stellar flux leaking
outward into the disk, the resulting field
loops being inflated by the differential
rotation (LRBK95), and the reconnection between
the open disk field and the closed stellar
field loops.
  The dashed lines marked by the letters
`n'  indicate neutral surfaces along which
reconnection occurs.

\bigskip\noindent
Fig.13

\smallskip
\noindent
{\it Effective} Alfv\'en radius
$R_A$ (normalized by the {\it nominal}
Alfv\'en radius $r_A$)
as a function of the normalized
corotation radius $r_{cr}/r_A$ for
$\bar{\alpha}=0.2$ obtained
from equation (\ref{rmg12}).
  We have
neglected the $\delta^{3/2}=(R_A/r_2)^{3/2}$
terms compared
with unity and taken the value ${\cal F}_0
=0.234$ from LBC.
  Only the part of the solid curve above the
dashed line  $R_A=r_{cr}$
is consistent with our assumptions.
The dashed horizontal line $r_{to}$ indicates
the ``turnover radius'' of the disk rotation
curve or effective Alfv\'en radius in the regime
discussed by LRBK95 where $\dot{M}_{accr}$ falls onto
the star and the star spins-up.
  For this line the turbulent magnetic diffusivity
of the disk is taken to be $\alpha D = 0.1$.
The points $a~b~c~d$ and the associated
vertical lines represent possible transitions
between spin-down and spin-up of the pulsar as
discussed in the text.

\bigskip\noindent
Fig. 14

\smallskip
\noindent
Comparision of the observational and computational
spectra of the
cyclotron line. The solid curve is the observational results taken from
%\cite{o},
Mihara et al. (1990),
the dot curve is the approximation
by the comptonized
spectrum, and feature (17) with
 $a=7 \cdot 10^{-4}$~${\rm\frac{eV \cdot s }{cm}}$,
 $\sigma=2 \cdot 10^{-4}$~${\rm\frac{eV \cdot s }{cm}}$; from Baushev and
Bisnovatyi-Kogan (1999).

\bigskip\noindent
Fig.15

\smallskip
\noindent
  Comparison of the observational and computational X-ray spectra
of Her X-1. The solid curve is the observational results taken from
%\cite{p}
McCray et al. (1982),
the dot curve is the approximation
by the comptonized
spectrum, and the cyclotron feature (\ref{cyc17}) with
$T_s=0.9$~KeV, $T_e=8$~KeV, $\tau_e=14$,
$a=7 \cdot 10^{-4}$~${\rm\frac{eV \cdot s }{cm}} $,
$\sigma=10^{-4}$~${\rm\frac{eV \cdot s }{cm}} $, $B=4 \times 10^{10}$ Gs;
from Baushev and Bisnovatyi-Kogan (1999).

\bigskip\noindent
Fig.16

\smallskip
\noindent
Schematic structure of the accretion column near the
magnetic pole of the neutron star (top), and its radiation spectrum (bottom);
from Bisnovatyi-Kogan (1999a).

\bigskip\noindent
Fig.17

\smallskip
\noindent
Sketch of the magnetic field threading
an accretion disk shown increase of
the field owing to flux freezing in
the accreting disk matter from
Bisnovatyi-Kogan and Ruzmaikin (1976).

\bigskip\noindent
Fig.18

\smallskip
\noindent
Sketch of the electromagnetic outflows from
the two sides of the disk owing to
the Faraday unipolar dynamo action
of a rotating magnetized disk (Lovelace 1976).

\bigskip\noindent
Fig.19

\smallskip
\noindent
Sketch of an accretion disk
threaded by a magnetic field for
conditions which may lead to hydromagnetic
jet formation.

\bigskip\noindent
Fig.20

\smallskip
\noindent
Simulations results for stationary MHD outflow
obtained in a sphe\-rical coordinate system (Ustyugova et al.
1999).
  The solid lines represent the
poloidal magnetic field,
and the arrows the velocity vectors.
  The dashed lines in the left-hand
plot represent the slow magnetosonic surface (the lowest dashed
line), the Alfv\'en surface (the middle line), and the fast
magnetosonic surface (the top line). In the right-hand panel, the
dashed lines are surfaces of constant toroidal current density,
while the lines on the outer boundary are the projections of the
fast magnetosonic Mach cones.

\bigskip\noindent
Fig.21

\smallskip
\noindent
Dependences of different
velocities on distance $s$ measured
in units of $r_i$
from the disk along  the second
 magnetic field line away from
the axis in Figure 20 (Ustyugova
{\it et al.} 1999).
  This field line ``starts'' from
the disk  at $r \approx 6r_i$
where it has an angle
 $\theta \approx 28^o$ relative to
the $z-$axis.
   The velocities are measured in
units of $\sqrt{GM/r_i}$.
   Here, $v_p$ is the poloidal velocity along
the field line and
$v_\perp$ is the poloidal velocity perpendicular
to the field line.
   Also, $v_{Ap}$  is the
poloidal Alfv\'en velocity,  $c_{fm}$
is the fast magnetosonic velocity, and
$v_{esc}$ is the local escape velocity.

\bigskip\noindent
Fig.22

\smallskip
\noindent
 ``Initial'' dipole-like
vacuum
magnetic field.
   In this and subsequent plots,
$r$ and $z$ are measured in units
of the radius $r_0$ of the $O-$point
in the disk plane.
   The solid lines are the magnetic
field lines for the case where
the flux function on the disk surface is
$\Psi(r,z=0)=
a^3r^2 K/(a^2+r^2)^{3/2}$ with
$a=r_0/\sqrt{2}$.
    In this and subsequent
plots $\Psi$ is measured in units
of $\Psi_0=r_0^2 K/3^{3/2}$.
Note that $B_z(0) =
6\sqrt{3}\Psi_0/r_0^2
\approx 10.4\Psi_0/r_0^2$.
   The dashed lines are the field lines
for the case where the outer boundaries
($R_{max}=10,~Z_{max}=10$)
are perfectly conducting;
for this case
$\Psi(r,z=0) \rightarrow
\Psi(r,0)[1-(r/r_0-1)^2/81]$ so that the
$O-$point is still at $r_0$ and $\Psi_0$
is unchanged.
     Because of  axisymmetry
and reflection symmetry about the $z=0$ plane,
the field need be shown in only one
quadrant.

\bigskip\noindent
Fig.23

\smallskip
\noindent
Poloidal field lines  for Poynting jet
case  for twist
$T =1.84$ rad. and $(-H)_{max} =1.13\Psi_0/r_0$
with $\Psi=$const contours measured in
units of $\Psi_0$ which is the maximum
value of $\Psi$.
  The outer boundaries at $r=R_{max}$
and $z=Z_{max}$ are perfectly conducting
and correspond to an external plasma.
   This external plasma will
expand in response to the magnetic
pressure of the jet so that $R_{max}$
and $Z_{max}$ will increase with time.
    The initial
poloidal magnetic field is shown by the
dashed lines in Figure 22.
  The dashed contour is the separatrix
with the X-point indicated.
   Note that the radial width
of the upgoing
field lines along the axis is about
one-half the width of downgoing field
lines at the outer wall as required
for equilibrium.

\bigskip\noindent
Fig.24

\smallskip
\noindent
 Three dimensional view of two field lines originating
from the disk at $x=\pm 0.32~r_0$ ($\Psi =0.4\Psi_0$) for the
Poynting jet of Figure 23.
  Each field line has a twist of
$\approx 8.22$ rad. or about $1.31$
rotations about the $z-$axis from
its beginning at $r_1$ and end at $r_2$.
   The $z-$axis is tilted towards
the viewer by $30^\circ$.

\bigskip\noindent
Fig.25

\smallskip
\noindent
Time evolution of dipole-like field threading the disk
from the initial configuration $t=0$ (bottom panels) to the final
quasi-stationary state $t=1.2 t_{out}$, where $t_{out}$ is the
rotation period of the disk at the outer radius $R_{max}$ of the
simulation region.
   The left-hand
panels show the poloidal field lines which
are the same as $\Psi(r,z)=$const lines;
$\Psi$ is
normalized by $\Psi_{max}$ and the
spacing between lines is $0.1$.
  The middle panels show the poloidal
velocity vectors ${\bf v}_p$.
   The right-hand panels show the constant
lines of $-rB_\phi(r,z)>0$ in
units of $\Psi_{max}/r_0$ and the spacing
between lines is $0.1$.

\end{document}